\documentclass[12pt,a4paper]{article}

\pdfoutput=1
\usepackage{jheplike,ifpdf,tocloft,array,amsmath,mathrsfs,mathtools,xcolor,shuffle}
\graphicspath{{figures/}}

\widowpenalty=500\clubpenalty=1000\unitlength=1mm\textheight=8.7in\hypersetup{pdftitle={},pdfcreator={},linkcolor=[rgb]{0.15,0.35,0.75},colorlinks=true,citecolor=[rgb]{0.675,0,0.2},urlcolor=[rgb]{0.15,0.35,0.65}}
\let\olditemize\itemize\renewcommand{\itemize}{\vspace{-2pt}\olditemize\setlength{\itemsep}{1pt}\setlength{\parskip}{0pt}\setlength{\parsep}{-0pt}}
\let\oldenumerate\enumerate\renewcommand{\enumerate}{\vspace{-4pt}\oldenumerate\setlength{\itemsep}{1pt}\setlength{\parskip}{0pt}\setlength{\parsep}{0pt}}
\makeatletter\renewcommand\section{\addtocontents{toc}{\protect\addvspace{-2.25\p@}}\@startsection {section}{1}{\z@}{0.5ex \@plus .2ex \@minus 0.2ex}{0.3ex \@plus.1ex\@minus .5ex}{\normalfont\large\bfseries}}
\renewcommand\subsection{\addtocontents{toc}{\protect\addvspace{-2.5\p@}}\@startsection {subsection}{1}{\z@}{0.5ex \@plus .2ex \@minus 0.2ex}{0.3ex \@plus.1ex\@minus .5ex}{\normalfont\bfseries}}
\renewcommand\subsubsection{\addtocontents{toc}{\protect\addvspace{-2.5\p@}}\@startsection {subsubsection}{1}{\z@}{0.5ex \@plus .2ex \@minus 0.2ex}{0.3ex \@plus.1ex\@minus .5ex}{\normalfont\bfseries}}
\DeclareMathOperator*{\Res}{\mathrm{Res}}
\newcommand{\eq}[1]{\vspace{-0.5pt}\begin{equation}#1\vspace{-0.5pt}\end{equation}}

\newcommand{\fwbox}[2]{\text{\makebox[#1][c]{$\hspace{-150pt}\displaystyle#2\hspace{-150pt}$}}}
\newcommand{\fwboxL}[2]{\text{\makebox[#1][l]{$#2$}}}
\newcommand{\fwboxR}[2]{\text{\makebox[#1][r]{$#2$}}}
\newcommand{\equivR}{\fwbox{14.5pt}{\hspace{-0pt}\fwboxR{0pt}{\raisebox{0.47pt}{\hspace{1.25pt}:\hspace{-4pt}}}=\fwboxL{0pt}{}}}
\newcommand{\equivL}{\fwbox{14.5pt}{\fwboxR{0pt}{}=\fwboxL{0pt}{\raisebox{0.47pt}{\hspace{-4pt}:\hspace{1.25pt}}}}}
\newcommand{\fig}[3]{\raisebox{#1}{\includegraphics[scale=#2]{#3}}}
\newcommand{\bigger}[1]{\raisebox{-0.95pt}{\scalebox{1.25}{$#1$}}}
\renewcommand{\phi}{\varphi}

\renewcommand{\tilde}{\widetilde}
\newcommand{\ab}[1]{\langle #1\rangle}
\renewcommand{\sb}[1]{[ #1]}
\newcommand{\lsLabel}{\mathfrak{f}}
\newcommand{\intLabel}{\mathcal{I}}
\newcommand{\edgeA}{\text{{\footnotesize$a$}}}\newcommand{\edgeB}{\text{{\footnotesize$b$}}}\newcommand{\edgeC}{\text{{\footnotesize$c$}}}\newcommand{\edgeD}{\text{{\footnotesize$d$}}}\newcommand{\edgeE}{\text{{\footnotesize$e$}}}\newcommand{\edgeF}{\text{{\footnotesize$f$}}}\newcommand{\edgeG}{\text{{\footnotesize$g$}}}\newcommand{\edgeH}{\text{{\footnotesize$h$}}}
\newcommand{\brFour}[4]{\mathbin{\hspace{-1.5pt}\big[\hspace{-3.5pt}\big[#1,\!#2,\!#3,\!#4\big]\hspace{-3.5pt}\big]\hspace{-1.75pt}}}
\newcommand{\brSix}[6]{\mathbin{\hspace{-1.5pt}\big[\hspace{-3.5pt}\big[#1,\!#2,\!#3,\!#4,\!#5,\!#6\big]\hspace{-3.5pt}\big]\hspace{-1.5pt}}}

\newcommand{\brEight}[8]{\mathbin{\hspace{-1.5pt}\big[\hspace{-3.5pt}\big[#1,\!#2,\!#3,\!#4,\!#5,\!#6,\!#7,\!#8\big]\hspace{-3.5pt}\big]\hspace{-1.5pt}}}
\newcommand{\br}[1]{\mathbin{\hspace{-1.5pt}\big[\hspace{-3.5pt}\big[#1\big]\hspace{-3.5pt}\big]\hspace{-1.5pt}}}

\definecolor{varcolor}{rgb}{0.08,0.44,0.2}
\definecolor{functioncolor}{rgb}{0.08,0.28,0.6}
\newcommand{\vardef}[1]{{\color{varcolor}{\sl #1}\rule[-1.05pt]{7.5pt}{.75pt}}}
\newcommand{\defn}[3]{~\\[-35pt]\begin{itemize}\item[]\indent\hspace{-21pt}$\bullet$\hspace{-.75pt} {\tt {\color{functioncolor}#1}\![}#2{\tt\,]\!:\hspace{2pt}}#3\end{itemize}\vspace{-10pt}}
\newcommand{\defnNA}[3]{~\\[-35pt]\begin{itemize}\item[]\indent\hspace{-21pt}$\bullet$\hspace{-.75pt} {\tt {\color{functioncolor}#1}\!}#2{\tt\,\!:\hspace{2pt}}#3\end{itemize}\vspace{-10pt}}
\newcommand{\var}[1]{{\tt{\color{varcolor}{\sl#1}}}}
\newcommand{\fun}[1]{{\color{functioncolor}#1}}
\newcommand{\uscore}{\rule[-1.05pt]{7.5pt}{.75pt}}
\definecolor{hblue}{rgb}{0,0,0.575}
\definecolor{hred}{rgb}{0.575,0.0,0.225}
\definecolor{dim}{rgb}{0.55,0.55,0.55}
\definecolor{deemph}{rgb}{0.25,0.25,0.25}

\title{\texorpdfstring{~\\[20pt]Prescriptive Unitarity for Non-Planar\\[-0pt]Six-Particle Amplitudes at Two Loops\\[-20pt]}{Prescriptive Unitarity for Non-Planar Six-Particle Amplitudes at Two Loops}}
\author[1,2]{~\\[-26pt]Jacob~L.~Bourjaily,}%\email{bourjaily@nbi.ku.dk}
\affiliation[1]{Niels Bohr International Academy and Discovery Center, Niels Bohr Institute,\\University of Copenhagen, Blegdamsvej 17, DK-2100, Copenhagen \O, Denmark}
\affiliation[2]{Center for the Fundamental Laws of Nature, Department of Physics\\ Jefferson Physical Laboratory, Harvard University, Cambridge, MA 02138, USA}
\author[3]{Enrico~Herrmann,}%\email{eh10@stanford.edu}
\affiliation[3]{SLAC National Accelerator Laboratory, Stanford University, Stanford, CA 94039, USA}
\author[4]{Cameron~Langer,}%\email{cklanger@ucdavis.edu}
\affiliation[4]{Center for Quantum Mathematics and Physics (QMAP),\\Department of Physics, University of California, Davis, CA 95616, USA}
\author[1]{Andrew~J.~McLeod,}%\email{amcleod@nbi.ku.dk}
\author[4]{Jaroslav~Trnka}%\email{trnka@ucdavis.edu}
\abstract{
We extend the applications of prescriptive unitarity beyond the planar limit to provide local, polylogarithmic, integrand-level representations of six-particle MHV scattering amplitudes in both maximally supersymmetric Yang-Mills theory and gravity. The integrand basis we construct is diagonalized on a spanning set of non-vanishing leading singularities that ensures the manifest matching of all soft-collinear singularities in both theories. As a consequence, this integrand basis naturally splits into infrared-finite and infrared-divergent parts, with hints toward an integrand-level exponentiation of infrared divergences. Importantly, we use the \emph{same} basis of integrands for both theories, so that the presence or absence of residues at infinite loop momentum becomes a feature detectable by inspecting the cuts of the theory. Complete details of our results are provided as ancillary files to this work's submission to the \texttt{arXiv}.

This work has been updated to take into account the results of \cite{Bourjaily:2019gqu}, which leads to a simpler and more uniform representation of these amplitudes. To be clear: the representation given in \texttt{v1} on the \texttt{arXiv} are correct---merely less optimal.\\
\indent (The version given in \texttt{v2} inherited certain typographical mistakes.)}
\preprint{}

\begin{document}
\maketitle

%================================================================================================================
%    1. Introduction 
%         
%================================================================================================================
\newpage
\setcounter{page}{1}\vspace{15pt}
%\pagenumbering{roman}%\clearpage
\pagenumbering{arabic}
\vspace{-0pt}
\section{Introduction and Overview}\label{sec:introduction}\vspace{-0pt}
%================================================================================================================

Scattering amplitudes constitute fundamental objects in any weakly coupled quantum field theory, where they describe the basic interactions of elementary particles. While Feynman diagrams provide an algorithmic and broadly applicable recipe for the computation of these quantities, the Feynman expansion suffers from serious computational and conceptual problems---most significantly, the large number of diagrams that contribute to each scattering process, which individually depend on unphysical (gauge) degrees of freedom that cancel in the full amplitude. This deeply obscures the symmetries and mathematical simplicity of many known $S$-matrices. 

In recent decades, many new methods for perturbation theory have been developed, leading to enormous progress in our ability to compute scattering amplitudes and enhancing our understanding of the mathematical structure that underlies quantum field theory. These developments include generalized unitarity~\cite{Bern:1994zx,Bern:1994cg,Bern:1997sc,Britto:2004nc,Bern:2007ct}, on-shell recursion relations \cite{BCF,BCFW,ArkaniHamed:2010kv}, geometric descriptions of on-shell physics \cite{ArkaniHamed:2009dg,ArkaniHamed:2009dn,ArkaniHamed:2009sx,ArkaniHamed:2009vw,ArkaniHamed:2010gg,ArkaniHamed:2012nw,Arkani-Hamed:2013jha}, color-kinematic duality \cite{BCJ,Bern:2019prr}, scattering equations \cite{Cachazo:2013gna,Cachazo:2013hca,Cachazo:2013iaa,Cachazo:2013iea}, various bootstrap methods \cite{Bourjaily:2011hi,Dixon:2013eka,Golden:2014pua,Drummond:2014ffa,Bourjaily:2015bpz,Bourjaily:2016evz,Caron-Huot:2016owq,Almelid:2017qju,Caron-Huot:2018dsv,Henn:2018cdp}, and on and on. Much of this progress has been driven by concrete calculations beyond the reach of recent imagination. Time and time again, such computations have yielded unanticipated surprises and often shockingly simple results~\cite{Parke:1986gb,Bern:2005iz,Drummond:2006rz,Drummond:2008vq,Goncharov:2010jf,CaronHuot:2011ky}. 

A key insight underpinning a number of those developments was the realization that the computation of perturbative scattering amplitudes can be divided into two steps: that of `summing Feynman diagrams' to obtain a representative loop {\it integrand}---a rational differential form on the space of internal loop momenta; and that of tackling the more difficult problem of loop integration. In particular, the vast difference in difficulty between `easy' and `hard' loop integrals has motivated the development of integrand techniques that lead to easier loop integrals. Good integrand choices can additionally make important physical properties manifest. In fact, the structure of loop integrands has historically been a rich source of insight into the nature of scattering amplitudes in various theories. For example, the dual-conformal symmetry \cite{Drummond:2006rz,Alday:2007hr} (and its extension to the Yangian symmetry~\cite{Drummond:2008vq,Drummond:2009fd}) of maximally supersymmetric Yang-Mills (`sYM') theory in the planar limit was discovered first as a symmetry of loop integrands \cite{Drummond:2006rz}.

Among the most important strategies for obtaining and representing amplitude integrands is {\it generalized unitarity} \cite{Bern:1994zx,Bern:1994cg,Bern:1997sc,Britto:2004nc,Bern:2007ct}. The basic idea underlying this method is that loop integrands, as rational functions, should be constructible in terms of their multidimensional residues. In particular, residues of loop integrands---which put some subset of Feynman propagators on-shell---must factorize into products of lower-loop amplitudes (ultimately, trees). This idea becomes a powerful tool once it is realized that loop integrands can be represented in terms of any sufficiently large \emph{basis} of rational functions. Such integrand bases can therefore be constructed, studied, and integrated independently of any particular theory or scattering process. 

It has historically proven difficult to construct bases of integrands large enough to represent scattering amplitudes at high loop order or particle multiplicity. Indeed, surprisingly few examples are known beyond the planar limit of any theory. Today, even for the arguably simplest four-dimensional quantum field theories~\cite{ArkaniHamed:2008gz}---$\mathcal{N}=4$ sYM theory and $\mathcal{N}=8$ supergravity (`SUGRA')---the present state of the art is only five loops for four particles \cite{Bern:2012uc,Bern:2017ucb} and two loops for five particles \cite{Carrasco:2011mn,Bern:2015ple}. Compare this with the case of planar sYM amplitudes, for which we now have four-particle integrands through ten loops \cite{Bourjaily:2016evz}, local integrands through three loops for arbitrary multiplicity and helicity \cite{Bourjaily:2013mma,Bourjaily:2015jna,Bourjaily:2017wjl}, function-level results through seven loops for six particles \cite{Caron-Huot:2019vjl,Caron-Huot:2019bsq}, and symbols for four-loop seven-particle amplitudes \cite{Dixon:2016nkn,Drummond:2018caf}.

In this work, we construct a four-dimensional integrand basis for six-particle scattering at two loops in (nonplanar) sYM and SUGRA.\footnote{Note that a global definition of `\emph{the} nonplanar integrand' is problematic due to the lack of global labels (see \cite{Ben-Israel:2018ckc,Tourkine:2019ukp} for recent attempts to remedy this problem). When we refer to the integrand in a nonplanar setting, we mean the knowledge of all coefficients in the decomposition (\ref{schematic_loop_integrand_formula}).} We use the strategy outlined in \cite{integrandBases}, but also take advantage of a crucial simplification that last occurs at this multiplicity---it is possible to construct a basis of local integrands that are manifestly polylogarithmic. In particular, we construct a basis of loop integrands $\{\mathcal{I}_i(\ell_1,\ell_2)\}$ large enough to represent both theories simultaneously, deriving a representation of (the integrand of) these amplitudes taking the form 
\vspace{-2pt}\eq{\mathcal{A}_{6,\text{MHV}}^{\mathcal{N}=4,8}=\sum_{i}{\color{hred}\lsLabel_i^{\text{\scriptsize$\,\mathcal{N}$}}}\mathcal{I}_{i}+\text{permutations}\,.\label{schematic_loop_integrand_formula}\vspace{-6pt}}
In both cases, the coefficients ${\color{hred}\lsLabel_i}$ correspond to the same set of cuts---just interpreted in two different theories. (We will be more explicit about what is meant by the sum over permutations in \mbox{section \ref{subsec:final_results}}.)

In addition to being individually polylogarithmic (i.e.~all integrands are individually `$d\!\log$' as defined in \cite{Arkani-Hamed:2014via,Bern:2014kca}), we have engineered our basis of integrands to be {\it pure} as defined in \cite{ArkaniHamed:2010gh}. That is, all basis elements have unit leading singularities on all co-dimension eight residues and are expected to evaluate to maximal transcendental weight functions after integration. Moreover, many of the integrands in our basis---those not directly defined to support residues associated with soft-collinear divergences---are free of any infrared divergences. As such, we expect this representation to be well suited for integration. 

Another desirable feature of our integrand basis $\{\mathcal{I}_i\}$ is that it is normalized and diagonalized according to a spanning set of `leading singularities' \cite{Britto:2004nc,Cachazo:2008vp}. In terms of such a basis, the coefficients needed to represent amplitudes in any theory are simply the relevant leading singularities ${\color{hred}\lsLabel_i^{\text{\scriptsize$\,\mathcal{N}$}}}$ in the corresponding theory. (For less supersymmetric theories, our basis would have to be extended; however, we can always choose the integrands presented in this paper to represent a subspace of the larger basis.) The spanning set of leading singularities we have chosen is enumerated in \mbox{table \ref{six_point_ls_table}}. Details and explicit definitions for each of these functions are collected in \mbox{appendix \ref{appendix:explicit_coefficients}}: the color-dressed on-shell functions in sYM are described in \ref{appendix:ls_of_sym}, and formulae for all SUGRA leading singularities are defined in \ref{appendix:ls_of_sugra}. 

As an example, integrand $\mathcal{I}_{10}$ in our basis involves a collection of Feynman propagators with the topology
\eq{\mathcal{I}_{10}\;\bigger{\Leftrightarrow}\fwbox{80pt}{\fig{-28.5pt}{1}{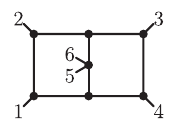}}\quad\text{and coefficient}\quad{\color{hred}\lsLabel_{10}^{\text{\scriptsize$\,\mathcal{N}$}}}\;\bigger{\Leftrightarrow}\fwbox{80pt}{\fig{-28.5pt}{1}{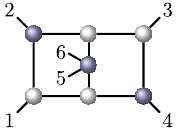}}\,.
\label{illustration_of_terms_in_rep}}
This basis element is theory-independent; it is normalized to have a co-dimension eight residue of unit magnitude associated with exactly one of the solutions to cutting all eight of its propagators, and to vanish on the defining cuts of the remainder of the integrand basis. Depending on the context, ${\color{hred}\lsLabel_{10}^{\text{\scriptsize$\,\mathcal{N}$}}}$ denotes the leading singularity in either sYM or SUGRA, built from the tree amplitudes in the corresponding theory. In the former case, these tree amplitudes are fully dressed by color-factors built from structure constants of the gauge group; in the latter case, the trees are simply those of supergravity. We will review these ingredients below, with full details provided in \mbox{appendix \ref{appendix:explicit_coefficients}}. 

Before moving on, it is worth highlighting several key aspects of the representation (\ref{schematic_loop_integrand_formula}). It has been conjectured that MHV amplitudes in sYM are polylogarithmic and free of poles at infinity to all multiplicity and loop-order \cite{Arkani-Hamed:2014via}; it is also known that starting at two loops, amplitudes in SUGRA have poles at infinity with degree growing with multiplicity, starting for six particles \cite{Bourjaily:2018omh}. The basis of integrands we construct here has term-wise support on poles at infinity (analogous to the representation of one-loop integrands in \mbox{ref.\ \cite{Bourjaily:2013mma}}); these residues at infinity cancel via global residues theorems in sYM, while the non-vanishing residues of SUGRA are indirectly matched by the \emph{same} global residue theorems (using the fact that we have matched all the residues at finite loop momentum). 

We should emphasize that the representation we have derived is strictly four-dimensional. If the infrared singularities are regulated by going to the Coulomb branch of the theory, then nothing about our representation would need to change (as the $\mathcal{O}(m^2)$ corrections to the integrand coefficients would vanish in the massless limit). This is not the case for dimensional regularization, however, where the basis must be enlarged \emph{and} the coefficients must be changed. There does not yet exist any \emph{simple} way to upgrade a four-dimensional integrand to find something suitable for dimensional regularization, but lessons from the current state-of-the-art, including the recent examples of two-loop five particle amplitudes in sYM and SUGRA \cite{Abreu:2018aqd,Chicherin:2018old,Chicherin:2018yne,Chicherin:2019xeg,Abreu:2019rpt}, suggest the usefulness of starting from a partial basis of integrands whose differential equations are in canonical form \cite{Henn:2013pwa,Henn:2014qga}. Thus, we expect our result to generalize to $4-2\epsilon$ dimensions in a similarly nice manner, but leave that to future work.

Our six-particle results represent a natural step forward in complexity, following recent developments at two loops for five particles \cite{Badger:2017jhb,Abreu:2017hqn,Abreu:2018jgq,Gehrmann:2015bfy,Badger:2018enw,Abreu:2018zmy,Abreu:2018aqd,Chicherin:2018yne,Chicherin:2018old,Chicherin:2019xeg,Abreu:2019rpt}. It furthermore opens up avenues for exploring a potential extension of dual conformal symmetry beyond the planar limit \cite{Arkani-Hamed:2014via,Bern:2014kca,Bern:2015ple,Bern:2017gdk,Bern:2018oao,Chicherin:2018wes} as well as an extension of the empirical second entry conditions for five-particle amplitudes \cite{Chicherin:2017dob}. \\

This paper is organized as follows. In \mbox{section \ref{sec:unitarity}} we review the basic principles of generalized unitarity and its refinement in the guise of \emph{prescriptive unitarity}. In particular, we discuss on-shell functions in \mbox{section \ref{subsec:general_cuts}}, and summarize how these can be used to fix the coefficients for a given integrand basis in the unitarity based expansion of amplitudes. The construction of unitarity bases for non-planar theories is somewhat beyond the scope of this work (see the forthcoming work of \mbox{ref.\ \cite{integrandBases}}), but we outline the broad themes in \mbox{section \ref{subsec:generalized_unitarity}}, and how the particular integrand basis needed for six-particle amplitudes in sYM and SUGRA was constructed in \mbox{section \ref{subsec:chiral_integrands}}. Our main results are summarized in \mbox{section \ref{sec:the_result}}, together with a discussion of its most salient features. More complete details of our results can be found in the appendices. Specifically, each of the explicit numerators defining the basis integrands are given in \mbox{appendix \ref{appendix:explicit_numerators}} and closed formulae are provided for all the required leading singularities of sYM and SUGRA in \mbox{appendix \ref{appendix:explicit_coefficients}}.

In addition to these details, we have included with our submission to the {\tt arXiv} ancillary files that provide fully usable expressions for all our ingredients. We have prepared plain-text definitions of each integrand and coefficient needed for the representation of two-loop six-point MHV amplitudes in sYM and SUGRA, and we have provided additional functionality for {\sc Mathematica} users. The detailed organization and content of these ancillary files are described in \mbox{appendix \ref{appendix:ancillary_files}}.

%================================================================================================================
\newpage\vspace{-0pt}\section{Principles of Generalized and Prescriptive Unitarity}\label{sec:unitarity}\vspace{-0pt}
%================================================================================================================

Our results for the two-loop six-particle MHV amplitudes in $\mathcal{N}=4$ sYM and $\mathcal{N}=8$ SUGRA follow from the basic principles of generalized unitarity \cite{Bern:1994cg,Bern:1997sc,Britto:2004nc,Bern:2007ct}, and its refined form of \emph{prescriptive} unitarity \cite{Bourjaily:2017wjl}. Although these methods are well known, it is worthwhile to briefly review both the basic ingredients and the philosophy guiding the prescriptive representation we have found for six-particle amplitudes. Readers already familiar with these ideas can safely skip to \mbox{section \ref{subsec:chiral_integrands}}, where we describe the essential strategy behind building prescriptive integrand bases for non-planar theories, and introduce some notational details. For a more thorough discussion of the ideas underlying generalized unitarity, we suggest e.g.~\mbox{refs.\ \cite{Dixon:1996wi,Bern:2011qt,Elvang:2013cua,Henn:2014yza,Dixon:2015der}}.

This section is organized as follows. After briefly introducing the starring characters in this story---\emph{on-shell functions}---in \mbox{section \ref{subsec:general_cuts}}, we outline the principles of generalized and prescriptive unitarity in \mbox{section \ref{subsec:generalized_unitarity}}, and discuss technicalities and strategies involved in applying these ideas to (especially MHV) amplitudes in non-planar quantum field theories in \mbox{section \ref{subsec:chiral_integrands}}.\\

%================================================================================================================
\vspace{-0pt}\subsection{\emph{On-Shell Functions}: the Residues of Loop Amplitudes}\label{subsec:general_cuts}\vspace{-0pt}
%================================================================================================================

The most important idea involved in unitarity-based methods is that, for any local quantum field theory, the only poles arising in the Feynman diagram expansion correspond to putting some states \emph{on-shell}. For any unitary quantum field theory, the residues on such poles are dictated by on-shell scattering processes, which involve only the physical states that can propagate along these on-shell lines, and in particular involves summing over a complete set of states that can be exchanged. This basic idea becomes a powerful tool when one realizes that, at the level of the integrand, such residues only involve lower-loop or lower-multiplicity scattering processes. 

Consider, for instance, the famous unitarity cut involving just two on-shell propagators \cite{Cutkosky:1960sp} where the loop amplitude factorizes into two lower-loop on-shell processes:
\vspace{-6pt}\eq{\underset{\scriptstyle (\ell{+}Q)^2{=}0 \atop\scriptstyle \ell^2{=}0}{\text{Res}}\left[\fig{-29pt}{1}{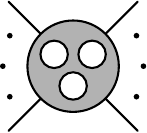}\right]=\sum_{\scriptstyle L{=}L_1{+}L_2{+}1 \atop \scriptstyle \text{states}}\fig{-32pt}{1}{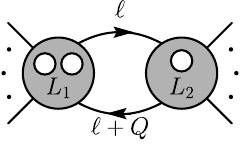}\vspace{-5pt}}
This cutting procedure need not stop here---further residues can be taken until all vertices correspond to trees.\footnote{This happens for any theory of massless particles in four dimensions, as the on-shell three point amplitude cannot be corrected in perturbation theory, and is therefore fixed at tree-level. Similar statements also exist for massless theories in any number of dimensions \cite{Baadsgaard:2015twa}.} For example, for a three-loop amplitude we could iterate the cutting procedure until we find diagrams such as
\vspace{-2pt}\eq{\fig{-17.125pt}{1}{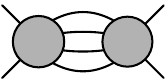}\;\;\; \fig{-17.125pt}{1}{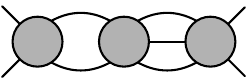}\;\;\;\fig{-17.125pt}{1}{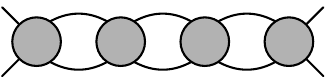}\;,\label{fig:basic_factorizations_egs}\vspace{-2pt}}
among others.

Pictures such as those in (\ref{fig:basic_factorizations_egs}), in which vertices represent amplitudes and every edge represents an on-shell physical state, are called \emph{on-shell functions}. They are defined to be the product of the amplitudes at each vertex, summed over all the internal states that can propagate at the edges. This state sum involves an integration over the on-shell phase space of the internal particles (often localized by momentum conservation at the vertex amplitudes). See \mbox{ref.\ \cite{ArkaniHamed:book}} for more details. Since all relevant building blocks entering the on-shell functions described above are tree-level amplitudes, they can in principle be computed in any quantum field theory. In particular, using modern tools such as BCFW recursion relations \cite{BCF,BCFW} (as implemented in e.g.\ \cite{Dixon:2010ik,Bourjaily:2010wh,Bourjaily:2012gy}), these can be computed relatively easily to high multiplicity. 

One aspect that is still quite poorly understood is the role of color factors in the classification of on-shell functions. To be clear: there is absolutely no obstruction (or even subtlety) in the definition of on-shell functions for amplitudes with non-trivial color (in any representation of any gauge group); however, the interplay between these color factors and the kinematic functions that they decorate is still largely uncharted territory (see, however, \cite{Ochirov:2016ewn,Ochirov:2019mtf} for recent work). 

In the current work, we take a pragmatic view of color factors, described in detail in \mbox{appendix \ref{appendix:ls_of_sym}}. Specifically, being interested in a theory with all adjoint states, we associate to each particle a color matrix $(T^a)_{bc}\equivR f^{abc}$ in the adjoint representation of some Lie algebra with structure constants $f^{abc}$, and simply trace over the color indices associated with all internal on-shell particles. This results in color factors that can be encoded as graphs built out of structure constants. Once a particular gauge group is specified, it is straightforward to expand these graphs into more familiar representations---such as the $1/N_c$ or multi-trace expansion of $SU(N_c)$ gauge theory \cite{Dixon:1996wi}. But this is by no means the only gauge group we might consider, and we are not limited to any particular order in the $1/N_c$ expansion (except insofar as this expansion truncates at any fixed loop order).\footnote{It is worth contrasting this situation with the very interesting work of \mbox{ref.\ \cite{Ben-Israel:2018ckc}} which constructs an interesting and novel representation of non-planar multiloop integrands for sYM, but very specifically within the multi-trace framework of the $1/N_c$ expansion.}

Finally, we should mention the special role played by a particular class of on-shell functions called \emph{leading singularities} \cite{Britto:2004nc,Cachazo:2008vp}. These on-shell functions correspond to maximal-codimension residues of loop amplitude integrands. In $d$ spacetime dimensions, these on-shell functions either involve $dL$ internal on-shell propagators, or $dL$ residues from localizing fewer propagators but cutting additional Jacobian poles (composite residues). Examples of these in four dimensions at one and two loops relevant to MHV amplitudes in four dimensions include:
\eq{\fig{-27.125pt}{1}{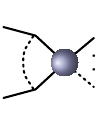}\hspace{40pt}\fig{-27.125pt}{1}{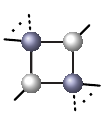}\hspace{40pt}\fig{-27.125pt}{1}{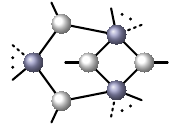}\label{example_ls}}
In the first of these, the dashed line signals that in addition to three internal propagators being cut, the momentum flowing through that edge should be set to zero (making the number of constraints equal to four) (see e.g.\ \cite{Buchbinder:2005wp}). For leading singularities supporting MHV helicity configurations, it is not hard to see that the only helicity amplitudes allowed at the vertices are either MHV or $\overline{\text{MHV}}$---the latter involving at most three particles. By convention, these are colored in our figures by blue and white vertices, respectively. 

A fair amount is known about the leading singularities of sYM beyond the planar limit, as well as in SUGRA. Like any other massless theory in four dimensions, on-shell functions have a representation in terms of the Grassmannian \cite{ArkaniHamed:2012nw,Arkani-Hamed:2014bca,Elvang:2014fja,Huang:2013owa,Bourjaily:2016mnp,Herrmann:2016qea}. This opens up certain geometric interpretations and powerful computational tools. Although much is known about how these functions are classified for MHV amplitudes in sYM \cite{Arkani-Hamed:2014bca}, considerably less is understood beyond this case (see, however,~\cite{Bourjaily:2016mnp}). 

It turns out that \emph{leading singularities} alone suffice for the representation of six-particle amplitudes in sYM and SUGRA for the simple reason that a complete basis of integrands for these processes exists that is purely polylogarithmic. Beyond six particles, however, this is not true, even in the MHV sectors of sYM and SUGRA. To understand this subtlety, it is worth briefly describing how integrand bases can be constructed, how the degrees of freedom in these basis can be fixed according to cuts, and when these cuts can be taken to be \emph{leading} singularities.

%================================================================================================================
\newpage\vspace{-0pt}\subsection{Generalized and Prescriptive Unitarity at Two Loops}\label{subsec:generalized_unitarity}\vspace{-0pt}
%================================================================================================================

Once it is realized that perturbative scattering amplitudes, viewed as loop \emph{integrands}, are rational differential forms of internal and external momenta, it becomes clear that they can be expanded into any sufficiently large but otherwise arbitrary \emph{basis} of loop integrands:
\eq{\mathcal{A} = \sum_i \mathfrak{c}_i\, \mathcal{I}^0_i\,.\label{schematic_unitarity}}
Such an expansion exists both before and after loop integration---where, in the latter case, it should really be viewed as a statement about a cohomology basis for rational forms on loop momenta \cite{Ita:2015tya,Georgoudis:2015hca}. We take the former view, in which the basis of integrands (\ref{schematic_unitarity}) is simply an independent set of rational functions. 

One way to construct such a basis $\{\mathcal{I}^0_i\}$ of integrands at $L$ loops is to simply write down all Feynman diagrams in a given theory, and consider the space spanned by the loop-dependent parts of each. (These integrands will rarely be independent.) Interestingly, the size of the space required to represent amplitudes in different theories varies considerably, and in general the problem of constructing a basis of integrands just large enough to represent the amplitudes in a given theory is an interesting and important one. However, a separate and arguably more important question is how to find a \emph{good} set of basis integrands. Let us briefly discuss the role of these questions in the context of sYM.

In the planar limit of sYM, there exists a simple guiding principle for constructing sufficiently large integrand bases: \emph{dual-conformal invariance} \cite{Drummond:2006rz,Bern:2006ew,Bern:2007ct,Alday:2007hr,Drummond:2008vq}. Dual conformal invariance is closely related to what is colloquially described as the `power-counting' of the theory: a planar loop integrand $\mathcal{I}$ is said to scale like a $p$-gon at infinity if
\eq{\lim_{\ell_i\to\infty}\mathcal{I}=\frac{1}{(\ell_i^2)^p}\Big(1+\mathcal{O}(1/\ell_i^2)\Big)\,,\label{planar_power_counting}}
for every loop momentum indexed by $i=1,\ldots,L$, when these loop momenta are defined by the planar dual of the integrand's graph of propagators. It is known that amplitudes in planar sYM are dual conformally invariant, and manifest dual conformal invariance requires (among other important conditions) that integrands have 4-gon (`box') power-counting. It is not terribly difficult to construct the space of all $L$-loop planar integrands with box power-counting, and to use this basis to represent amplitudes in planar sYM which led to results up to ten loops \cite{Bourjaily:2016evz}.

Beyond the planar limit, however, no simple definition of `power-counting' exists owing to the fact that there is no preferred \emph{routing} of loop momenta---no preferred choices of origins for the $L$ internal loop momenta, or for how these loop momenta should flow through the graph. 

As some of the authors will describe in a forthcoming work \cite{integrandBases}, there does exist a graph-theoretic definition of power-counting that allows us to enumerate a rigorously well-defined subspace of Feynman integrands (which may or may not be large enough to represent amplitudes in a given theory). The essential idea is to reinterpret (\ref{planar_power_counting}) graph-theoretically: a pair of loop integrands have the same `power-counting' if their leading terms behave graph-isomorphically as all their loop momenta are taken to infinity. With this, we may define a basis of non-planar integrands with `$p$-gon power-counting' to be the space of all integrands that behave like a specified set of `scalar $p$-gon integrands' at infinity (or better). The precise definitions of these bases are somewhat involved and go beyond the needs of this work; we defer a more thorough discussion to \mbox{ref.\ \cite{integrandBases}}. But let us qualitatively describe the space of two-loop integrands required to represent six-particle amplitudes in sYM and SUGRA. 

At two loops, a basis of `triangle power-counting' integrands may be defined as the space of all loop integrands that scale like one of the following scalar triangle power-counting integrands
\vspace{-2pt}\eq{\Bigg\{\fig{-24.5pt}{1}{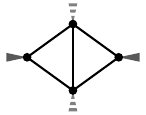}\;,\fig{-24.5pt}{1}{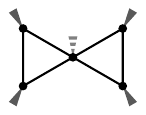}\Bigg\}\label{scalar_triangles_at_two_loops}\vspace{-2pt}}
(or better) at infinite loop momentum. For example, the integrand 
\vspace{-2pt}\eq{\fig{-24.5pt}{1}{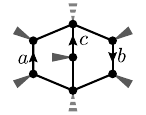}\;\; \text{with numerator}\;\;\; \mathfrak{n}=(a+q_1)^2(b+q_2)^2(c+q_3)^2\,,\vspace{-2pt}}
scales like the first scalar integral in (\ref{scalar_triangles_at_two_loops}), for \emph{any} loop-independent momenta $q_i$. It is straightforward to generalize this example to construct spaces of numerators for arbitrary graphs that contain either of the scalar triangles in (\ref{scalar_triangles_at_two_loops}) as contact terms.

Notice that our definition of power-counting is entirely graph-theoretic (and thus independent of how the internal loop momenta are routed through the graph) as well as dimensionally agnostic. The spacetime dimension appears only in the question of how many independent degrees of freedom exist for these numerators---and how many of these degrees of freedom may be spanned by their contact terms.

Specializing to four dimensions, we can compute the ranks of all these vector spaces of numerators and enumerate the irreducible degrees of freedom that must be fixed. The complete list of two-loop integrand topologies with irreducible triangle power-counting numerators is enumerated in \mbox{table~\ref{four_dim_triangle_power_counting_basis}}. 

Before we describe the much harder problem of finding (reasonably) good representatives of each required integrand, it is worth briefly discussing why we have chosen to use a \emph{triangle} power-counting basis to represent six-point amplitudes in sYM---as, na\"{i}vely, one would expect the theory to be expressible in terms of integrands with box power-counting.

\begin{table}[b!]\centering\vspace{-30pt}$$\vspace{0pt}\fig{-140pt}{1.65}{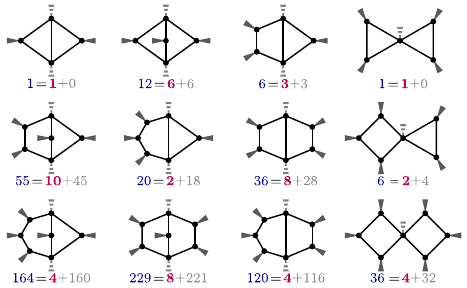}$$\vspace{-28pt}\caption{A complete list of irreducible (in $4d$) integrand topologies consistent with `triangle power-counting' as defined and described in \mbox{ref.\ \cite{integrandBases}} and the ranks of the degrees of freedom for their numerators. In {\color{hblue} blue} are listed the total rank of numerators, and in {\color{hred} red} are listed just the rank spanned by numerators \emph{modulo contact-term degrees of freedom}.}\label{four_dim_triangle_power_counting_basis}\vspace{-26pt}\end{table}
%

%============================================================================
\vspace{-5pt}\paragraph{Why have we chosen our basis to have \emph{triangle} power-counting?}~\\[-16pt]
%============================================================================

Considering that we know that MHV \cite{Bourjaily:2018omh} amplitudes in sYM are free of poles at infinity through three loops (and suspect this to hold to all loops), it may be surprising that we have chosen \emph{triangle} power-counting integrands to form our basis, as such integrands almost always have support on poles at infinity. There are two main reasons why we have made this choice. 

The first reason is that we would like to be able to have a basis large enough to represent \emph{both} sYM \emph{and} SUGRA, and we know that, starting at six points, amplitudes in SUGRA \emph{do} support residues at infinity \cite{Bourjaily:2018omh}. Thus, at least some integrands with triangle power-counting are required in our basis (and completeness requires that we consider them all). 

Although the above motivation is a good one, it may not actually be the most important to us. Indeed, we strongly suspect that it will be much easier to represent amplitudes in sYM in terms of integrands with triangle power-counting than using those which behave like boxes (at least until a better definition of a `box power-counting' basis exists). The reason for this is a bit subtle and involves two unfortunate aspects of (any existing definition of) box power-counting in four dimensions. Luckily, both problems are already familiar at one loop where, in retrospect, these problems were the motivation behind the introduction of triangles into the `chiral box' expansion described in \mbox{ref.\ \cite{Bourjaily:2013mma}}. 

The two main problems with constructing a basis of integrands with box power-counting in four dimensions are that: first, integrands with more than $4L$ propagators are required to form a basis---and such integrands typically have many more leading singularities than they have degrees of freedom (leading to many delicate choices, see e.g. \cite{Bourjaily:2017wjl}); and second, such an integrand basis is topologically over-complete. By this we mean that not all integrands of a given topology are independent. At one loop, for example, there are six pentagon integrands, but these pentagons are related due to the existence of Gram determinant constraints in four dimensions. Thus, the construction of a (not over-)complete basis requires choosing only five of the six pentagons to appear. These types of choices can of course be made, but lead to additional undesirable complexity in the expansion coefficients.

%\newpage
%============================================================================
\vspace{-5pt}\paragraph{Choosing Bases and Finding Integrand Coefficients}~\\[-16pt]
%============================================================================

Let us suppose that some initial (perhaps far from optimal) basis of integrands is known. Provided the theory is cut-constructible \cite{Bern:1996je} (which is true for all theories in dimensional regularization via $d$-dimensional cuts \cite{Anastasiou:2006jv}), then all the coefficients $\mathfrak{c}_i$ in (\ref{schematic_unitarity}) would be determined by a linear algebra problem of matching cuts with field theory. Depending on the initial choice of basis, the coefficients $\mathfrak{c}_i$ could easily be large sums of on-shell functions with (in general, algebraic) prefactors arising from the Jacobians of the basis integrands on those cuts. 

A minimal set of cuts sufficient to determine all coefficients is called a \emph{spanning set} of cuts. Given any spanning set of cuts, it is possible to diagonalize the integrand basis, resulting in (typically) a \emph{much} better one. The virtues of such bases were described in \mbox{ref.\ \cite{Bourjaily:2017wjl}}, where they were called `prescriptive' due to the fact that in such a basis the coefficients $\mathfrak{c}_i$ are trivially just the cuts used to define the basis.

Whenever there exists a spanning set of \emph{leading singularities}---residues of maximal codimension---then the amplitude has a $d\log$ form. Note that integrals in $d\log$ forms are conjectured \cite{ArkaniHamed:2012nw} to be associated with polylogarithmic functions of maximal weight (see e.g.~\cite{Herrmann:2019upk} for recent progress on integrating $d\log$ forms). A natural question is: when can this happen? Can cuts always be chosen to be leading singularities? The answer, unfortunately, is: no. Consider for example any integrand basis which includes the scalar double-box:
\vspace{-16pt}\eq{\mathcal{I}_{\text{db}}\equivR\hspace{-10pt}\fig{-29.25pt}{1}{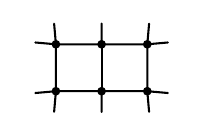}\,.\vspace{-16pt}\label{elliptic_double_box}}
Such an integrand is required to represent scalar $\varphi^4$-theory, but it also represents a component amplitude in planar sYM \cite{CaronHuot:2012ab}. How is its coefficient matched? One can show that no leading singularities of $\mathcal{I}_{\text{db}}$ exist, as cutting all 7 propagators (in $d{=}4$) results in an elliptic integral \cite{CaronHuot:2012ab,Bourjaily:2017bsb} over the remaining degree of freedom `$x$':
\vspace{-16pt}\eq{\fig{-29.25pt}{1}{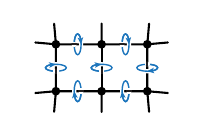}\hspace{-10pt}=\int\!\!\!\frac{dx}{y}\,,\quad\text{where}\quad y^2=Q(x)\vspace{-16pt}}
and $Q(x)$ is an irreducible quartic.

To be clear, the existence of non-polylogarithmic integrals (those without leading singularities) does not in any way prevent us from including integrands such as (\ref{elliptic_double_box}) in bases, nor from matching their coefficients. Indeed, as motivated (and exploited) in \mbox{ref.\ \cite{Bourjaily:2015jna}}, these degrees of freedom can simply be fixed by matching the integrand at any sufficient number of points along the cut surface. 

In recent years, a number of researchers have started exploring amplitudes and individual integrals outside the world of polylogarithms (see, for example, \cite{Adams:2013kgc,Bloch:2014qca,Bloch:2016izu,Adams:2016xah,Hidding:2017jkk,Adams:2018bsn,Broedel:2019hyg,Broedel:2019kmn}).  Associated to these cases are integrands without leading singularities, which turn out to be a ubiquitous feature of almost all amplitudes in almost all quantum field theories \cite{Sabry,Broadhurst:1987ei,Laporta:2004rb,Czakon:2008ii,Bourjaily:2017bsb,Bourjaily:2018ycu,Bourjaily:2018yfy} for high enough loop order, massive particles, or sufficiently many external legs. In massless theories, these issues arise earlier beyond the planar limit. To see this, consider the two loop cut surfaces associated with the integrands
\vspace{-3pt}\eq{\fig{-32.125pt}{1}{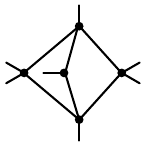}\quad\text{and}\quad\fig{-32.125pt}{1}{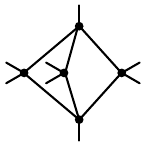}\label{polylog_obstructions}\vspace{-3pt}}
for seven and eight particles, respectively. While the first case is infrared divergent, the maximal co-dimension residue of the integral is an elliptic integral, suggesting some degree of rigidity (in the sense of \mbox{refs.\ \cite{Bourjaily:2017bsb,Bourjaily:2018ycu,Bourjaily:2018yfy}}); in the latter case, the integral is known to be finite, and to involve an integral over a K3 surface---a Calabi-Yau two-fold \cite{Bourjaily:2018yfy}. Careful readers will notice that neither of the maximal cuts of (\ref{polylog_obstructions}) have MHV support; as such, these cuts should vanish for all MHV amplitudes. Nevertheless, these cut surfaces have irremovable support within integrands
\eq{\fig{-32.125pt}{1}{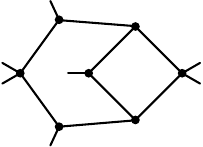}\;\bigger{\supset}\fig{-32.125pt}{1}{7pt_tardigrade}\;,\quad\fig{-32.125pt}{1}{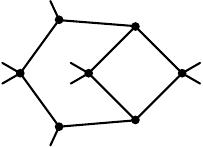}\;\bigger{\supset}\fig{-32.125pt}{1}{8pt_tardigrade}\,;\label{integrands_which_contain_cys}}
that {\it do} have cuts which are non-vanishing for even MHV amplitudes. We must therefore include terms such as (\ref{polylog_obstructions}) in any basis large enough to represent amplitudes in sYM (or any less supersymmetric theory, for that matter). From this argument, we expect that, beyond the planar limit and at two loops or higher, six particle amplitudes are the last for which we can maintain manifest polylogarithmicity together with a local basis of integrands.

There is one final subtlety regarding the integrand basis construction that should be considered. Even for integrands without elliptic curves or K3 surfaces in their cut structure, it may not be possible to find a spanning set of leading singularities to fix their coefficients. For six particles at two loops, the scalar integral (loop-momentum independent numerator)
\eq{\fig{-24.725pt}{1}{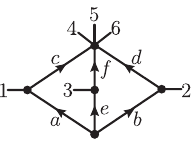}\;\,,\label{222example}}
is a simple example of an integrand without any leading singularities. Consider the collinear cut parameterized by $a=\alpha\,p_1$ and $b=\beta\,p_2$, 
\eq{\fig{-24.725pt}{1}{int_222_example}\;\longmapsto\;\fig{-24.725pt}{1}{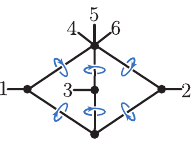}\;= \frac{1}{\alpha\beta(\alpha s_{13} + \beta s_{23} + \alpha\beta s_{12})}\,.\label{hexacut_of_222_example}}
Upon taking any further residue we find a double pole. Indeed, this arises in all possible cut-pathways, signaling that the integral (\ref{222example}) does not have \emph{any} leading singularities. The existence of such double poles is a signal that the integral (\ref{222example}) shows a transcendental weight drop, confirming the known result \cite{Gehrmann:2001ck,Gehrmann:1999as}. 

Again, we could fix the coefficient of such an integral by matching field theory anywhere along the codimension-six cut surface of (\ref{hexacut_of_222_example}). But we can in fact do better: knowing (or expecting) that MHV amplitudes in sYM should be maximal weight, we may simply cull such integrands from our basis. Note that the integral topology (\ref{hexacut_of_222_example}) with a loop-momentum dependent numerator is part of our result.

Given the above considerations, it is clear that much thought must go into the choice of a `good' integrand basis. In the case of six-particle MHV amplitudes, there still remains the question of the correct choice of spanning cuts. There are many choices available, and these can be tuned to expose different aspects of field theory.

%================================================================================================================
\vspace{-0pt}\subsection{A Good Start: \emph{Chiral} Numerators for Non-Planar Integrands}\label{subsec:chiral_integrands}\vspace{-0pt}
%================================================================================================================
As described above, there is a substantial difference between constructing \emph{some} integrand basis, and a \emph{good} integrand basis. One well-motivated strategy would be to choose an integrand basis such that a maximal number of integrands have vanishing coefficients (for some amplitudes of interest). This strategy is indeed a reasonable one, but it would lead to a very different representation of amplitudes than what we derive here. For example, one could choose as many defining contours as possible to involve poles at infinity; such integrands would necessarily have vanishing coefficients in sYM (and fewer vanishing coefficients in SUGRA). In such a basis, relatively few leading singularities would be matched directly, and all the others would follow from completeness or residue theorems. This is not the approach that we describe here. 

Rather, our guiding principle will be to choose integrands that are naturally engineered to match as many physical singularities as possible. In this basis, the composite leading singularities such as that on the left of (\ref{example_ls})---those directly responsible for the soft-collinear divergences of loop amplitudes---are matched \emph{directly}, by infrared-divergent integrals. Upon diagonalizing the integrand basis, this tends to render a maximal subset of integrands to be infrared finite (a feature we consider quite valuable). Soft-collinear divergences are associated with parity-even contours in loop-momentum space; as such, \emph{chiral} integrands tend to vanish on these cuts. 

Before we describe how these integrands may be constructed, it will be useful to define one recurring bit of notation for these numerators. Specifically, we find it useful to define a `bracket' built out of traces of (pairs of) four-momenta expressed in terms of $2\times2$ matrices (by dotting any $p^\mu$ into the Pauli matrices $\sigma_\mu^{\alpha\,\dot\alpha}$):
\eq{\brEight{a_1}{a_2}{b_1}{b_2}{\cdot}{\cdot}{c_1}{c_2}\equivR\Big[(a_1\!\cdot\!a_2)^{\alpha}_{\phantom{\alpha}\beta}(b_1\!\cdot\!b_2)^{\beta}_{\phantom{\beta}\gamma}\cdots(c_1\!\cdot\!c_2)^{\delta}_{\phantom{\gamma}\alpha}\Big]\label{definition_of_br}}
where $(a_1\!\cdot\!a_2)^{\alpha}_{\phantom{\alpha}\beta}\equivR a_1^{\alpha\,\dot{\alpha}}\epsilon_{\dot{\alpha}\dot{\gamma}}a_2^{\dot{\gamma}\gamma}\epsilon_{\gamma\beta}$.
Elsewhere, this bracket has been called a `Dirac trace', denoted `$\text{tr}_{+}[\cdots]$'$\equivR\br{\cdots}$ (e.g.~\cite{Badger:2013gxa,Badger:2015lda,Badger:2016ozq}). Further details---including the bracket's symmetries and degenerations---are discussed in \mbox{appendix \ref{bracket_details}}.

%===============================
\paragraph{Constructing Chiral Numerators for Box Integrands at One-Loop}~\\[-12pt]
%===============================

In order to familiarize ourselves with the notation introduced above and to better understand its salient features, let us first look at some examples of chiral integrands at one loop. Perhaps the simplest example relevant to our discussion is the `two-mass-easy' \emph{scalar} box integral, which (prior to any normalization) we denote as
\eq{\fig{-38.5pt}{1}{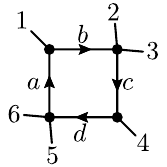} \bigger{\Leftrightarrow}\;\;\frac{1}{a^2\,b^2\,c^2\,d^2}\,,\label{eq:scalar_2me_box}}
where $a,b,c,d$ represent the momenta flowing through the propagators of the graph. Of course, momentum conservation requires (in all-incoming conventions) that 
\eq{a+p_1=b,\quad b+p_{23}=c,\quad c+p_4=d,\quad d+p_{56}=a,}
where $p_{i\cdots j}\equivR(p_i+\ldots+ p_j)$. Momentum conservation can always be solved for example by declaring that $a\equivL\ell$, but it will be extremely useful for us to \emph{not} require any particular solution to momentum conservation (especially for non-planar integrands beyond one loop). In particular, we have used this graph-theoretic description of all loop-dependent numerators of our basis of integrands described in \mbox{appendix \ref{appendix:explicit_numerators}}. 

If we want to define an integrand like that of (\ref{eq:scalar_2me_box}) but with `triangle' power counting, we follow the definition of power-counting discussed in the previous section (and that of \mbox{ref.\ \cite{Bourjaily:2017wjl}}), which leads to a numerator of the form 
\eq{\frac{\mathfrak{n}(\ell)}{a^2\,b^2\,c^2\,d^2}\quad\text{with}\quad \mathfrak{n}(\ell)\in\text{span}_{q}\!\big\{\!(a+q)^2\!\big\}\,.\label{space_of_triangle_power_counting_box_numerators}}
Notice that $q$ can be any spacetime vector. In particular, as the propagators are simple translations of each other,
\eq{\text{span}_{q}\!\big\{\!(a+q)^2\!\big\}\simeq\text{span}_{q}\!\big\{\!(b+q)^2\!\big\}\simeq\text{span}_{q}\!\big\{\!(c+q)^2\!\big\}\simeq\text{span}_{q}\!\big\{\!(d+q)^2\!\big\}\,.}
Moreover, one can show that this space is spanned by a Lorentz-invariant degree-two polynomial in $\ell$, 
\eq{\text{span}_{q}\!\big\{\!(a+q)^2\!\big\}\simeq\text{span}\!\big\{\ell^2,\ell^\mu,1\big\}\,,}
where we have ignored overall factors of mass-dimension, and used a spanning set of coordinate vectors $e_\mu$ to express each component of $\ell^\mu\equivR(\ell\!\cdot\!e^\mu)$. From this, one learns that the rank of such numerators is $(d+2)$ for (integer) $d$ spacetime dimensions. 

A very natural subspace of numerators for the integral (\ref{eq:scalar_2me_box}) would correspond to taking $q\in\{0,p_1,p_{123},p_{1234}\}$. These elements of the numerator basis are called \emph{contact terms}. After removing the contact terms from the space of numerators, there are $(d-2)$ degrees of freedom remaining. Let us now describe a very natural choice for the two remaining numerators in four dimensions. 

In the language above, we are interested in finding two numerators of the form $(a+q)^2\!\notin\!\text{span}\!\big\{a^2,b^2,c^2,d^2\big\}$. Notice that all the contact terms vanish on solutions to the so-called `quadruple cut' $a^2=b^2=c^2=d^2=0$. In four dimensions, there are precisely two solutions to the cut equations,  
\eq{\fig{-38.5pt}{1}{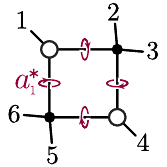}\bigger{\Leftrightarrow}\;{\color{hred}a_1^*}\equivR\frac{\lambda_1\big(|p_{56}|4\rangle\big)}{\ab{14}}\,, \quad\fig{-38.5pt}{1}{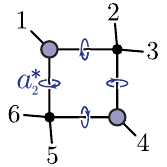}\bigger{\Leftrightarrow}\;{\color{hblue}a_2^*}\equivR \frac{\big([4|p_{56}|\big)\widetilde{\lambda}_1}{\sb{14}}\,.\label{quad_cuts_of_2me}}
The cuts above have been decorated in order to emphasize that ${\color{hred}a_1^*}$ is the solution for which the $\lambda$'s of $\{a,b,p_1\}$ and $\{c,d,p_4\}$ are mutually parallel (and thus would support non-vanishing $\overline{\text{MHV}}$ amplitudes at those vertices), while the solution ${\color{hblue}a_2^*}$ is a solution for which the $\widetilde{\lambda}$'s of $\{a,b,p_1\}$ and $\{c,d,p_4\}$ are mutually parallel (and thus would support non-vanishing MHV amplitudes at those vertices). It will be useful later that considerations of helicity flow ensure that MHV amplitudes have support \emph{exclusively} on the cut ${\color{hred}a_1^*}$ and vanish on the cut ${\color{hblue}a_2^*}$. 

We'd like to choose our remaining numerators to be \emph{chiral}: to have support on one solution or the other, but not both. This is easy to arrange: take, for example, 
\eq{\mathfrak{n}_1\equivR s_{14}(a-{\color{hblue}a_2^*})^2,\quad\text{and}\quad\mathfrak{n}_2\equivR s_{14}(a-{\color{hred}a_1^*})^2\,,}
where we have included the normalization $s_{14}\equivR(p_1+p_4)^2$ to normalize the residues on (\ref{quad_cuts_of_2me}) to be unit in magnitude. Notice that the requirement of chirality would be left invariant by the addition of any combination of contact terms to $\mathfrak{n}_i$. In terms of the `brackets' defined in (\ref{definition_of_br}), these numerators could be written as
\eq{\mathfrak{n}_1=\br{1,b,c,4}= s_{14}(a-{\color{hblue}a_2^*})^2,\quad\text{and}\quad\mathfrak{n}_2=\br{b,c,4,1}=s_{14}(a-{\color{hred}a_1^*})^2\,.}

In earlier literature, these numerators were referred to as (`magic') `wavy-line' and `dashed-line' numerators, respectively (see e.g.\ \cite{ArkaniHamed:2010kv, Drummond:2010mb}). For example, in \mbox{ref.\ \cite{Bourjaily:2013mma}}, the integral with numerator $\mathfrak{n}_1$ would have been drawn as the first figure in:
\eq{\frac{\br{1,b,c,4}}{a^2b^2c^2d^2}\equivL\fig{-38.5pt}{1}{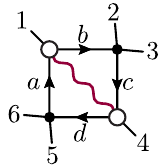}\equivL\fig{-38.5pt}{1}{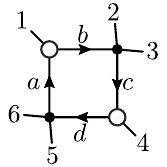}\,.\label{eg_chiral_box_one_loop}}
We will adopt the notation in the second figure, in which empty vertices signals that the numerator has been chosen to have support \emph{only} on cuts for which the so-decorated corners are in the $\overline{\text{MHV}}$ configuration---with the $\lambda$'s of all momenta involved at this vertex parallel.

The bracket notation is easily seen to encode such constraints. For example, 
\begin{itemize}
\item $\brFour{1}{b}{\cdot}{\cdot}$ enforces the numerator to vanish when $\widetilde{\lambda}_b\propto\widetilde{\lambda}_{1}$,
\item $\brFour{\cdot}{\cdot}{c}{4}$ enforces the numerator to vanish when $\widetilde{\lambda}_c\propto \widetilde{\lambda}_{4}$,
\item $\brFour{\cdot}{b}{c}{\cdot}$ enforces the numerator to vanish when $\lambda_{b}\propto\lambda_c$,
\end{itemize}
and so on. These examples make it clear, however, that the bracket notation is considerably more versatile than it may appear at first. For example, \emph{any} numerator involving one or more of the ingredients above would be `chiral' and vanish on the quad-cut ${\color{hblue}a_2^*}$. (Notice also that this implies that numerators as different from each other as $\br{1,b,2,3}$ and $\br{1,b,c,4}$, for example, differ only by contact terms; this fact is considerably more surprising.)

\newpage
%================================================================================================================
\vspace{-0pt}\section[Six-Point Amplitude Integrands of Maximal sYM and SUGRA]{Six-Point Amplitude Integrands of sYM and SUGRA}\label{sec:the_result}\vspace{-0pt}
%================================================================================================================
In this section we discuss the main results of this paper: the four-dimensional two-loop, six-particle MHV amplitude integrands in $\mathcal{N}=4$ super Yang-Mills and $\mathcal{N}=8$ supergravity. The explicit details of the result are presented more thoroughly in the appendices and ancillary files; here, we highlight the result's most important features, and discuss some of the particulars involved in its derivation. 

%===============================
\paragraph{Comparison with sYM in the Planar Limit}~\\[-12pt]
%===============================

In \mbox{ref.\ \cite{ArkaniHamed:2010kv}}, a closed formula was guessed (and checked against recursion) for the two-loop, $n$-point MHV amplitude integrand in planar sYM. This formula was later justified and refined in \cite{ArkaniHamed:2010gh}; it can be expressed in terms of a deceptively simple sum:
\eq{A_n^{\text{MHV}}=\bigger{\displaystyle\sum_{1\leq a<b<c<d<n+a}}\hspace{-10pt}\fig{-34.925pt}{1}{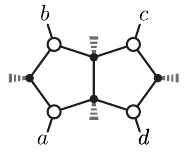}\,.\label{double_pentagon_formula_for_mhv}}
This formula is `deceptively' simple in that it includes several `boundary cases':
\eq{\fig{-34.925pt}{1}{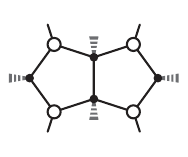}\in\left\{\rule{0pt}{32pt}\right.\fig{-34.925pt}{1}{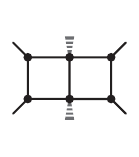},\fig{-34.925pt}{1}{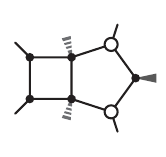},\fig{-34.925pt}{1}{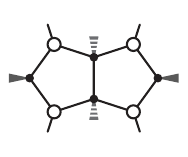}\left.\rule{0pt}{32pt}\right\}\,.\label{planar_mhv_breakdown}}
In retrospect, these boundary cases are precisely those needed to match the soft-collinear divergences of the theory. By construction, none of the integrands in (\ref{planar_mhv_breakdown}) have support at infinity---a fact which is closely related to their power-counting (and that these integrands were engineered to be manifestly dual-conformally invariant).

The generalization of equation (\ref{double_pentagon_formula_for_mhv}) to all helicity amplitudes in planar sYM was derived in \mbox{ref.\ \cite{Bourjaily:2015jna}} and extended to all \emph{three} loop integrands in planar sYM in \mbox{ref.\ \cite{Bourjaily:2017wjl}}. There, it started to become clear that there was a tension between box power-counting (or dual-conformal invariance) and having a `nice' integrand basis. Indeed, it was found that (as is more familiar at one loop), three-loop integrands with box power-counting are topologically overcomplete---requiring either choices to be made regarding which integrands to include, or for magic to be found among an over-constrained system of equations. (The latter was found in \mbox{ref.\ \cite{Bourjaily:2017wjl}}, but we have every expectation that this will be increasingly difficult to find at higher loop orders.) 

As discussed in \mbox{section \ref{subsec:generalized_unitarity}}, one (surprisingly easy) way to relieve this tension is to allow for integrands with triangle power-counting (or worse) in four dimensions. This immediately renders the basis of integrands topologically complete (and not over-complete), making every degree of freedom fixable in terms of evaluations or residues taken along cut surfaces. 

It turns out that for six particles the basis of integrands with triangle power-counting is fully polylogarithmic; as such (once all integrands with double poles have been removed), it can be fully diagonalized on leading singularities. Some of the leading singularities in the integrand basis support non-vanishing coefficients for MHV amplitudes, while others do not. Let us briefly describe the choices we have made for the leading singularities on which to normalize our integrand basis, and the form of the representation that results for the integrand of the amplitude.

\begin{table}[b!]\vspace{-22pt}$$\fig{-100pt}{1}{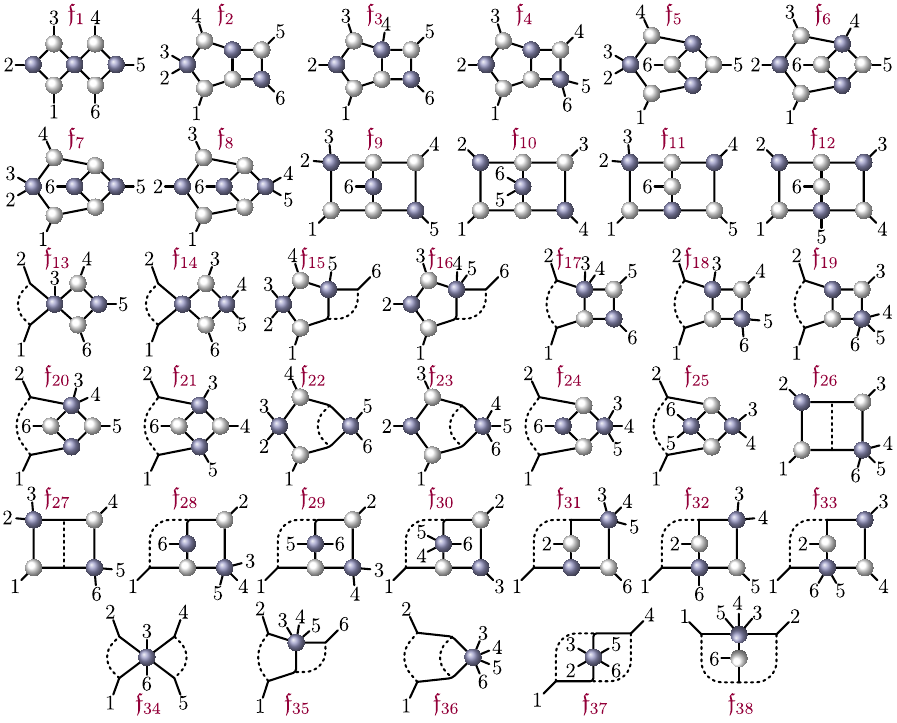}$$\vspace{-22pt}
\centering\vspace{-2pt}\caption{A spanning set of (non-vanishing) six-particle leading singularities for maximally supersymmetric $(\mathcal{N}=4,8)$ Yang-Mills theory and gravity. These cuts are not enough to span the basis of integrands with triangle power-counting; but they correspond to the subspace of integrands with non-vanishing coefficients.}\label{six_point_ls_table}\vspace{-26pt}
\end{table}
%
%================================================================================================================
\subsection{Prescriptive Integrands for a Spanning Set of Leading Singularities}
%================================================================================================================

The construction of chiral integrands is still more of an art form than an algorithmic procedure. However, knowing the number of (representatives of each) independent numerator degree of freedom turns out to be extremely valuable data to guide their construction. In particular, it tells us at the outset the number of degrees of freedom we should use to match field theory manifestly. 

The list of MHV-supported leading singularities we would like to match manifestly is given in \mbox{table \ref{six_point_ls_table}}. The figures here have multiple meanings: they represent leading singularities for six-point processes in any quantum field theory where the vertices of these graphs represent tree-amplitudes in that theory. Secondly, these pictures represent particular contour integrals on which we choose to normalize (and diagonalize) a subset of the integrands in our basis. The eight conditions defining each contour correspond to putting all propagators of the graph on-shell, and also setting the momentum of any dashed edge in the figure to zero. 

The cuts in \mbox{table \ref{six_point_ls_table}} are not by themselves sufficient to fix all integrand degrees of freedom in our basis. However, the remaining degrees of freedom fall into one of two categories: they cannot be normalized on any maximal co-dimension residue (as in the example (\ref{hexacut_of_222_example}) discussed above), in which case we may declare their coefficients to be zero; or they can be normalized on cuts that vanish for MHV amplitudes. 

To illustrate the second possibility, consider the integrands with `kissing-box' topology---depicted by the bottom-right graph in \mbox{table \ref{four_dim_triangle_power_counting_basis}}. As indicated in that table, there are 4 irreducible degrees of freedom assigned to the numerator of integrands with this topology. These four degrees of freedom can easily be seen to correspond to the 4 solutions to the cut equations (two solutions for each box). Using the same notation as in (\ref{quad_cuts_of_2me}), these numerators can be seen to support cuts with \emph{particular} (and distinct) configurations of helicity amplitudes:
\eq{\begin{split}&\raisebox{-28pt}{\includegraphics[scale=1]{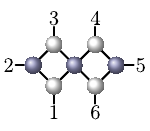}}  \underset{\fwbox{0pt}{\raisebox{-12pt}{\text{\footnotesize{MHV}}}}}{\longleftrightarrow} \raisebox{-28pt}{\includegraphics[scale=1]{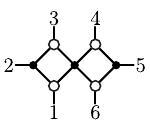}}\,,
\quad\raisebox{-28pt}{\includegraphics[scale=1]{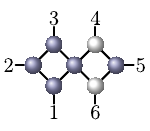}} \underset{\fwbox{0pt}{\raisebox{-12pt}{\text{\footnotesize{NMHV}}}}}{\longleftrightarrow} \raisebox{-28pt}{\includegraphics[scale=1]{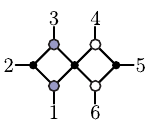}} \phantom{\,.} \\
&\raisebox{-28pt}{\includegraphics[scale=1]{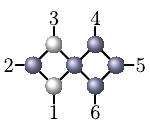}}  \underset{\fwbox{0pt}{\raisebox{-12pt}{\text{\footnotesize{NMHV}}}}}{\longleftrightarrow} \raisebox{-28pt}{\includegraphics[scale=1]{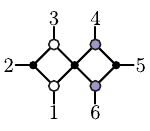}}\,,\quad\raisebox{-28pt}{\includegraphics[scale=1]{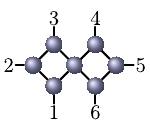}}  \underset{\fwbox{0pt}{\raisebox{-12pt}{\text{\footnotesize{N$^2$MHV}}}}}{\longleftrightarrow} \raisebox{-28pt}{\includegraphics[scale=1]{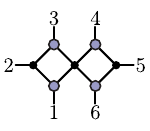}}\,.\end{split}\label{kissing_box_integrands_and_cuts}}
Numerators which are correctly normalized for these four cases are very easy to construct by analogy to the example (\ref{eg_chiral_box_one_loop}). For the purpose of constructing MHV amplitudes, only the first integrand in (\ref{kissing_box_integrands_and_cuts}) is needed---the others correspond to integrands in our basis that will be assigned vanishing coefficients.

As can be seen in \mbox{table \ref{four_dim_triangle_power_counting_basis}}, in the basis of integrands with triangle power-counting, all integrands with eight propagators at two loops have precisely as many numerator degrees of freedom as solutions to their cut equations. Thus, we can similarly restrict our attention to only those cuts that have MHV support. These (non-composite) leading singularities correspond to the first 12 entries in \mbox{table \ref{six_point_ls_table}}.

Including also the cuts with non-MHV (but otherwise non-vanishing) support will fix most of the remaining degrees of freedom in our basis, but not all. The remaining set of cuts needed to specify our basis (and represent amplitudes) are those which have no support in sYM (or SUGRA). For the result described in this present work, we have absorbed all remaining degrees of freedom of the basis into residues at infinite loop momentum for which we know that both sYM \emph{and} SUGRA have vanishing support. (Had we chosen to normalize some element of our integrand basis on a pole at infinity for which SUGRA had not vanished, we would have been forced to either add this term to SUGRA relative to sYM in our representation (\ref{schematic_loop_integrand_formula}) or include a term whose coefficient in sYM would have been zero.)

Once a spanning set of cuts has been chosen and tentative numerators chosen to normalize these integrands on these cuts, it is still necessary to diagonalize the basis. Luckily, because graph inclusion is a triangular system, any starting initial basis of integrand normalized on a spanning set of contours will automatically be triangular in cuts; and diagonalization, therefore, becomes a simple matter of iterative subtractions.

These subtractions are easiest when starting with chiral numerators. Consider for example our basis integrands $\mathcal{I}_1$ and $\mathcal{I}_9$ as given in \mbox{appendix \ref{appendix:explicit_numerators}}:
\eq{\hspace{20pt}\begin{array}{@{}c@{}}\fwboxR{0pt}{\mathcal{I}_1\!:\;\; }\fig{-39.5pt}{1}{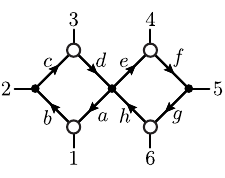}\\\fwboxR{0pt}{\mathfrak{n}(\mathcal{I}_1)= }\brFour{1}{\edgeB}{\edgeC}{3}\brFour{4}{\edgeF}{\edgeG}{6}\end{array}\qquad\quad\raisebox{7.5pt}{\text{or}}\qquad\quad\qquad\begin{array}{@{}c@{}}\fwboxR{0pt}{\mathcal{I}_9\!:\;\; }\fig{-39.5pt}{1}{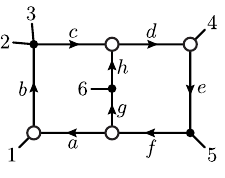}\\\fwboxR{0pt}{\mathfrak{n}(\mathcal{I}_9)= }-\brEight{1}{\edgeB}{\edgeC}{\edgeH}{\edgeG}{\edgeF}{\edgeE}{4}\end{array}\,.\hspace{-20pt}\label{example_integrands}}
These integrands include as sub-topologies many of the other graphs appearing in \mbox{table \ref{six_point_ls_table}}. However, all of the integrals in our basis associated with these sub-topologies involve a soft internal momentum which is parity even; thus, the chirality of the numerators above automatically ensures that $\mathcal{I}_1$ and $\mathcal{I}_9$ vanish on the other cuts. 

This is not always the case, however. Consider for example integrand $\mathcal{I}_2$,
\eq{\hspace{10pt}\fwboxR{0pt}{\mathcal{I}_2\!: }\fig{-39.5pt}{1}{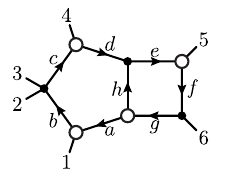}\;\text{with}\quad\mathfrak{n}(\mathcal{I}_2)=  -\brFour{1}{\edgeB}{\edgeC}{4}\brFour{5}{\edgeF}{\edgeG}{\edgeA}-\frac{1}{2}\brEight{1}{\edgeB}{\edgeC}{4}{5}{\edgeF}{\edgeG}{\edgeA}\,.\label{int2_example}}
the second term in the numerator vanishes on the `defining cut' for this element of the basis; to see this, one can easily verify that
\eq{\brEight{1}{\edgeB}{\edgeC}{4}{5}{\edgeF}{\edgeG}{\edgeA}=\edgeB^2\brSix{4}{\edgeC}{\edgeA}{\edgeG}{\edgeF}{5}-\edgeA^2\brSix{4}{\edgeC}{\edgeB}{\edgeG}{\edgeF}{5}\,.\label{pentabox_correction_expanded}}
The reader may therefore naturally wonder the role it plays. It turns out that this term is fixed via diagonalization---in this case, diagonalization with respect to an integrand in our basis with vanishing coefficient. That is, this second term in the integrand can be understood as representing otherwise unenumerated basis elements: e.g.\ the second in (\ref{pentabox_correction_expanded}) corresponding to a seven-propagator integral with numerator $\brSix{4}{\edgeC}{\edgeA}{\edgeG}{\edgeF}{5}$. This integrand element is normalized to have unit reside `at infinity' (in both loops) as follows:
\vspace{-10pt}\eq{\fig{-34pt}{1}{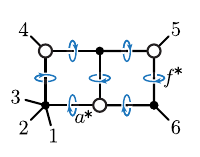}\quad\text{with}\quad a^*\!(z)\equivR\frac{\big(\lambda_6+z\lambda_5\big)\big(|p_{123}|4\rangle\big)}{z\ab{45}+\ab{46}},\quad f^*\!(z)\equivR z\lambda_5\widetilde{\lambda}_6\;.\label{heptacut_at_infty_example}\vspace{-6pt}}
On this cut, both terms in the numerator of ${\cal I}_2$ have a pole that sends the propagators of each loop to infinity. In the parametrization given above, this would correspond to setting $z\!=\!-\ab{46}/\ab{45},\infty$. The diagonalization of our integrand basis thus requires that $\mathcal{I}_2$ vanish on this symmetric sum, which fixes the relative coefficient of the monomials in (\ref{int2_example}).

%============================================================
\subsection{Putting Everything Together: MHV Amplitudes in sYM and SUGRA}\label{subsec:final_results}
%============================================================

After choosing a complete integrand basis, normalized on a spanning set of cuts that include those enumerated in \mbox{table \ref{six_point_ls_table}}, only the 38 integrands depicted there will have non-vanishing coefficients for MHV amplitudes in sYM and SUGRA. Thus, we find that we may write
\vspace{-0pt}\eq{\mathcal{A}_{6,\text{MHV}}^{\mathcal{N}=4,8}=\sum_{i}{\color{hred}\lsLabel_i^{\text{\scriptsize$\,\mathcal{N}$}}}\mathcal{I}_{i}+\text{permutations}\,.\label{loop_integrand_formula}\vspace{-6pt}}
The meaning of the sum over permutations should now be clear: as we are required to match each of the leading singularities of \mbox{table \ref{six_point_ls_table}} once \emph{and only once}, we must sum over all permutations \emph{without duplication}. Duplication here means with respect to specific field-theory cuts, and corresponds to automorphisms of the on-shell diagrams that preserve both the white/blue decorations of the vertices (the parity of the cut solution), and also map soft momenta to soft momenta. (Remember that amplitudes in sYM and SUGRA are always permutation invariant, and thus the ordering of external legs attached to the same vertex never matter.) Thus, we must sum over all re-labelings of the external legs that result in genuinely distinct cut configurations. 

In \mbox{appendix \ref{explicit_integrand_numerator_tables}}, explicit numerators are given for all 38 integrands (with non-vanishing coefficients) in our basis. (The notation used is described in some detail in \mbox{appendix \ref{bracket_details}}.) The coefficients (leading singularities) ${\color{hred}\lsLabel_i^{\text{\scriptsize$\,\mathcal{N}$}}}$ for both $\mathcal{N}=4$ sYM and $\mathcal{N}=8$ SUGRA are given explicitly in appendices \ref{appendix:ls_of_sym} and \ref{appendix:ls_of_sugra}, respectively. Further details, including machine-readable formulae, can be found in the ancillary files attached to this work, and are documented in \mbox{appendix \ref{appendix:ancillary_files}}. 

%===============================
\vspace{-0pt}\paragraph{Explicit Checks and Eliminating Sign Ambiguities}~\\[-16pt]
%===============================

One rather subtle aspect of any representation of an amplitude such as (\ref{loop_integrand_formula}) involves the \emph{signs} of the terms in the sum. It could be argued that the overall sign of any particular on-shell function \emph{in isolation} is not well defined (or arbitrary), as on-shell residues \emph{necessarily} involve an ordering of the propagators being cut via a Jacobian. As such, the reader may wonder for example, why the numerator of $\mathcal{I}_9$ in (\ref{example_integrands}) has a minus sign in front of it? After all, we could have defined the integral's coefficient with a different sign. But---more intrinsically---if its coefficient's sign should be understood as conventional, how is this sign determined at all?

Even if we concede that the sign of any term is arbitrary, the \emph{relative} signs between terms are fully determined by Cauchy's residue theorem (or its multi-dimensional manifestation, the `Global Residue Theorem' (GRT) \cite{GriffithsHarris}). Specifically, although we have chosen some residues at infinity to vanish manifestly---such as that of (\ref{heptacut_at_infty_example}), by having integrands normalized on these contours in our basis (with coefficient zero) and diagonalizing the rest of our basis integrands (so that they all individually vanish)---there are many other poles at infinity whose residues vanish in a very non-trivial way via Cauchy's residue theorem. Checking that all poles at infinity vanish for sYM therefore not only provides a very serious check of our result, but also fully determines the relative signs between all terms. 

Having fixed all relative signs, and checked that all poles at infinity vanish, there still remains one \emph{overall} sign ambiguity. This final ambiguity was fixed by comparing the expression (\ref{loop_integrand_formula}) to the planar limit in equation (\ref{planar_mhv_breakdown}). Specifically, we verified that, upon summing all integrands together with their permutations in (\ref{loop_integrand_formula}), color-decomposing every on-shell function of sYM, extracting the coefficient of the single-trace term (at leading order in $1/N_c$), and converting each of these integrands to dual-momentum coordinates (valid in the planar limit), that the resulting rational loop-momentum integrand \emph{exactly} matches that of (\ref{planar_mhv_breakdown}). Beyond merely fixing the final overall sign ambiguity, this comparison amounts to a very strong test of our result's consistency. 

%===============================
\vspace{-2pt}\paragraph{Infrared Structure: Finiteness and Divergences}~\\[-16pt]
%===============================

A nice feature of our result is that many of the integrands in our basis vanish explicitly in the soft/collinear regions responsible for infrared divergences as a result of choosing chiral numerators. In general, infrared divergences can arise in massless scattering amplitudes where loop momenta become either soft or collinear to external momenta. 
%
%\begin{figure}
%\centering
% \includegraphics[scale=.2]{./figures/soft_collinear_regions}
%\caption{\label{fig:soft_col_reg}Potential collinear and soft divergent regions of one-loop integrals.}
%\end{figure}
%
Enumerating all potentially infrared-divergent regions can become quite intricate for higher-loop, large-multiplicity amplitudes. (For a recent discussion in the context of QCD amplitudes, see e.g.~\cite{Anastasiou:2018rib} and the references therein.) Furthermore, the question whether or not there will be an infrared singularity after loop integration crucially depends on whether the numerator of each diagram softens or fully cancels the divergence. In lieu of the power-counting analyses of the infrared-divergent regions performed in e.g.~\cite{Anastasiou:2018rib}, we can probe potentially dangerous infrared regions at the integrand level via unitarity cuts. In the context of planar $\mathcal{N}=4$ sYM, such an analysis has been performed in \mbox{ref.\ \cite{Bourjaily:2015jna}}; however, similar analyses can be extended beyond the planar limit without much effort. In the language of unitarity cuts, infrared divergences are linked to integrands that have support on certain composite residues. Collinear composit residues can be taken whenever three-particle vertices involve a single external (massless) leg. Since one of the three legs is already massless, cutting either of the internal propagators factorizes the second propagator into a pair of simple poles---corresponding to the MHV and $\overline{\text{MHV}}$ cut solutions, respectively. Taking both residues thus restricts us to the intersection of these two cut surfaces, where all three momenta are collinear and one expects a $1/\epsilon$ divergence upon integration. Similarly, when two adjacent three-particle vertices (each involving a single external leg) are both made collinear in this way, there is no longer any momentum flow through the propagator connected to both vertices. This tells us we are in a soft-collinear limit, and should expect a $1/\epsilon^2$ divergence upon integration.%\footnote{This way of identifying the leading collinear and soft regions essentially coincides with the soft/collinear power-counting procedure encountered in other contexts (for example~\cite{Anastasiou:2018rib}).}

Diagram numerators can potentially cancel the composite poles described above. Whenever this happens, we determine that the resulting integrand does not have the corresponding infrared singularity. Prime examples of this effect are the chiral integrals defined e.g. in \cite{ArkaniHamed:2010gh}. In our setup here, it is easy to confirm that the integrands associated with non-composite leading singularities---namely, $\{\mathcal{I}_1,\ldots,\mathcal{I}_{12}\}$, as well as those integrands not specifically engineered to match a soft-collinear divergence, $\{\mathcal{I}_{22},\mathcal{I}_{23},\mathcal{I}_{26},\mathcal{I}_{27}\}$---vanish in all leading infrared-divergent regions. This happens because each of their numerators is constructed to vanish on the MHV cut solutions for these vertices. The rest of the integrands in the basis are infrared divergent by construction. 

This partitioning of our basis into explicitly infrared-finite and infrared-divergent subspaces seems important, rather than a mere curiosity. Like seen at two loops in the planar sector of sYM in \mbox{ref.\ \cite{Bourjaily:2015jna}}, this hints toward the exponentiation of infrared divergences at the integrand-level. Could we use these representations to build manifestly finite integrands for remainder functions or ratio functions? We leave this question to future work.

%================================================================================================================
%    4. Conclusions
%================================================================================================================
\vspace{-0pt}\section{Conclusions and Future Directions}\label{sec:conclusions}\vspace{-0pt}
%================================================================================================================

In this paper we have  constructed the four-dimensional integrands for two-loop six-point MHV amplitudes in maximally supersymmetric Yang-Mills theory and supergravity. Our integrand representation includes only pure, unit-leading-singularity integrals. We obtained these results via prescriptive unitarity, where each basis integrand directly matches a specific, physical leading singularity.

The six-point amplitude is the largest multiplicity for which both sYM and SUGRA can be expanded into the same power-counting basis. Notably, the results in both theories take exactly the same form and only differ by the interpretation of the corresponding on-shell functions. Moreover, the choice of chiral numerators for individual basis elements nicely delineates integrals that are infrared finite and divergent, matching the infrared properties of their associated physical singularities.

Going forward, there are several natural pathways towards a better understanding of perturbation theory beyond the planar limit. Although we have shown that even for MHV amplitudes, integrands will require non-polylogarithmic integrands beyond 6-points, the basis we have constructed is sufficient for  6-point NMHV amplitudes in sYM and SUGRA---allowing us to probe the first intrinsically infrared-finite function beyond the planar limit: the 6-point ratio function. In a similar vein, it is natural to try and generalize these results to higher multiplicity or loop orders. 

We expect that all amplitude integrands in sYM will be free of poles at infinity to all orders \cite{Arkani-Hamed:2014via,Bern:2014kca}, and should therefore be expressible in the basis of triangle power-counting integrands (with some definition of `triangle power-counting'). In contrast, it is known that amplitudes in supergravity require larger integrand bases in general due to their worse behavior at infinity for large multiplicity. Moreover, the behavior of supergravity amplitudes at infinity appears to be in conflict with cut-constructibility in four dimensions \cite{Bourjaily:2018omh,Edison:2019ovj}, and suggests that additional (and novel) conditions must be found to construct amplitude integrands in supergravity. 

Another natural path forward would be to integrate the six-point MHV amplitudes we have found. This will not be easy, but may lead to the discovery of new simplicities such as that hinted by the `non-planar dual-conformal symmetry' described in \mbox{refs.\ \cite{Bern:2017gdk,Bern:2018oao,Chicherin:2018wes}}, as well as being a natural testing ground for questions about the restrictions on symbol entries~\cite{Chicherin:2017dob}, for example. Integrating the amplitudes as we have represented them, however, may require upgrading them to $4-2\epsilon$ dimensions in order to exploit the loop-integration technology available in dimensional regularization. We hope that the framework of differential equations may help guide this line of further investigation.

%================================================================================================================
%    Acknowledgements 
%================================================================================================================
\vspace{-4pt}
\section*{Acknowledgements}%
\vspace{-0pt}
\noindent This work was performed in part at the Aspen Center for Physics, which is supported by National Science Foundation grant PHY-1607611, and the Harvard Center of Mathematical Sciences and Applications. 
This project has been supported by an ERC Starting Grant \mbox{(No.\ 757978)} and a grant from the Villum Fonden (JLB); by a grant from the Simons Foundation (341344, LA) (JLB); and supported by a Carlsberg Postdoctoral Fellowship (CF18-0641) (AJM).
The research of J.T. and C.L.\ is supported in part by U.S. Department of Energy grant DE-SC0009999 and by the funds of University of California. E.H.\ is supported by the U.S. Department of Energy under contract DE-AC02-76SF00515.

%================================================================================================================
%    Appendix A:  
%================================================================================================================
\newpage\appendix\vspace{-0pt}\section[\mbox{Explicit Numerators for Six-Point Integrands}]{\mbox{\hspace{0pt}Explicit Numerators for Six-Point Integrands}}\label{appendix:explicit_numerators}\vspace{-0pt}
\subsection{Representing (Functions of) Four-Momenta}\label{bracket_details}

The basis of integrands we construct involve `chiral' numerators built from the trace-like bracket introduced in \mbox{section \ref{subsec:chiral_integrands}}. We consider all four-momenta---including loop momenta---to be expressed (without any loss of generality) as $2\times2$ matrices constructed using the Pauli matrices
\vspace{-2pt}\eq{p^\mu\mapsto p^{\alpha\,\dot{\alpha}}\equivR p^\mu\sigma_\mu^{\alpha\,\dot\alpha}=\left(\begin{array}{@{}c@{$\;\;$}c@{}}\fwboxL{12pt}{p^0}\fwbox{8pt}{+}\fwboxR{16pt}{p^3}&\fwboxL{12pt}{p^1}\fwbox{8pt}{-}\fwboxR{16pt}{ip^2}\\\fwboxL{12pt}{p^1}\fwbox{10pt}{+}\fwboxR{16pt}{ip^2}&\fwboxL{12pt}{p^0}\fwbox{10pt}{-}\fwboxR{16pt}{p^3}\end{array}\right)\,.\label{matrix_momenta_defn_appendix}\vspace{-2pt}}
Notice that the indices $\alpha,\dot\alpha$ of $p$ in (\ref{matrix_momenta_defn_appendix}) correspond to $SU(2)_{L,R}$ indices of the local (complexified) Lorentz group, respectively, and can be raised and lowered using the Levi-Civita symbol. In terms of these, the trace-like `bracket' is defined to be
\eq{
\brEight{a_1}{a_2}{b_1}{b_2}{\cdot}{\cdot}{c_1}{c_2}
\equivR
\Big[(a_1\!\cdot\!a_2)^{\alpha}_{\phantom{\alpha}\beta}(b_1\!\cdot\!b_2)^{\beta}_{\phantom{\beta}\gamma}\cdots(c_1\!\cdot\!c_2)^{\delta}_{\phantom{\gamma}\alpha}\Big]\label{definition_of_br_appendix}\vspace{-2pt}}
where
\vspace{-4pt}\eq{(a_1\!\cdot\!a_2)^{\alpha}_{\phantom{\alpha}\beta}\equivR a_1^{\alpha\,\dot{\alpha}}\epsilon_{\dot{\alpha}\dot{\gamma}}a_2^{\dot{\gamma}\gamma}\epsilon_{\gamma\beta}\,.}
This definition is reminiscent of a trace for good reason: it can alternatively be written as the `Dirac trace' $\mathrm{tr}_+(a_1,a_2,b_1,b_2,\dots,c_1,c_2)$; we choose to introduce new notation here because it makes clearer that it is not cyclically symmetric, and because in this instance we happen to prefer Pauli's $\sigma$'s over Dirac's $\gamma$'s. 

Notice that the norm squared of any momentum is equal to the determinant \mbox{$p^2\equivR p^\mu p_\mu=\det(p^{\alpha\,\dot\alpha})$}. Thus, when a momentum $p$ is null, the matrix $p^{\alpha\,\dot\alpha}$ has less than full rank and can therefore be expressed as an outer product $p^{\alpha\,\dot\alpha}\equivL\lambda^{\alpha}\widetilde{\lambda}^{\dot\alpha}$ of spinor-helicity variables \cite{vanderWaerden:1929}. When discussing the momentum $p_a$ of a massless external state labelled by index $a$, we will occasionally refer to the spinor variables for $p_a$ by $\lambda_a\equivL|a\rangle$ and $\widetilde{\lambda}_a\equivL|a]$. In terms of these, we can define the $SL(2)$ invariants \mbox{$\ab{ab}\equivR\det\big(\lambda_a,\lambda_b\big)$} and \mbox{$[ab]\equivR\det\big(\widetilde{\lambda}_a,\widetilde{\lambda}_b\big)$}, as well as objects such as $[a|q|b\rangle\equivR\widetilde{\lambda}_a^{\dot\alpha}q_{\alpha\dot\alpha}\lambda_b^{\alpha}$ for (not necessarily massless) momenta $q$. This notation is somewhat more cumbersome, but unquestionably more familiar than the new notation in (\ref{definition_of_br_appendix}). An example of where this more familiar notation may arise is for brackets involving at least one massless particle. For example, if $a_1$ were massless then
\vspace{-2pt}\eq{\brEight{a_1}{a_2}{b_1}{b_2}{\cdot}{\cdot}{c_1}{c_2}\equivL\big[a_1\big|a_2\big|\cdots\big|c_2\big|a_1\rangle\,.\vspace{-2pt}}
The generalized square-angle bracket on the right hand side above should be familiar notation to most amplitudes researchers; but for our purposes we may consider it to be defined in terms of the bracket (\ref{definition_of_br_appendix}) on the left. 

One final, potential simplification worth mentioning is that whenever a bracket involves consecutively repeated indices, they factor out: 
\vspace{-2pt}\eq{\br{\cdots {\color{hblue}x},\!{\color{hred}q},\!{\color{hred}q},{\color{hblue}y},\!\cdots}={\color{hred}q^2}\br{\cdots,\!{\color{hblue}x},\!{\color{hblue}y},\cdots}\,.\vspace{-2pt}}
(\emph{Nota bene}: the above identity implies that $\br{q,q}=q^2\br{}=2q^2$---as an `empty' trace should be understood as being taken over the $2\times2$ identity matrix.)

\newpage
%===============================================================
\subsection{Explicit Expressions for the Numerators of Basis Integrands}\label{explicit_integrand_numerator_tables}
%===============================================================
%
\vspace{-20pt}\eq{\fwboxL{380pt}{\hspace{-7pt}\text{\underline{\rule{15pt}{0pt}integrand\rule{15pt}{0pt}}}\rule{13.4pt}{0pt}\text{\underline{\rule{10pt}{0pt}numerator\phantom{g}\rule{223pt}{0pt}}}}\nonumber\vspace{-20pt}}
\eq{\fwboxL{380pt}{\hspace{-10pt}\intLabel_{1}\!\!:\fig{-26.75pt}{0.7}{int1}\;\;\text{{\small$\phantom{-}\brFour{1}{\edgeB}{\edgeC}{3}\brFour{4}{\edgeF}{\edgeG}{6}$}}}\tag{$\intLabel.{1}$}\label{int_1_num}}%
\eq{\fwboxL{380pt}{\hspace{-10pt}\intLabel_{2}\!\!:\fig{-26.75pt}{0.7}{int2}\;\;\text{{\small$-\brFour{1}{\edgeB}{\edgeC}{4}\brFour{5}{\edgeF}{\edgeG}{\edgeA}+\frac{1}{2}\brEight{1}{\edgeB}{\edgeC}{4}{5}{\edgeF}{\edgeG}{\edgeA}$}}}\tag{$\intLabel.{2}$}\label{int_2_num}}%
\eq{\fwboxL{380pt}{\hspace{-10pt}\intLabel_{3}\!\!:\fig{-26.75pt}{0.7}{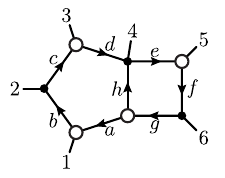}\;\;\text{{\small$-\brFour{1}{\edgeB}{\edgeC}{3}\brFour{5}{\edgeF}{\edgeG}{\edgeA}+\frac{1}{2}\brEight{1}{\edgeB}{\edgeC}{3}{5}{\edgeF}{\edgeG}{\edgeA}$}}}\label{int_3_num}\tag{$\intLabel.{3}$}}%
\eq{\fwboxL{380pt}{\hspace{-10pt}\intLabel_{4}\!\!:\fig{-26.75pt}{0.7}{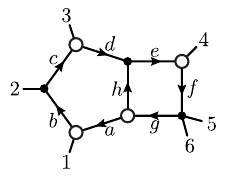}\;\;\text{{\small$-\brFour{1}{\edgeB}{\edgeC}{3}\brFour{4}{\edgeF}{\edgeG}{\edgeA}+\frac{1}{2}\brEight{1}{\edgeB}{\edgeC}{3}{4}{\edgeF}{\edgeG}{\edgeA}$}}}\label{int_4_num}\tag{$\intLabel.{4}$}}%
\eq{\fwboxL{380pt}{\hspace{-10pt}\intLabel_{5}\!\!:\fig{-26.75pt}{0.7}{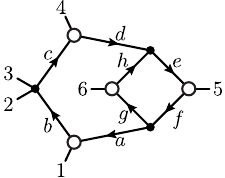}\;\;\text{{\small$\phantom{-}\brFour{1}{\edgeB}{\edgeC}{4}\brFour{5}{\edgeF}{\edgeH}{6}$}}}\label{int_5_num}\tag{$\intLabel.{5}$}}%
\eq{\fwboxL{380pt}{\hspace{-10pt}\intLabel_{6}\!\!:\fig{-26.75pt}{0.7}{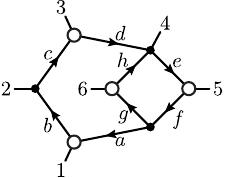}\;\;\text{{\small$\phantom{-}\brFour{1}{\edgeB}{\edgeC}{3}\brFour{5}{\edgeF}{\edgeH}{6}$}}}\label{int_6_num}\tag{$\intLabel.{6}$}}%
\eq{\fwboxL{380pt}{\hspace{-10pt}\intLabel_{7}\!\!:\fig{-26.75pt}{0.7}{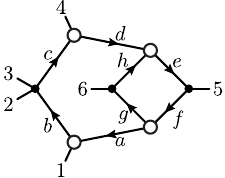}\;\;\text{{\small$-\brFour{1}{\edgeB}{\edgeC}{4}\Big(\brFour{\edgeA}{\edgeG}{\edgeH}{\edgeD}-\edgeA^2\edgeH^2-\edgeD^2\edgeG^2\Big)$}}}\label{int_7_num}\tag{$\intLabel.{7}$}}%
\eq{\fwboxL{380pt}{\hspace{-10pt}\intLabel_{8}\!\!:\fig{-26.75pt}{0.7}{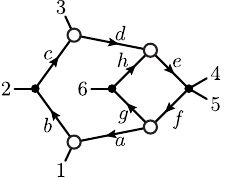}\;\;\text{{\small$-\brFour{1}{\edgeB}{\edgeC}{3}\Big(\brFour{\edgeA}{\edgeG}{\edgeH}{\edgeD}-\edgeA^2\edgeH^2-\edgeD^2\edgeG^2\Big)$}}}\label{int_8_num}\tag{$\intLabel.{8}$}}%
\eq{\fwboxL{380pt}{\hspace{-10pt}\intLabel_{9}\!\!:\fig{-26.75pt}{0.7}{int9}\;\;\text{{\small$-\brEight{1}{\edgeB}{\edgeC}{\edgeH}{\edgeG}{\edgeF}{\edgeE}{4}$}}}\label{int_9_num}\tag{$\intLabel.{9}$}}%
\vspace{-4pt}\eq{\fwboxL{380pt}{\hspace{-10pt}\intLabel_{10}\!\!:\fig{-26.75pt}{0.7}{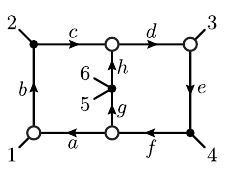}\;\;\text{{\small$-\brEight{1}{\edgeB}{\edgeC}{\edgeH}{\edgeG}{\edgeF}{\edgeE}{3}$}}}\label{int_10_num}\tag{$\intLabel.{10}$}}%
\vspace{-4pt}\eq{\fwboxL{380pt}{\hspace{-10pt}\intLabel_{11}\!\!:\fig{-26.75pt}{0.7}{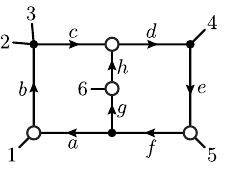}\;\;\text{{\small$\phantom{-}\frac{1}{2}\Big(\!\brEight{1}{\edgeB}{\edgeC}{\edgeD}{\edgeE}{5}{\edgeG}{\edgeH}-\brEight{5}{\edgeE}{\edgeD}{\edgeC}{\edgeB}{1}{\edgeG}{\edgeH}\!\Big)$}}}\label{int_11_num}\tag{$\intLabel.{11}$}}%
\vspace{-4pt}\eq{\fwboxL{380pt}{\hspace{-10pt}\intLabel_{12}\!\!:\fig{-26.75pt}{0.7}{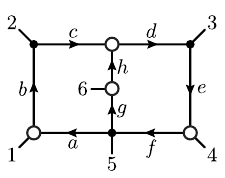}\;\;\text{{\small$\phantom{-}\frac{1}{2}\Big(\!\brEight{1}{\edgeB}{\edgeC}{\edgeD}{\edgeE}{4}{\edgeG}{\edgeH}-\brEight{4}{\edgeE}{\edgeD}{\edgeC}{\edgeB}{1}{\edgeG}{\edgeH}\!\Big)$}}}\label{int_12_num}\tag{$\intLabel.{12}$}}%
\vspace{-4pt}\eq{\fwboxL{380pt}{\hspace{-10pt}\intLabel_{13}\!\!:\fig{-26.75pt}{0.7}{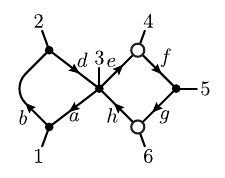}\;\;\text{{\small$\phantom{-}s_{12}\brFour{4}{\edgeF}{\edgeG}{6}$}}}\label{int_13_num}\tag{$\intLabel.{13}$}}%
\vspace{-4pt}\eq{\fwboxL{380pt}{\hspace{-10pt}\intLabel_{14}\!\!:\fig{-26.75pt}{0.7}{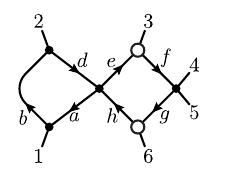}\;\;\text{{\small$\phantom{-}s_{12}\brFour{3}{\edgeF}{\edgeG}{6}$}}}\label{int_14_num}\tag{$\intLabel.{14}$}}%
\vspace{-4pt}\eq{\fwboxL{380pt}{\hspace{-10pt}\intLabel_{15}\!\!:\fig{-26.75pt}{0.7}{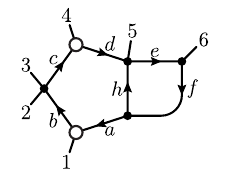}\;\;\text{{\small$
-\br{6,\!\edgeA}\brFour{1}{\edgeB}{\edgeC}{4}+\frac{1}{2}\Big(\!\brSix{1}{\edgeB}{\edgeC}{4}{6}{\edgeA}-\edgeC^2\brFour{1}{\edgeB}{\edgeD}{6}+\edgeD^2\brFour{1}{\edgeB}{\edgeC}{6}\!\Big)
$}}}\label{int_15_num}\tag{$\intLabel.{15}$}}%
\vspace{-4pt}\eq{\fwboxL{380pt}{\hspace{-10pt}\intLabel_{16}\!\!:\fig{-26.75pt}{0.7}{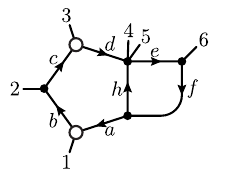}\;\;\text{{\small$-\br{6,\!\edgeA}\brFour{1}{\edgeB}{\edgeC}{3}+\frac{1}{2}\Big(\!\brSix{1}{\edgeB}{\edgeC}{3}{6}{\edgeA}-\edgeC^2\brFour{1}{\edgeB}{\edgeD}{6}+\edgeD^2\brFour{1}{\edgeB}{\edgeC}{6}\!\Big)
$}}}\label{int_16_num}\tag{$\intLabel.{16}$}}%
\eq{\fwboxL{380pt}{\hspace{-10pt}\intLabel_{17}\!\!:\fig{-26.75pt}{0.7}{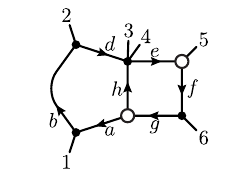}\;\;\text{{\small$-s_{12}\brFour{5}{\edgeF}{\edgeG}{\edgeA}+\frac{1}{2}\Big(\!\brSix{1}{2}{5}{\edgeF}{\edgeG}{\edgeA}+\edgeD^2\brFour{5}{\edgeF}{\edgeG}{1}\!\Big)$}}}\label{int_17_num}\tag{$\intLabel.{17}$}}%
\eq{\fwboxL{380pt}{\hspace{-10pt}\intLabel_{18}\!\!:\fig{-26.75pt}{0.7}{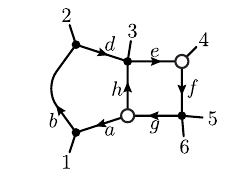}\;\;\text{{\small$-s_{12}\brFour{4}{\edgeF}{\edgeG}{\edgeA}+\frac{1}{2}\Big(\!\brSix{1}{2}{4}{\edgeF}{\edgeG}{\edgeA}+\edgeD^2\brFour{4}{\edgeF}{\edgeG}{1}\!\Big)$}}}\label{int_18_num}\tag{$\intLabel.{18}$}}%
\eq{\fwboxL{380pt}{\hspace{-10pt}\intLabel_{19}\!\!:\fig{-26.75pt}{0.7}{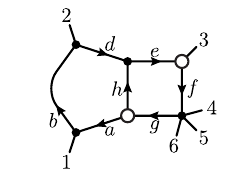}\;\;\text{{\small$-s_{12}\brFour{3}{\edgeF}{\edgeG}{\edgeA}+\frac{1}{2}\Big(\!\brSix{1}{2}{3}{\edgeF}{\edgeG}{\edgeA}+\edgeD^2\brFour{3}{\edgeF}{\edgeG}{1}\!\Big)$}}}\label{int_19_num}\tag{$\intLabel.{19}$}}%
\eq{\fwboxL{380pt}{\hspace{-10pt}\intLabel_{20}\!\!:\fig{-26.75pt}{0.7}{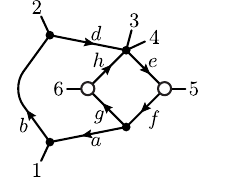}\;\;\text{{\small$\phantom{-}s_{12}\brFour{5}{\edgeF}{\edgeG}{6}$}}}\label{int_20_num}\tag{$\intLabel.{20}$}}%
\eq{\fwboxL{380pt}{\hspace{-10pt}\intLabel_{21}\!\!:\fig{-26.75pt}{0.7}{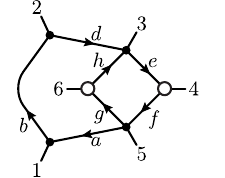}\;\;\text{{\small$\phantom{-}s_{12}\brFour{4}{\edgeF}{\edgeG}{6}$}}}\label{int_21_num}\tag{$\intLabel.{21}$}}%
\eq{\fwboxL{380pt}{\hspace{-10pt}\intLabel_{22}\!\!:\fig{-26.75pt}{0.7}{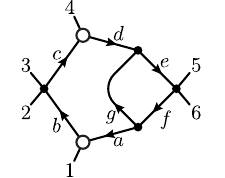}\;\;\text{{\small$\phantom{-}\brFour{1}{\edgeB}{\edgeC}{4}s_{56}$}}}\label{int_22_num}\tag{$\intLabel.{22}$}}%
\eq{\fwboxL{380pt}{\hspace{-10pt}\intLabel_{23}\!\!:\fig{-26.75pt}{0.7}{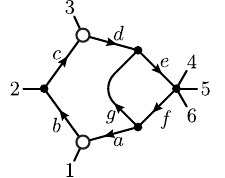}\;\;\text{{\small$\phantom{-}\brFour{1}{\edgeB}{\edgeC}{3}s_{456}$}}}\label{int_23_num}\tag{$\intLabel.{23}$}}%
\eq{\fwboxL{380pt}{\hspace{-10pt}\intLabel_{24}\!\!:\fig{-26.75pt}{0.7}{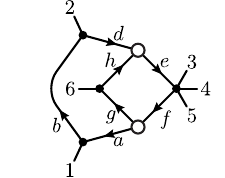}\;\;\text{{\small$-s_{12}\Big(\brFour{\edgeA}{\edgeG}{\edgeH}{\edgeD}-\edgeA^2\edgeH^2-\edgeD^2\edgeG^2\Big)$}}}\label{int_24_num}\tag{$\intLabel.{24}$}}%
\eq{\fwboxL{380pt}{\hspace{-10pt}\intLabel_{25}\!\!:\fig{-26.75pt}{0.7}{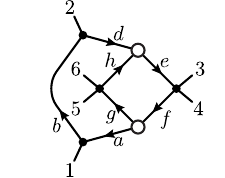}\;\;\text{{\small$-s_{12}\Big(\brFour{\edgeA}{\edgeG}{\edgeH}{\edgeD}-\edgeA^2\edgeH^2-\edgeD^2\edgeG^2\Big)$}}}\label{int_25_num}\tag{$\intLabel.{25}$}}%
\eq{\fwboxL{380pt}{\hspace{-10pt}\intLabel_{26}\!\!:\fig{-26.75pt}{0.7}{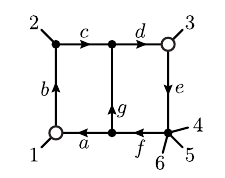}\;\;\text{{\small$-\brSix{1}{\edgeB}{\edgeC}{\edgeF}{\edgeE}{3}$}}}\label{int_26_num}\tag{$\intLabel.{26}$}}%
\eq{\fwboxL{380pt}{\hspace{-10pt}\intLabel_{27}\!\!:\fig{-26.75pt}{0.7}{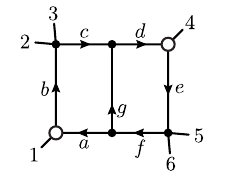}\;\;\text{{\small$-\brSix{1}{\edgeB}{\edgeC}{\edgeF}{\edgeE}{4}$}}}\label{int_27_num}\tag{$\intLabel.{27}$}}%
\vspace{-4pt}\eq{\fwboxL{380pt}{\hspace{-10pt}\intLabel_{28}\!\!:\fig{-26.75pt}{0.7}{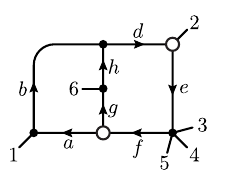}\;\;\text{{\small$-\brSix{1}{\edgeH}{\edgeG}{\edgeF}{\edgeE}{2}+\frac{1}{2}\big(\edgeG^2+\edgeH^2\big)\brFour{2}{\edgeE}{\edgeF}{1}
$}}}\label{int_28_num}\tag{$\intLabel.{28}$}}%
\vspace{-4pt}\eq{\fwboxL{380pt}{\hspace{-10pt}\intLabel_{29}\!\!:\fig{-26.75pt}{0.7}{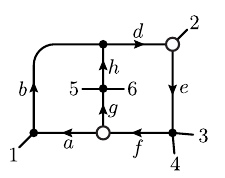}\;\;\text{{\small$-\brSix{1}{\edgeH}{\edgeG}{\edgeF}{\edgeE}{2}+\frac{1}{2}\big(\edgeG^2+\edgeH^2\big)\brFour{2}{\edgeE}{\edgeF}{1}$}}}\label{int_29_num}\tag{$\intLabel.{29}$}}%
\vspace{-4pt}\eq{\fwboxL{380pt}{\hspace{-10pt}\intLabel_{30}\!\!:\fig{-26.75pt}{0.7}{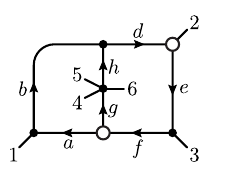}\;\;\text{{\small$-\brSix{1}{\edgeH}{\edgeG}{\edgeF}{\edgeE}{2}+\frac{1}{2}\big(\edgeG^2+\edgeH^2\big)\brFour{2}{\edgeE}{\edgeF}{1}$}}}\label{int_30_num}\tag{$\intLabel.{30}$}}%
\vspace{-4pt}\eq{\fwboxL{380pt}{\hspace{-10pt}\intLabel_{31}\!\!:\fig{-26.75pt}{0.7}{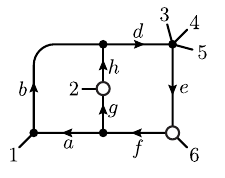}\;\;\text{{\small$\phantom{-}\frac{1}{2}\Big(\!\brSix{1}{\edgeD}{\edgeE}{6}{\edgeG}{\edgeH}-\brSix{6}{\edgeE}{\edgeD}{1}{\edgeG}{\edgeH}\!\Big)$}}}\label{int_31_num}\tag{$\intLabel.{31}$}}%
\eq{\fwboxL{380pt}{\hspace{-10pt}\intLabel_{32}\!\!:\fig{-26.75pt}{0.7}{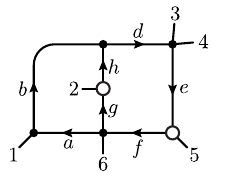}\;\;\text{{\small$\phantom{-}\frac{1}{2}\Big(\!\brSix{1}{\edgeD}{\edgeE}{5}{\edgeG}{\edgeH}-\brSix{5}{\edgeE}{\edgeD}{1}{\edgeG}{\edgeH}\!\Big)$}}}\label{int_32_num}\tag{$\intLabel.{32}$}}%
\eq{\fwboxL{380pt}{\hspace{-10pt}\intLabel_{33}\!\!:\fig{-26.75pt}{0.7}{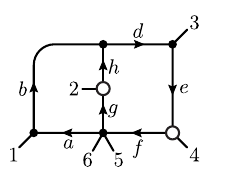}\;\;\text{{\small$\phantom{-}\frac{1}{2}\Big(\!\brSix{1}{\edgeD}{\edgeE}{4}{\edgeG}{\edgeH}-\brSix{4}{\edgeE}{\edgeD}{1}{\edgeG}{\edgeH}\!\Big)$}}}\label{int_33_num}\tag{$\intLabel.{33}$\vspace{-5pt}}}%
\vspace{-8pt}\eq{\fwboxL{380pt}{\hspace{-10pt}\intLabel_{34}\!\!:\fig{-26.75pt}{0.7}{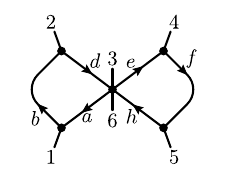}\;\;\text{{\small$\phantom{-}s_{12}s_{45}$}}}\label{int_34_num}\tag{$\intLabel.{34}$}\vspace{-5pt}}%
\vspace{-5pt}\eq{\fwboxL{380pt}{\hspace{-10pt}\intLabel_{35}\!\!:\fig{-26.75pt}{0.7}{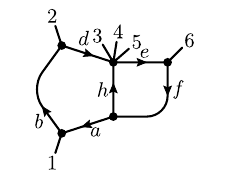}\;\;\text{{\small$-s_{12}\br{6,\!\edgeA}+\frac{1}{2}\Big(\!\brFour{1}{2}{6}{\edgeA}-\brFour{1}{\edgeB}{\edgeD}{6}+\edgeD^2s_{16}\!\Big)$}}}\label{int_35_num}\tag{$\intLabel.{35}$}\vspace{-5pt}}%
\vspace{-2pt}\eq{\fwboxL{380pt}{\hspace{-10pt}\intLabel_{36}\!\!:\fig{-26.75pt}{0.7}{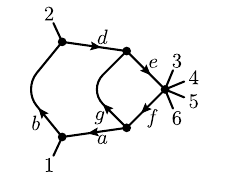}\;\;\text{{\small$\phantom{-}s_{12}^2$}}}\label{int_36_num}\tag{$\intLabel.{36}$}\vspace{-5pt}}%
\vspace{-8pt}\eq{\fwboxL{380pt}{\hspace{-10pt}\intLabel_{37}\!\!:\fig{-26.75pt}{0.7}{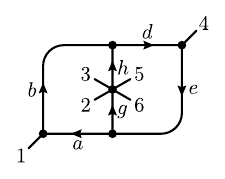}\;\;\text{{\small$-\brFour{1}{\edgeH}{\edgeG}{4}+\frac{1}{2}\big(\edgeG^2+\edgeH^2\big)s_{14}$}}}\label{int_37_num}\tag{$\intLabel.{37}$}\vspace{-5pt}}
\vspace{-8pt}\eq{\fwboxL{380pt}{\hspace{-10pt}\intLabel_{38}\!\!:\fig{-26.75pt}{0.7}{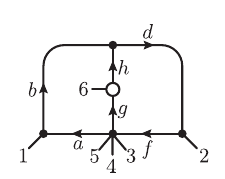}
\;\;\text{{\small$\frac{1}{2}\Big(\!\brFour{1}{2}{\edgeG}{\edgeH}-\brFour{2}{1}{\edgeG}{\edgeH}\!\Big)$}}}\label{int_38_num}\tag{$\intLabel.{38}$}\vspace{-5pt}}

%================================================================================================================
%    Appendix B:  
%================================================================================================================
\newpage\vspace{-0pt}\section[\mbox{Explicit Leading Singularities for sYM and SUGRA}]{\mbox{\hspace{0pt}Explicit Leading Singularities for sYM and SUGRA}}\label{appendix:explicit_coefficients}\vspace{-0pt}

In this appendix, we give explicit representations of all the leading singularities enumerated in \mbox{table~\ref{six_point_ls_table}} for six-point MHV amplitudes in sYM and SUGRA. We will start with Yang-Mills theory because the presence of non-kinematic color-factors requires a bit more technical machinery to be developed. Formulae for SUGRA are comparatively simple, and are enumerated in \ref{appendix:ls_of_sugra}.

%======================================================
\subsection{Color-Dressed Leading Singularities of MHV Amplitudes in \textbf{sYM}}\label{appendix:ls_of_sym}
%======================================================

Let us start with a discussion of how Bose-symmetric (tree) amplitudes involving gluons may be represented and sewn together into unambiguous color-dressed leading singularities in gauge theory. Let us use a calligraphic $\mathcal{A}_{n}^{\text{YM}}$ to represent the color-dressed amplitude. We may disentangle the color structure from the kinematics (non-uniquely) in the way described by  Del Duca, Dixon and Maltoni (DDM) in \mbox{ref.\ \cite{DelDuca:1999rs}}. (See also \cite{Ochirov:2016ewn}.) From the $n$ external particles of an amplitude $\mathcal{A}_{n}^{\text{YM}}$, two arbitrary legs ${\color{hred}\alpha}$ and ${\color{hred}\beta}$ are chosen; in terms of these, we can represent the amplitude as 
\eq{\mathcal{A}_{n}^{\text{YM}}({\color{hred}\alpha},{\color{black}A},{\color{hred}\beta})=\sum_{{\color{hblue}\vec{a}}\in\mathfrak{S}\hspace{-1pt}(\hspace{-1pt}{\color{black}A}\hspace{-0.75pt})}f_{{\color{hred}\alpha\;\beta}}^{\;\,{\color{hblue}\vec{a}}}{\color{hblue}A_n^{\text{YM}}}({\color{hred}\alpha},{\color{hblue}\vec{a}},{\color{hred}\beta})\,,
\label{ddmification}}
where the sum is taken over the \mbox{$(n-2)!$} permutations ${\color{hblue}\vec{a}}\equivL({\color{hblue}a_1},{\color{hblue}\ldots},{\color{hblue}a_{\text{-}1}})$ of the (unordered) set ${\color{black}A}\equivR[n]\backslash\{{\color{hred}\alpha},{\color{hred}\beta}\}$, denoted $\mathfrak{S}(A)$. Notice that (following notational conventions familiar to {\sc Mathematica} users) we have used `$\text{-}1$' to denote the final entry of an ordered list. The color-factor appearing in (\ref{ddmification}) denotes 
\eq{f_{{\color{hred}\alpha\;\beta}}^{\;\,{\color{hblue}\vec{a}}}\equivR\sum_{e_i}f^{{\color{hred}\alpha}\,{\color{hblue}a_1}\,e_1}f^{e_1\,{\color{hblue}a_2}\,e_2}\cdots f^{e_{\text{-}1}\,{\color{hblue}a_{\text{-}1}}\,{\color{hred}\beta}}\equivL\fig{-15pt}{1}{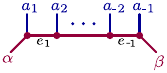}\equivL\fig{-15pt}{1}{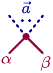}\label{notational_conventions_for_ddm_fs}}
where the $f^{a\,b\,c}$ are structure constants in some Lie algebra. Notice that the graphical notation of (\ref{notational_conventions_for_ddm_fs}) should be understood with some care: ${\color{hblue}\vec{a}}$ is absolutely ordered, and (unless ${\color{hblue}\vec{a}}$ consists of a single leg) $f_{{\color{hred}\alpha\;\beta}}^{\;\,{\color{hblue}\vec{a}}}$ is not cyclically symmetric. Finally, notice that we may liberally extend (\ref{notational_conventions_for_ddm_fs}) to include the case ${\color{hblue}\vec{a}}={\color{hblue}\{\}}$, where $f_{{\color{hred}\alpha}\;{\color{hred}\beta}}^{\;{\color{hblue}\{\}}}\equivR\delta_{{\color{hred}\alpha}\,{\color{hred}\beta}}$.

As already seen above, we have used `spherical' vertices to denote fully Bose-symmetric tree amplitudes in both sYM and SUGRA. The color-ordered partial tree-amplitudes appearing in the DDM expansion (\ref{ddmification}), ${\color{hblue}A_{n}^{\text{YM}}}$, will be denoted by `flat' vertices according to the usual conventions. Thus, the expansion (\ref{ddmification}) becomes graphically, 
\vspace{-5pt}\eq{\fig{-22.125pt}{1.25}{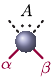}=\sum_{{\color{hblue}\vec{a}}\in\mathfrak{S}\hspace{-1pt}(\hspace{-1pt}{\color{black}A}\hspace{-0.75pt})}\;\;\fig{-22.125pt}{1.25}{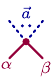}\hspace{-5pt}\times\hspace{-5pt}\fig{-22.125pt}{1.25}{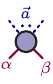}\,.\label{graphical_ddmification}\vspace{-5pt}}
The careful reader will notice that we are consistently coloring \emph{{\color{hblue}ordered}} lists of external legs in {\color{hblue}blue}, while \emph{unordered} sets of legs are indicated in black. Also following familiar conventions, the blue vertices indicating amplitudes in (\ref{graphical_ddmification}) will always be taken to mean MHV amplitudes, while white vertices will be taken to mean $\overline{\text{MHV}}$ (three-point) amplitudes. 

This graphical notation can be used to represent color-dressed leading singularities in sYM as (sums of) color factors times kinematic factors; for instance:
\eq{\hspace{0pt}\fwbox{200pt}{
\hspace{-15pt}\fwbox{120pt}{
%\hbAfig
\fig{-35pt}{1}{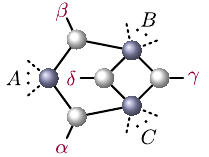}
}\hspace{-10pt}%
=\hspace{5pt}\displaystyle\sum_{\hspace{-13.5pt}\substack{\fwboxR{30pt}{{\color{hblue}\vec{a}}\!\in\!\mathfrak{S}\hspace{-1pt}(\hspace{-1pt}A\hspace{-0.75pt})}\\[-2pt]
\fwboxR{30pt}{(\hspace{-1pt}{\color{hblue}\vec{b}_1}\hspace{-1pt}|\hspace{-0.5pt}{\color{hblue}\vec{b}_2}\hspace{-1pt})\!\!\in\!\mathfrak{S}\hspace{-1pt}(\hspace{-1pt}B\hspace{-0.75pt})}\\
\fwboxR{30pt}{(\hspace{-1pt}{\color{hblue}\vec{c}_1}\hspace{-1pt}|\hspace{-0.5pt}{\color{hblue}\vec{c}_2}\hspace{-1pt})\!\!\in\!\mathfrak{S}\hspace{-1pt}(\hspace{-1pt}C\hspace{-0.75pt})}\\}\hspace{-7.5pt}}\fig{-35pt}{1}{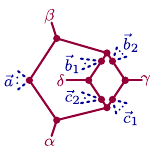}\times\fig{-35pt}{1}{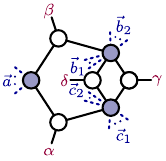}\;.}
\label{ddm_detail_of_hexaboxA}}
Let's discuss each of these ingredients in turn. 

\subsubsection{\texorpdfstring{Color Factors at Two Loops: \emph{Diagrammatics \& Notation}}{Color Factors at Two Loops: Diagrammatics & Notation}}

It is not hard to see that color structures with the topological structure of those appearing in the example (\ref{ddm_detail_of_hexaboxA}) are sufficient to represent all contributions at two loops. We could define a (maximally, manifestly) symmetric color-factor
\eq{\tilde{f}\bigger{\big[}{\color{hblue}\vec{a}},{\color{hblue}\vec{b}},{\color{hblue}\vec{c}}\bigger{\big]}\equivR\fig{-17.125pt}{1}{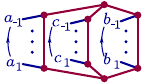}=\fig{-17.125pt}{1}{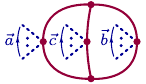}=\sum_{{\color{hred}e_i}}f^{{\color{hred}e_1\,e_3\,e_5}}f_{{\color{hred}e_1\;e_2}}^{\;\;\,\,{\color{hblue}\vec{a}}}f_{{\color{hred}e_3\;e_4}}^{\;\;\,\,{\color{hblue}\vec{c}}}f_{{\color{hred}e_5\;e_6}}^{\;\;\,\,{\color{hblue}\vec{b}}}f^{{\color{hred}e_2\,e_6\,e_4}}
\label{defn_of_symmetric_color_factor}}
where ${\color{hblue}\vec{a}}\equivR({\color{hblue}a_1},{\color{hblue}\ldots},{\color{hblue}a_{\text{-}1}})$ is an ordered list of external leg labels, and similarly for ${\color{hblue}\vec{b}}$ and ${\color{hblue}\vec{c}}$. These `symmetric' color-factors enjoy $\mathfrak{S}_3\times\mathbb{Z}_2$ symmetry generated by the equivalences
\eq{\tilde{f}\bigger{\big[}{\color{hblue}\vec{a}},{\color{hblue}\vec{b}},{\color{hblue}\vec{c}}\bigger{\big]}=\tilde{f}\bigger{\big[}{\color{hblue}\vec{b}},{\color{hblue}\vec{a}},{\color{hblue}\vec{c}}\bigger{\big]}=\tilde{f}\bigger{\big[}{\color{hblue}\vec{c}},{\color{hblue}\vec{b}},{\color{hblue}\vec{a}}\bigger{\big]}=\tilde{f}\bigger{\big[}{\color{hblue}\vec{a}}^{\,R},{\color{hblue}\vec{b}}^{\,R},{\color{hblue}\vec{c}}^{\,R}\bigger{\big]}\,,\label{symmetries_of_symmetric_fs}
}
where ${\color{hblue}\vec{a}}^{\,R}\equivR({\color{hblue}a_{\text{-}1}},{\color{hblue}\ldots},{\color{hblue}a_1})$ is the `reversal' of ${\color{hblue}\vec{a}}$. 

For our own purposes, we find it more convenient to introduce a (notationally) less-manifestly-symmetric color-factor function
\eq{\fwbox{0pt}{\hspace{-30pt}f\!\bigger{\big[}{\color{hblue}\vec{a}},{\color{hblue}\vec{b}},{\color{hblue}\vec{c}}\bigger{\big]}\equivR(\text{-}1)^{|{\color{hblue}\vec{b}}|}\tilde{f}\bigger{\big[}{\color{hblue}\vec{a}},{\color{hblue}\vec{b}}^{\,R},{\color{hblue}\vec{c}}\bigger{\big]}\!=\fig{-17.125pt}{1}{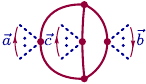}
=\sum_{{\color{hred}e_i}}f^{{\color{hred}e_1\,e_3\,e_6}}f_{{\color{hred}e_1\;e_2}}^{\;\;\,\,{\color{hblue}\vec{a}}}f_{{\color{hred}e_3\;e_4}}^{\;\;\,\,{\color{hblue}\vec{c}}}f_{{\color{hred}e_5\;e_6}}^{\;\;\,\,{\color{hblue}\vec{b}}}f^{{\color{hred}e_2\,e_5\,e_4}} .}\label{our_color_factors}}
These color factors enjoy the same symmetries as $\tilde{f}\bigger{\big[}{\color{hblue}\vec{a}},{\color{hblue}\vec{b}},{\color{hblue}\vec{c}}\bigger{\big]}$ in (\ref{symmetries_of_symmetric_fs}), but with a bit more notational complication (for which we apologize):
\eq{f\!\bigger{\big[}{\color{hblue}\vec{a}},{\color{hblue}\vec{b}},{\color{hblue}\vec{c}}\bigger{\big]}=f\!\bigger{\big[}{\color{hblue}\vec{c}},{\color{hblue}\vec{b}},{\color{hblue}\vec{a}}\bigger{\big]}=(\text{-}1)^{|{\color{hblue}\vec{a}}|+|{\color{hblue}\vec{b}}|}f\!\bigger{\big[}{\color{hblue}\vec{b}}^{\,R},{\color{hblue}\vec{a}}^{\,R},{\color{hblue}\vec{c}}\bigger{\big]}=f\!\bigger{\big[}{\color{hblue}\vec{a}}^{\,R},{\color{hblue}\vec{b}}^{\,R},{\color{hblue}\vec{c}}^{\,R}\bigger{\big]}\,.\label{symmetries_of_our_fs}
}
In terms of the color-factor function defined in (\ref{our_color_factors}), it is easy to see that the color-factor graph appearing in (\ref{ddm_detail_of_hexaboxA}) would be expressed
\eq{\fig{-35pt}{1}{ddmification_of_hexaboxA_colors}=f\!\bigger{\big[}({\color{hred}\alpha}|{\color{hblue}\vec{a}}|{\color{hred}\beta}),({\color{hblue}\vec{b}_2}|{\color{hred}\gamma}|{\color{hblue}\vec{c}_1}),({\color{hblue}\vec{c}_2}|{\color{hred}\delta}|{\color{hblue}\vec{b}_1})\bigger{\big]}\,,}
where `$(\cdot|\cdots|\cdot)$' denotes the concatenation of ordered lists. For any particular Lie algebra, the structure constants defining these color factors can be combined and summed, and decomposed however one may wish. For the case of $SU(N_c)$ it is extremely straightforward to express each of these in terms of multi-trace terms in the $1/N_c$ expansion, if desired~\cite{Dixon:1996wi}.

%============================================================================
\subsubsection{\texorpdfstring{Kinematics: On-Shell Functions of \emph{Ordered} (MHV) Amplitudes}{Kinematics: On-Shell Functions of Ordered (MHV) Amplitudes}}
%============================================================================

Let us now discuss the kinematic parts of leading singularities such as those appearing in the summand of the example (\ref{ddm_detail_of_hexaboxA}) above:
\eq{\fig{-35pt}{1}{ddmification_of_hexaboxA_kinematics}\equivL\Gamma\!\bigger{\big[}({\color{hred}\alpha}|{\color{hblue}\vec{a}}|{\color{hred}\beta}),({\color{hred}\delta}|{\color{hblue}\vec{b}_1}|{\color{hred}\beta}|{\color{hblue}\vec{b}_2}|{\color{hred}\gamma}),({\color{hred}\gamma}|{\color{hblue}\vec{c}_1}|{\color{hred}\alpha}|{\color{hblue}\vec{c}_2}|{\color{hred}\delta})\bigger{\big]}\,.\label{gamma_defined_by_example}}
We would like to define a function $\Gamma$ which depends on the (clockwise) cyclic ordering of external legs involved in each MHV (blue) vertex of the graph (including any external legs connected to it via a chain of $\overline{\text{MHV}}$ vertices). For two-loop MHV leading singularities, $\Gamma$ will always depend on exactly three (cyclically) ordered lists $\Gamma\!\big[{\color{hblue}\vec{a}},{\color{hblue}\vec{b}},{\color{hblue}\vec{c}}\big]$. 

Notice that, unlike $f\!\big[{\color{hblue}\vec{a}},{\color{hblue}\vec{b}},{\color{hblue}\vec{c}}\big]$, $\Gamma$'s arguments include repetition. For example, in (\ref{gamma_defined_by_example}), the leg labelled ${\color{hred}\alpha}$ appears in its first and third entries, as both of these MHV vertices involve the leg ${\color{hred}\alpha}$ through the $\overline{\text{MHV}}$ vertex. Also notice that the ordering of the arguments of $\Gamma$ has absolutely no meaning, and so $\Gamma$ is fully $\mathfrak{S}_3$ invariant. Finally, as already emphasized above, the arguments of $\Gamma$ are \emph{separately} cyclically invariant.\footnote{Being built from ordered MHV amplitudes, they also enjoy a dihedral symmetry, but at the possible cost of an overall sign.} An explicit definition of $\Gamma$ was given in \mbox{ref.\ \cite{Bourjaily:2018omh}}, to which we refer the reader for a more thorough discussion; however, it is worth quoting the two primary formulae derived there. Recall the famous Parke-Taylor formula \cite{Parke:1986gb} for color-ordered MHV amplitudes\footnote{We are neglecting the factor for supermomentum conservation $\delta^{2\times2}\!\big(\sum_a p_a\big)\delta^{2\times4}\!\big(\sum_{a}\lambda_a\widetilde{\eta}_a\big)$.} involving a cyclically-ordered set of legs ${\color{hblue}\vec{\sigma}}$ 
\eq{\text{PT}({\color{hblue}\vec{\sigma}})=\text{PT}({\color{hblue}\sigma_1},{\color{hblue}\ldots},{\color{hblue}\sigma_{\text{-}1}})\equivR\frac{1}{\ab{{\color{hblue}\sigma_1}\,{\color{hblue}\sigma_2}}\cdots\ab{{\color{hblue}\sigma_{\text{-}1}\sigma_1}}}\,.}
The first and perhaps most conceptually simple formula for $\Gamma\!\big[{\color{hblue}\vec{a}},{\color{hblue}\ldots},{\color{hblue}\vec{c}}\big]$ is that it is equal to the sum of $\text{PT}\big({\color{hblue}\vec{\sigma}}\big)$ for all cyclic permutations ${\color{hblue}\vec{\sigma}}$ consistent with the \emph{cyclic} orderings of each of the arguments ${\color{hblue}\vec{a}},{\color{hblue}\ldots},{\color{hblue}\vec{c}}$. More formally, we could write
\vspace{-0pt}\eq{\Gamma\big[{\color{hblue}\vec{a}},{\color{hblue}\ldots},{\color{hblue}\vec{c}}\big]=\sum_{\fwboxL{30pt}{\hspace{-5pt}{\color{hblue}\vec{\sigma}}\!\in\!{\color{hblue}\vec{a}}\!\shuffle\!{\color{hblue}\vec{b}}\!\shuffle\!\cdots\!\shuffle\!{\color{hblue}\vec{c}}}}\text{PT}({\color{hblue}\vec{\sigma}})\,,\label{shuffle_sum_formulae}\vspace{-0pt}}
where the symbol `$\shuffle$' is being used (somewhat unusually) to denote what was described above: namely, all `cyclic shuffles' of each list. Notice that, unlike other uses and definitions of `shuffle', duplicated elements of distinct lists may appear; they must be aligned between terms and are not repeated in the shuffle. Although this usage of shuffle and the symbol `$\shuffle$' is admittedly a bit unfamiliar, we believe it to be in the same spirit of ordinary shuffle sums, and therefore have allowed ourselves this slight abuse of notation. Using (\ref{shuffle_sum_formulae}), we could express (\ref{gamma_defined_by_example}) as
\vspace{-2pt}\eq{\begin{split}~\\[-28pt]\fwbox{0pt}{\hspace{-5pt}\Gamma\!\bigger{\big[}({\color{hred}\alpha}|{\color{hblue}\vec{a}}|{\color{hred}\beta}),({\color{hred}\delta}|{\color{hblue}\vec{b}_1}|{\color{hred}\beta}|{\color{hblue}\vec{b}_2}|{\color{hred}\gamma}),({\color{hred}\gamma}|{\color{hblue}\vec{c}_1}|{\color{hred}\alpha}|{\color{hblue}\vec{c}_2}|{\color{hred}\delta})\bigger{\big]}=\fig{-35pt}{1}{ddmification_of_hexaboxA_kinematics}\hspace{-5pt}=\sum_{\hspace{-0pt}{\color{hblue}\widehat{a}}\in{\color{hblue}\vec{a}}\shuffle({\color{hblue}\vec{c}_2}|{\color{hred}\delta}|{\color{hblue}\vec{b}_1})\hspace{-30pt}}\text{PT}({\color{hred}\alpha},{\color{hblue}\widehat{a}},{\color{hred}\beta},{\color{hblue}\vec{b}_2},{\color{hred}\gamma},{\color{hblue}\vec{c}_1})\,}\\[10pt]~\end{split}\vspace{-35pt}}
\vspace{-40pt}

\noindent where as before `$|$' denotes the concatenation of ordered lists.

While the representation (\ref{shuffle_sum_formulae}) for $\Gamma$ makes it clear that $\Gamma$ only ever has single poles in the external kinematics (and is therefore `local' in the sense discussed in \mbox{ref.\ \cite{Arkani-Hamed:2014bca}}), a more compact (and algorithmically more efficient) formula for $\Gamma$ also exists. This representation is simply
\eq{\Gamma\big[{\color{hblue}\vec{a}},{\color{hblue}\ldots},{\color{hblue}\vec{c}}\big]=\mathfrak{J}^2\times \text{PT}({\color{hblue}\vec{a}})\text{PT}({\color{hblue}\vec{b}})\text{PT}({\color{hblue}\vec{c}})\,;\label{direct_formula_for_gamma}}
the definition of the Jacobian factor $\mathfrak{J}$ is somewhat involved, and so we refer the reader to \mbox{ref.\ \cite{Bourjaily:2018omh}} for details. It is worth remarking that, for the sake of computational efficiency, only the compact formula (\ref{direct_formula_for_gamma}) is used to compute the leading singularities of sYM in the ancillary files for this work.

%============================================================================
\subsubsection{Closed Formulae for All Two-Loop MHV Leading Singularities}
%============================================================================

There are only six classes of leading singularities for two-loop MHV amplitudes at all multiplicity. It is not hard to give a closed formula for each. We colloquially refer to them as `kissing boxes' (\ref{general_kissing_box_ls}), `pentaboxes' (\ref{general_pentabox_ls}), `hexaboxes' A (\ref{general_hexaboxA_ls}) and B (\ref{general_hexaboxB_ls}), and `(non-planar) double pentagons' A  (\ref{general_doublePentagonA_ls}) and B (\ref{general_doublePentagonB_ls}):
\vspace{-5pt}\eq{\hspace{-115pt}\fwboxL{300pt}{\hspace{-15pt}\fwbox{120pt}{
\fig{-35pt}{1}{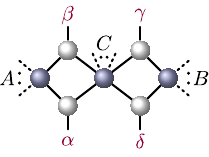}
}\hspace{-10pt}%
=\hspace{5pt}\displaystyle\sum_{\hspace{-13.5pt}\substack{\fwboxR{30pt}{{\color{hblue}\vec{a}}\!\in\!\mathfrak{S}\hspace{-1pt}(\hspace{-1pt}A\hspace{-0.75pt})}\\[-2pt]
\fwboxR{30pt}{{\color{hblue}\vec{b}}\!\in\!\mathfrak{S}\hspace{-1pt}(\hspace{-1pt}B\hspace{-0.75pt})}\\
\fwboxR{30pt}{(\hspace{-1pt}{\color{hblue}\vec{c}_1}\hspace{-1pt}|\hspace{-0.5pt}{\color{hblue}\vec{c}_2}\hspace{-1pt}|\hspace{-0.5pt}{\color{hblue}\vec{c}_3}\hspace{-1pt})\!\!\in\!\mathfrak{S}\hspace{-1pt}(\hspace{-1pt}C\hspace{-0.75pt})}}\hspace{-7.5pt}}
\;\raisebox{-10pt}{$\Bigg($}\hspace{-8pt}\begin{array}{@{$\!$}l@{}}~\\[4pt]\phantom{\times}
f\!\bigger{\big[}({\color{hred}\alpha}|{\color{hblue}\vec{a}}|{\color{hred}\beta}),({\color{hblue}\vec{c}_3}|{\color{hred}\gamma}|{\color{hblue}\vec{b}}|{\color{hred}\delta}|\vec{c_1}),{\color{hblue}\vec{c}_2}\bigger{\big]}\times\\[3pt]
\phantom{\times}\Gamma\!\bigger{\big[}({\color{hred}\alpha}|{\color{hblue}\vec{a}}|{\color{hred}\beta}),({\color{hred}\gamma}|{\color{hblue}\vec{b}}|{\color{hred}\delta}),({\color{hred}\delta}|{\color{hblue}\vec{c}_1}|{\color{hred}\alpha}|{\color{hblue}\vec{c}_2}|{\color{hred}\beta}|{\color{hblue}\vec{c}_3}|{\color{hred}\gamma})\bigger{\big]}\end{array}\,\raisebox{-8pt}{$+({\color{hred}\alpha}\leftrightarrow{\color{hred}\beta})$}\hspace{-2.5pt}\raisebox{-10pt}{$\Bigg)\,$}}\label{general_kissing_box_ls}\tag{KB}}
\vspace{-5pt}\eq{\hspace{-115pt}\fwboxL{300pt}{
\hspace{-15pt}\fwbox{120pt}{
\fig{-35pt}{1}{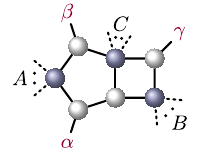}
}\hspace{-10pt}%
=\hspace{5pt}\displaystyle\sum_{\hspace{-13.5pt}\substack{\fwboxR{30pt}{{\color{hblue}\vec{a}}\!\in\!\mathfrak{S}\hspace{-1pt}(\hspace{-1pt}A\hspace{-0.75pt})}\\[-2pt]
\fwboxR{30pt}{{\color{hblue}\vec{b}}\!\in\!\mathfrak{S}\hspace{-1pt}(\hspace{-1pt}B\hspace{-0.75pt})}\\
\fwboxR{30pt}{(\hspace{-1pt}{\color{hblue}\vec{c}_1}\hspace{-1pt}|\hspace{-0.5pt}{\color{hblue}\vec{c}_2}\hspace{-1pt})\!\!\in\!\mathfrak{S}\hspace{-1pt}(\hspace{-1pt}C\hspace{-0.75pt})}}\hspace{-7.5pt}}
\;\raisebox{-10pt}{$\Bigg($}\hspace{-8pt}\begin{array}{@{$\!$}l@{}}~\\[4pt]\phantom{\times}
f\!\bigger{\big[}({\color{hred}\alpha}|{\color{hblue}\vec{a}}|{\color{hred}\beta}),({\color{hblue}\vec{c}_2}|{\color{hred}\gamma}|{\color{hblue}\vec{b}}),{\color{hblue}\vec{c}_1}\bigger{\big]}\times\\[3pt]
\phantom{\times}\Gamma\!\bigger{\big[}({\color{hred}\alpha}|{\color{hblue}\vec{a}}|{\color{hred}\beta}),({\color{hred}\gamma}|{\color{hblue}\vec{b}}|{\color{hred}\alpha}),({\color{hred}\alpha}|{\color{hblue}\vec{c}_1}|{\color{hred}\beta}|{\color{hblue}\vec{c}_2}|{\color{hred}\gamma})\bigger{\big]}\end{array}\raisebox{-10pt}{$\Bigg)\,$}}\label{general_pentabox_ls}\tag{PB}}
\vspace{-5pt}\eq{\hspace{-115pt}\fwboxL{300pt}{
\hspace{-15pt}\fwbox{120pt}{
\fig{-35pt}{1}{hexaboxA_ls}
}\hspace{-10pt}%
=\hspace{5pt}\displaystyle\sum_{\hspace{-13.5pt}\substack{\fwboxR{30pt}{{\color{hblue}\vec{a}}\!\in\!\mathfrak{S}\hspace{-1pt}(\hspace{-1pt}A\hspace{-0.75pt})}\\[-2pt]
\fwboxR{30pt}{(\hspace{-1pt}{\color{hblue}\vec{b}_1}\hspace{-1pt}|\hspace{-0.5pt}{\color{hblue}\vec{b}_2}\hspace{-1pt})\!\!\in\!\mathfrak{S}\hspace{-1pt}(\hspace{-1pt}B\hspace{-0.75pt})}\\
\fwboxR{30pt}{(\hspace{-1pt}{\color{hblue}\vec{c}_1}\hspace{-1pt}|\hspace{-0.5pt}{\color{hblue}\vec{c}_2}\hspace{-1pt})\!\!\in\!\mathfrak{S}\hspace{-1pt}(\hspace{-1pt}C\hspace{-0.75pt})}}\hspace{-7.5pt}}
\;\raisebox{-10pt}{$\Bigg($}\hspace{-8pt}\begin{array}{@{$\!$}l@{}}~\\[4pt]\phantom{\times}
f\!\bigger{\big[}({\color{hred}\alpha}|{\color{hblue}\vec{a}}|{\color{hred}\beta}),({\color{hblue}\vec{b}_2}|{\color{hred}\gamma}|{\color{hblue}\vec{c}_1}),({\color{hblue}\vec{c}_2}|{\color{hred}\delta}|{\color{hblue}\vec{b}_1})\bigger{\big]}\times\\[3pt]
\phantom{\times}\Gamma\!\bigger{\big[}({\color{hred}\alpha}|{\color{hblue}\vec{a}}|{\color{hred}\beta}),({\color{hred}\delta}|{\color{hblue}\vec{b}_1}|{\color{hred}\beta}|{\color{hblue}\vec{b}_2}|{\color{hred}\gamma}),({\color{hred}\gamma}|{\color{hblue}\vec{c}_1}|{\color{hred}\alpha}|{\color{hblue}\vec{c}_2}|{\color{hred}\delta})\bigger{\big]}\end{array}\raisebox{-10pt}{$\Bigg)\,$}}\label{general_hexaboxA_ls}\tag{HBa}}
\vspace{-5pt}\eq{\hspace{-115pt}\fwboxL{300pt}{
\hspace{-15pt}\fwbox{120pt}{
\fig{-35pt}{1}{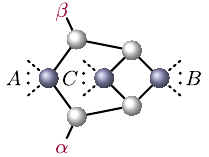}
}\hspace{-10pt}%
=\hspace{5pt}\displaystyle\sum_{\hspace{-13.5pt}\substack{
\fwboxR{30pt}{{\color{hblue}\vec{a}}\!\in\!\mathfrak{S}\hspace{-1pt}(\hspace{-1pt}A\hspace{-0.75pt})}\\[-2pt]
\fwboxR{30pt}{{\color{hblue}\vec{b}}\!\in\!\mathfrak{S}\hspace{-1pt}(\hspace{-1pt}B\hspace{-0.75pt})}\\
\fwboxR{30pt}{{\color{hblue}\vec{c}}\!\in\!\mathfrak{S}\hspace{-1pt}(\hspace{-1pt}C\hspace{-0.75pt})}}\hspace{-7.5pt}}
\;\raisebox{-10pt}{$\Bigg($}\hspace{-8pt}\begin{array}{@{$\!$}l@{}}~\\[4pt]\phantom{\times}
f\!\bigger{\big[}({\color{hred}\alpha}|{\color{hblue}\vec{a}}|{\color{hred}\beta}),{\color{hblue}\vec{b}},{\color{hblue}\vec{c}}\bigger{\big]}\times\\[3pt]
\phantom{\times}\Gamma\!\bigger{\big[}({\color{hred}\alpha}|{\color{hblue}\vec{a}}|{\color{hred}\beta}),({\color{hred}\beta}|{\color{hblue}\vec{b}}|{\color{hred}\alpha}),({\color{hred}\alpha}|{\color{hblue}\vec{c}}|{\color{hred}\beta})\bigger{\big]}\end{array}\raisebox{-10pt}{$\Bigg)\,$}}\label{general_hexaboxB_ls}\tag{HBb}}
\vspace{-5pt}\eq{\hspace{-115pt}\fwboxL{300pt}{
\hspace{-15pt}\fwbox{120pt}{
\fig{-35pt}{1}{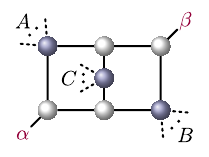}
}\hspace{-10pt}%
=\hspace{5pt}\displaystyle\sum_{\hspace{-13.5pt}\substack{
\fwboxR{30pt}{{\color{hblue}\vec{a}}\!\in\!\mathfrak{S}\hspace{-1pt}(\hspace{-1pt}A\hspace{-0.75pt})}\\[-2pt]
\fwboxR{30pt}{{\color{hblue}\vec{b}}\!\in\!\mathfrak{S}\hspace{-1pt}(\hspace{-1pt}B\hspace{-0.75pt})}\\
\fwboxR{30pt}{{\color{hblue}\vec{c}}\!\in\!\mathfrak{S}\hspace{-1pt}(\hspace{-1pt}C\hspace{-0.75pt})}}\hspace{-7.5pt}}
\;\raisebox{-10pt}{$\Bigg($}\hspace{-8pt}\begin{array}{@{$\!$}l@{}}~\\[4pt]\phantom{\times}
f\!\bigger{\big[}({\color{hred}\alpha}|{\color{hblue}\vec{a}}),({\color{hred}\beta}|{\color{hblue}\vec{b}}),{\color{hblue}\vec{c}}\bigger{\big]}\times\\[3pt]
\phantom{\times}\Gamma\!\bigger{\big[}({\color{hred}\alpha}|{\color{hblue}\vec{a}}|{\color{hred}\beta}),({\color{hred}\beta}|{\color{hblue}\vec{b}}|{\color{hred}\alpha}),({\color{hred}\alpha}|{\color{hblue}\vec{c}}|{\color{hred}\beta})\bigger{\big]}\end{array}\raisebox{-10pt}{$\Bigg)\,$}}\label{general_doublePentagonA_ls}\tag{DPa}}
\vspace{-5pt}\eq{\hspace{-115pt}\fwboxL{300pt}{
\hspace{-15pt}\fwbox{120pt}{
\fig{-35pt}{1}{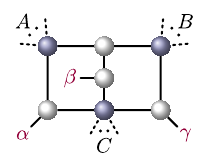}
}\hspace{-10pt}%
=\hspace{5pt}\displaystyle\sum_{\hspace{-13.5pt}\substack{
\fwboxR{30pt}{{\color{hblue}\vec{a}}\!\in\!\mathfrak{S}\hspace{-1pt}(\hspace{-1pt}A\hspace{-0.75pt})}\\[-2pt]
\fwboxR{30pt}{{\color{hblue}\vec{b}}\!\in\!\mathfrak{S}\hspace{-1pt}(\hspace{-1pt}B\hspace{-0.75pt})}\\
\fwboxR{30pt}{(\hspace{-1pt}{\color{hblue}\vec{c}_1}\hspace{-1pt}|\hspace{-0.5pt}{\color{hblue}\vec{c}_2}\hspace{-1pt})\!\!\in\!\mathfrak{S}\hspace{-1pt}(\hspace{-1pt}C\hspace{-0.75pt})}}\hspace{-7.5pt}}
\;\raisebox{-10pt}{$\Bigg($}\hspace{-8pt}\begin{array}{@{$\!$}l@{}}~\\[4pt]\phantom{\times}
f\!\bigger{\big[}({\color{hred}\alpha}|{\color{hblue}\vec{a}}),({\color{hblue}\vec{b}}|{\color{hred}\gamma}|{\color{hblue}\vec{c}_1}),({\color{hblue}\vec{c}_2}|{\color{hred}\beta})\bigger{\big]}\times\\[3pt]
\phantom{\times}\Gamma\!\bigger{\big[}({\color{hred}\alpha}|{\color{hblue}\vec{a}}|{\color{hred}\beta}),({\color{hred}\beta}|{\color{hblue}\vec{b}}|{\color{hred}\gamma}),({\color{hred}\gamma}|{\color{hblue}\vec{c}_1}|{\color{hred}\alpha}|{\color{hblue}\vec{c}_2}|{\color{hred}\beta})\bigger{\big]}\end{array}\raisebox{-10pt}{$\Bigg)\,$}}\label{general_doublePentagonB_ls}\tag{DPb}}

%============================================================================
\paragraph{Smooth Degeneration to Composite Leading Singularities}~\\[-12pt]
%============================================================================

An important and useful feature of the formulae enumerated above is that \emph{dashed leg ranges are allowed to be empty}. What does an `empty' leg range imply physically or functionally? It turns out to be easier to understand than may be expected. Consider for example `hexabox B' (\ref{general_hexaboxB_ls})
\eq{\fig{-35pt}{1}{hexaboxB_ls}\bigger{\underset{C\to\{\}}{\longmapsto}}\fig{-35pt}{1}{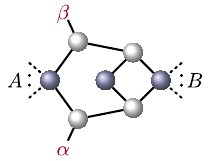}\,.}
A few moments thought will demonstrate that the formula (\ref{general_hexaboxB_ls}) is entirely well behaved in this limit, and in fact reduces to a simpler case:
\eq{\hspace{-115pt}\fwboxL{200pt}{
\hspace{-15pt}\fwbox{120pt}{
\fig{-35pt}{1}{hexaboxB_emptyC}
}\hspace{-10pt}
=\hspace{5pt}\displaystyle\sum_{\hspace{-13.5pt}\substack{
\fwboxR{30pt}{{\color{hblue}\vec{a}}\!\in\!\mathfrak{S}\hspace{-1pt}(\hspace{-1pt}A\hspace{-0.75pt})}\\[-2pt]
\fwboxR{30pt}{{\color{hblue}\vec{b}}\!\in\!\mathfrak{S}\hspace{-1pt}(\hspace{-1pt}B\hspace{-0.75pt})}}\hspace{-7.5pt}}
\;\raisebox{-10pt}{$\Bigg($}\hspace{-8pt}\begin{array}{@{$\!$}l@{}}~\\[4pt]\phantom{\times}
f\!\bigger{\big[}({\color{hred}\alpha}|{\color{hblue}\vec{a}}|{\color{hred}\beta}),{\color{hblue}\vec{b}},{\color{hblue}\{\}}\bigger{\big]}\times\\[3pt]
\phantom{\times}\Gamma\!\bigger{\big[}({\color{hred}\alpha}|{\color{hblue}\vec{a}}|{\color{hred}\beta}),({\color{hred}\beta}|{\color{hblue}\vec{b}}|{\color{hred}\alpha}),({\color{hred}\alpha},{\color{hred}\beta})\bigger{\big]}\end{array}\raisebox{-10pt}{$\Bigg)\,.$}}\label{hexaboxB_in_emptyC_limit}}
It is very easy to see that the kinematic factor is equivalent to a one-loop leading singularity:
\eq{\fwbox{0pt}{\hspace{-29pt}\Gamma\!\bigger{\big[}\hspace{-1pt}(\hspace{-1pt}{\color{hred}\alpha}|{\color{hblue}\vec{a}}|{\color{hred}\beta}\hspace{-1pt})\!,\!(\hspace{-1pt}{\color{hred}\beta}|{\color{hblue}\vec{b}}|{\color{hred}\alpha}\hspace{-1pt})\!,\!(\hspace{-1pt}{\color{hred}\alpha},\!{\color{hred}\beta}\hspace{-1pt})\hspace{-1pt}\bigger{\big]}\hspace{-3pt}=\hspace{-6pt}\fig{-35pt}{1}{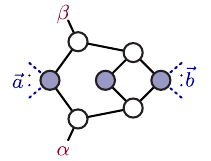}\hspace{-6pt}\simeq\hspace{-4pt}\fig{-35pt}{1}{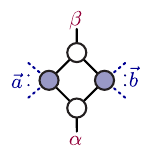}\hspace{-7pt}=\hspace{-1pt}\Gamma\!\bigger{\big[}\hspace{-1pt}(\hspace{-1pt}{\color{hred}\alpha}|{\color{hblue}\vec{a}}|{\color{hred}\beta}\hspace{-1pt})\!,\!(\hspace{-1pt}{\color{hred}\beta}|{\color{hblue}\vec{b}}|{\color{hred}\alpha}\hspace{-1pt})\hspace{-1pt}\bigger{\big]}\!.}}
If this is not entirely obvious, recall the definition of $\Gamma\!\big[{\color{hblue}\vec{a}},\ldots,{\color{hblue}\vec{c}}\big]$ in terms of sums of Parke-Taylor factors consistent with the cyclic ordering of each ordered vector ${\color{hblue}\vec{a}},\ldots,{\color{hblue}\vec{c}}$ (\ref{shuffle_sum_formulae}). A two element list is cyclically consistent with all permutations, and thus has no affect on the summand. 

Due to this aspect of the kinematic part of (\ref{hexaboxB_in_emptyC_limit}), the only difference between this degenerate case (\ref{hexaboxB_in_emptyC_limit}) and an \emph{actual} one-loop leading singularity is the color factor. This is a good thing, as two-loop amplitudes have different powers of the coupling relative to one-loop amplitudes. Indeed, the color factor is easily seen to be \emph{exactly} what would be expected if one approached the composite singularity
\eq{\fig{-35pt}{1}{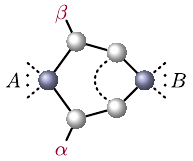}\equivR\hspace{-5pt}\fig{-35pt}{1}{hexaboxB_emptyC}}
where the momentum flowing through the dashed line vanishes. Thus, all the composite leading singularities needed in the representation of six particle amplitudes in sYM are simply special cases of the six general topologies enumerated above. Indeed, in \mbox{table \ref{six_point_ls_table}}, we drew each figure in a way to emphasize this fact: replace any dashed line with a bivalent (`empty') MHV vertex, and apply the appropriate formula from the general cases enumerated above.

\newpage
%============================================================================
\subsection{Two-Loop Leading Singularities of MHV Amplitudes in \textbf{SUGRA}}\label{appendix:ls_of_sugra}
%============================================================================

There does not yet exist any truly satisfactory approach to the computation of general on-shell functions in SUGRA. Although (as with all massless quantum field theories in four dimensions) there exists suggestive representations of on-shell functions in terms of Grassmannian integrals \cite{Herrmann:2016qea}, the actual expressions found for leading singularities in SUGRA still surprise us with their simplicity. 

Indeed, on some level, the only known way to compute the leading singularities of SUGRA is the general way: evaluate all the tree amplitudes involved at the vertices on the solutions to the cut equations, and sum over all states that can be exchanged. This definitional approach is guaranteed to always work (up to perhaps an overall sign\footnote{In the expressions given below, the signs of the were determined via consistency with sYM.}). As the leading singularities needed for MHV amplitudes involve only MHV (and $\overline{\text{MHV}}$) amplitudes at the vertices, we can make use of the tree-level formulae derived by Hodges in \cite{Hodges:2011wm,Hodges:2012ym} to obtain analytic expressions. 

The only ingredients needed for us are the six-point MHV tree amplitude in SUGRA \cite{Hodges:2012ym}, and the one-loop leading singularities, 
\eq{\fwboxL{260pt}{\hspace{-50pt}\fig{-28.5pt}{1}{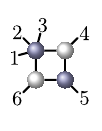}=\phantom{-}\frac{\big(\langle4|(5\text{+}6)|1|3|2|6\rangle-\langle4|(5\text{+}6)|2|3|1|6\rangle\big)\sb{45}\sb{56}}{\ab{12}\ab{13}\ab{14}\ab{16}\ab{23}\ab{24}\ab{26}\ab{34}\ab{36}\ab{45}\ab{56}}}\vspace{-8pt}\label{sugra_ls_11_formula}\tag{LS1}}
and
\eq{\fwboxL{260pt}{\hspace{-50pt}\fig{-28.5pt}{1}{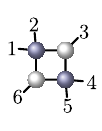}=\phantom{-}\frac{[3|(4\text{+}5)|6|(5\text{+}4)|3\rangle\sb{12}\sb{45}}{\ab{12}\ab{13}\ab{16}\ab{23}\ab{26}\ab{34}\ab{35}\ab{45}\ab{46}\ab{56}}\,.}\vspace{-4pt}\label{sugra_ls_12_formula}\tag{LS2}}
As for sYM, we have systematically neglected the ubiquitous factor imposing super-momentum conservation, $\delta^{2\times2}\big(\lambda\!\cdot\!\widetilde{\lambda}\big)\delta^{2\times8}\big(\lambda\!\cdot\!\widetilde{\eta}\big)$. With these, we find the following expressions for the leading singularities of \mbox{table \ref{six_point_ls_table}} in the case of $\mathcal{N}=8$ supergravity:
\vspace{-5pt}\eq{\fwboxL{260pt}{\hspace{-80pt}\fwboxL{12.5pt}{{\color{hred}\lsLabel_{1}^{8}}}=\fig{-28.5pt}{1}{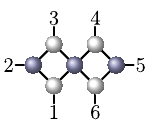}=\phantom{-}\frac{s_{456}\sb{12}\sb{23}\sb{45}\sb{56}}{\ab{12}\ab{14}\ab{16}\ab{23}\ab{34}\ab{36}\ab{45}\ab{56}}}\vspace{-4pt}\label{sugra_ls_1_formula}\tag{${\color{hred}\lsLabel_{1}^{8}}$}}
\vspace{-4pt}\eq{\fwboxL{260pt}{\hspace{-80pt}\fwboxL{12.5pt}{{\color{hred}\lsLabel_{2}^{8}}}=\fig{-28.5pt}{1}{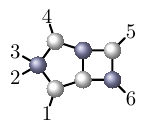}=\phantom{-}\frac{[1|(2\text{+}3)|4\rangle[4|(2\text{+}3)|1\rangle\sb{56}^{2}\sb{23}}{\ab{12}\ab{13}\ab{14}\ab{16}\ab{23}\ab{24}\ab{34}\ab{45}\ab{56}}}\vspace{-4pt}\label{sugra_ls_2_formula}\tag{${\color{hred}\lsLabel_{2}^{8}}$}}
\vspace{-4pt}\eq{\fwboxL{260pt}{\hspace{-80pt}\fwboxL{12.5pt}{{\color{hred}\lsLabel_{3}^{8}}}=\fig{-28.5pt}{1}{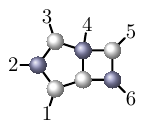}=\phantom{-}\frac{[4|(2\text{+}3)|1\rangle[6|(4\text{+}5)|3\rangle\sb{12}\sb{23}\sb{56}}{\ab{12}\ab{13}\ab{14}\ab{16}\ab{23}\ab{34}\ab{35}\ab{45}\ab{56}}}\vspace{-4pt}\label{sugra_ls_3_formula}\tag{${\color{hred}\lsLabel_{3}^{8}}$}}
\vspace{-4pt}\eq{\fwboxL{260pt}{\hspace{-80pt}\fwboxL{12.5pt}{{\color{hred}\lsLabel_{4}^{8}}}=\fig{-28.5pt}{1}{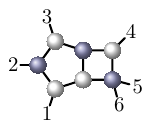}=\phantom{-}\frac{s_{456}[4|(2\text{+}3)|1\rangle\sb{12}\sb{23}\sb{56}}{\ab{12}\ab{13}\ab{15}\ab{16}\ab{23}\ab{34}\ab{45}\ab{46}\ab{56}}}\vspace{-4pt}\label{sugra_ls_4_formula}\tag{${\color{hred}\lsLabel_{4}^{8}}$}}
\vspace{-4pt}\eq{\fwboxL{260pt}{\hspace{-80pt}\fwboxL{12.5pt}{{\color{hred}\lsLabel_{5}^{8}}}=\fig{-28.5pt}{1}{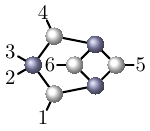}=-\frac{[1|(2\text{+}3)|4\rangle[4|(2\text{+}3)|1\rangle\sb{56}^{2}\sb{23}}{\ab{12}\ab{13}\ab{15}\ab{16}\ab{23}\ab{24}\ab{34}\ab{45}\ab{46}}}\vspace{-4pt}\label{sugra_ls_5_formula}\tag{${\color{hred}\lsLabel_{5}^{8}}$}}
\vspace{-4pt}\eq{\fwboxL{260pt}{\hspace{-80pt}\fwboxL{12.5pt}{{\color{hred}\lsLabel_{6}^{8}}}=\fig{-28.5pt}{1}{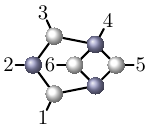}=\phantom{-}\frac{[4|(2\text{+}3)|1\rangle[5|(1\text{+}2)|3\rangle[6|(1\text{+}2)|3\rangle\sb{12}\sb{23}}{\ab{12}\ab{13}\ab{15}\ab{16}\ab{23}\ab{34}\ab{35}\ab{36}\ab{45}\ab{46}}}\vspace{-4pt}\label{sugra_ls_6_formula}\tag{${\color{hred}\lsLabel_{6}^{8}}$}}
\vspace{-4pt}\eq{\fwboxL{260pt}{\hspace{-80pt}\fwboxL{12.5pt}{{\color{hred}\lsLabel_{7}^{8}}}=\fig{-28.5pt}{1}{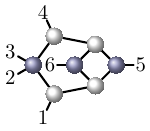}=-\frac{[1|(2\text{+}3)|4\rangle[4|(2\text{+}3)|1\rangle\sb{56}^{2}\sb{23}}{\ab{12}\ab{13}\ab{15}\ab{16}\ab{23}\ab{24}\ab{34}\ab{45}\ab{46}}}\vspace{-4pt}\label{sugra_ls_7_formula}\tag{${\color{hred}\lsLabel_{7}^{8}}$}}
\vspace{-4pt}\eq{\fwboxL{260pt}{\hspace{-80pt}\fwboxL{12.5pt}{{\color{hred}\lsLabel_{8}^{8}}}=\fig{-28.5pt}{1}{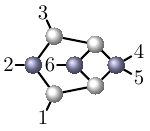}=\phantom{-}\frac{[6|(2\text{+}3)|1\rangle[6|(4\text{+}5)|3\rangle\sb{12}\sb{23}\sb{45}}{\ab{12}\ab{14}\ab{15}\ab{16}\ab{23}\ab{34}\ab{35}\ab{36}\ab{45}}}\vspace{-4pt}\label{sugra_ls_8_formula}\tag{${\color{hred}\lsLabel_{8}^{8}}$}}
\vspace{-4pt}\eq{\fwboxL{260pt}{\hspace{-80pt}\fwboxL{12.5pt}{{\color{hred}\lsLabel_{9}^{8}}}=\fig{-28.5pt}{1}{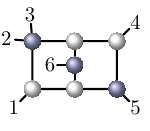}=-\frac{[1|(2\text{+}3)|4\rangle[6|(4\text{+}5)|1\rangle\sb{23}\sb{45}\sb{56}}{\ab{12}\ab{13}\ab{15}\ab{16}\ab{23}\ab{24}\ab{34}\ab{45}\ab{46}}}\vspace{-4pt}\label{sugra_ls_9_formula}\tag{${\color{hred}\lsLabel_{9}^{8}}$}}
\vspace{-4pt}\eq{\fwboxL{260pt}{\hspace{-80pt}\fwboxL{12.5pt}{{\color{hred}\lsLabel_{10}^{8}}}=\fig{-28.5pt}{1}{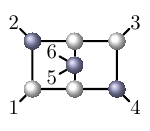}=-\frac{[2|(5\text{+}6)|1\rangle[4|(1\text{+}2)|3\rangle\sb{12}\sb{34}\sb{56}}{\ab{12}\ab{14}\ab{15}\ab{16}\ab{23}\ab{34}\ab{35}\ab{36}\ab{56}}}\vspace{-4pt}\label{sugra_ls_10_formula}\tag{${\color{hred}\lsLabel_{10}^{8}}$}}
\vspace{-4pt}\eq{\fwboxL{260pt}{\hspace{-80pt}\fwboxL{12.5pt}{{\color{hred}\lsLabel_{11}^{8}}}=\fig{-28.5pt}{1}{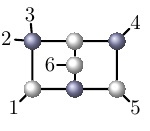}=\phantom{-}\frac{[1|(2\text{+}3)|6\rangle[4|(2\text{+}3)|1\rangle[6|(4\text{+}5)|6\rangle\sb{23}\sb{45}}{\ab{12}\ab{13}\ab{15}\ab{16}\ab{23}\ab{26}\ab{36}\ab{45}\ab{46}\ab{56}}}\vspace{-4pt}\label{sugra_ls_11_formula}\tag{${\color{hred}\lsLabel_{11}^{8}}$}}
\vspace{-4pt}\eq{\fwboxL{260pt}{\hspace{-80pt}\fwboxL{12.5pt}{{\color{hred}\lsLabel_{12}^{8}}}=\fig{-28.5pt}{1}{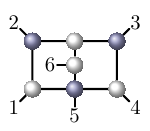}=\phantom{-}\frac{[5|(1\text{+}2)|6\rangle\sb{12}\sb{23}\sb{34}\big(\ab{12}\ab{46}\sb{26}\text{-}\ab{16}\ab{34}\sb{36}\big)}{\ab{12}\ab{14}\ab{15}\ab{16}\ab{26}\ab{34}\ab{36}\ab{45}\ab{46}\ab{56}}}\vspace{-4pt}\label{sugra_ls_12_formula}\tag{${\color{hred}\lsLabel_{12}^{8}}$}}

\eq{\fwboxL{260pt}{\hspace{-80pt}\fwboxL{12.5pt}{{\color{hred}\lsLabel_{13}^{8}}}=\fig{-28.5pt}{1}{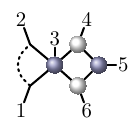}=\phantom{-}\fig{-28.5pt}{1}{ls1_1_123456}\hspace{-4pt}\times s_{12}}\vspace{-4pt}\label{sugra_ls_13_formula}\tag{${\color{hred}\lsLabel_{13}^{8}}$}}
\vspace{-4pt}\eq{\fwboxL{260pt}{\hspace{-80pt}\fwboxL{12.5pt}{{\color{hred}\lsLabel_{14}^{8}}}=\fig{-28.5pt}{1}{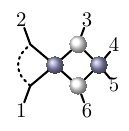}=\phantom{-}\fig{-28.5pt}{1}{ls1_2_123456}\hspace{-4pt}\times s_{12}}\vspace{-4pt}\label{sugra_ls_14_formula}\tag{${\color{hred}\lsLabel_{14}^{8}}$}}
\vspace{-4pt}\eq{\fwboxL{260pt}{\hspace{-80pt}\fwboxL{12.5pt}{{\color{hred}\lsLabel_{15}^{8}}}=\fig{-28.5pt}{1}{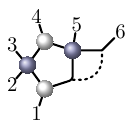}=\phantom{-}\fig{-28.5pt}{1}{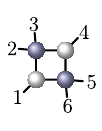}\hspace{-4pt}\times \frac{\langle1|6|5|4\rangle}{\ab{14}}}\vspace{-4pt}\label{sugra_ls_15_formula}\tag{${\color{hred}\lsLabel_{15}^{8}}$}}
\vspace{-4pt}\eq{\fwboxL{260pt}{\hspace{-80pt}\fwboxL{12.5pt}{{\color{hred}\lsLabel_{16}^{8}}}=\fig{-28.5pt}{1}{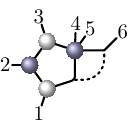}=\phantom{-}\fig{-28.5pt}{1}{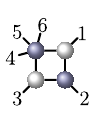}\hspace{-4pt}\times \frac{\langle1|6|(4\text{+}5)|3\rangle}{\ab{13}}}\vspace{-4pt}\label{sugra_ls_16_formula}\tag{${\color{hred}\lsLabel_{16}^{8}}$}}
\vspace{-4pt}\eq{\fwboxL{260pt}{\hspace{-80pt}\fwboxL{12.5pt}{{\color{hred}\lsLabel_{17}^{8}}}=\fig{-28.5pt}{1}{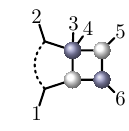}=\phantom{-}\fig{-28.5pt}{1}{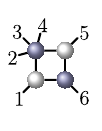}\hspace{-4pt}\times s_{12}}\vspace{-4pt}\label{sugra_ls_17_formula}\tag{${\color{hred}\lsLabel_{17}^{8}}$}}
\vspace{-4pt}\eq{\fwboxL{260pt}{\hspace{-80pt}\fwboxL{12.5pt}{{\color{hred}\lsLabel_{18}^{8}}}=\fig{-28.5pt}{1}{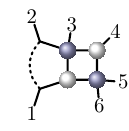}=\phantom{-}\fig{-28.5pt}{1}{ls1_2_234561}\hspace{-4pt}\times s_{12}}\vspace{-4pt}\label{sugra_ls_18_formula}\tag{${\color{hred}\lsLabel_{18}^{8}}$}}
\vspace{-4pt}\eq{\fwboxL{260pt}{\hspace{-80pt}\fwboxL{12.5pt}{{\color{hred}\lsLabel_{19}^{8}}}=\fig{-28.5pt}{1}{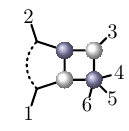}=\phantom{-}\fig{-28.5pt}{1}{ls1_1_456123}\hspace{-4pt}\times s_{12}}\vspace{-4pt}\label{sugra_ls_19_formula}\tag{${\color{hred}\lsLabel_{19}^{8}}$}}
\vspace{-4pt}\eq{\fwboxL{260pt}{\hspace{-80pt}\fwboxL{12.5pt}{{\color{hred}\lsLabel_{20}^{8}}}=\fig{-28.5pt}{1}{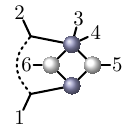}=\phantom{-}\fig{-28.5pt}{1}{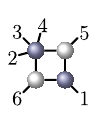}\hspace{-4pt}\times s_{12}}\vspace{-4pt}\label{sugra_ls_20_formula}\tag{${\color{hred}\lsLabel_{20}^{8}}$}}
\vspace{-4pt}\eq{\fwboxL{260pt}{\hspace{-80pt}\fwboxL{12.5pt}{{\color{hred}\lsLabel_{21}^{8}}}=\fig{-28.5pt}{1}{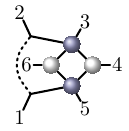}=\phantom{-}\fig{-28.5pt}{1}{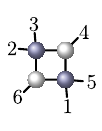}\hspace{-4pt}\times s_{12}}\vspace{-4pt}\label{sugra_ls_21_formula}\tag{${\color{hred}\lsLabel_{21}^{8}}$}}
\vspace{-4pt}\eq{\fwboxL{260pt}{\hspace{-80pt}\fwboxL{12.5pt}{{\color{hred}\lsLabel_{22}^{8}}}=\fig{-28.5pt}{1}{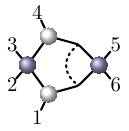}=\phantom{-}\fig{-28.5pt}{1}{ls1_2_234561}\hspace{-4pt}\times s_{56}}\vspace{-4pt}\label{sugra_ls_22_formula}\tag{${\color{hred}\lsLabel_{22}^{8}}$}}
\vspace{-4pt}\eq{\fwboxL{260pt}{\hspace{-80pt}\fwboxL{12.5pt}{{\color{hred}\lsLabel_{23}^{8}}}=\fig{-28.5pt}{1}{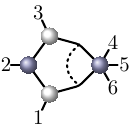}=\phantom{-}\fig{-28.5pt}{1}{ls1_1_456123}\hspace{-4pt}\times s_{123}}\vspace{-4pt}\label{sugra_ls_23_formula}\tag{${\color{hred}\lsLabel_{23}^{8}}$}}
\vspace{-4pt}\eq{\fwboxL{260pt}{\hspace{-80pt}\fwboxL{12.5pt}{{\color{hred}\lsLabel_{24}^{8}}}=\fig{-28.5pt}{1}{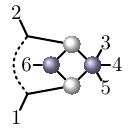}=\phantom{-}\fig{-28.5pt}{1}{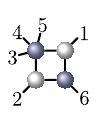}\hspace{-4pt}\times s_{12}}\vspace{-4pt}\label{sugra_ls_24_formula}\tag{${\color{hred}\lsLabel_{24}^{8}}$}}
\vspace{-4pt}\eq{\fwboxL{260pt}{\hspace{-80pt}\fwboxL{12.5pt}{{\color{hred}\lsLabel_{25}^{8}}}=\fig{-28.5pt}{1}{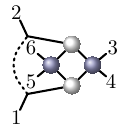}=\phantom{-}\fig{-28.5pt}{1}{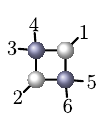}\hspace{-4pt}\times s_{12}}\vspace{-4pt}\label{sugra_ls_25_formula}\tag{${\color{hred}\lsLabel_{25}^{8}}$}}
\vspace{-4pt}\eq{\fwboxL{260pt}{\hspace{-80pt}\fwboxL{12.5pt}{{\color{hred}\lsLabel_{26}^{8}}}=\fig{-28.5pt}{1}{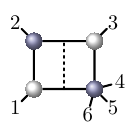}=\phantom{-}\fig{-28.5pt}{1}{ls1_1_456123}\hspace{-4pt}\times s_{12}}\vspace{-4pt}\label{sugra_ls_26_formula}\tag{${\color{hred}\lsLabel_{26}^{8}}$}}
\vspace{-4pt}\eq{\fwboxL{260pt}{\hspace{-80pt}\fwboxL{12.5pt}{{\color{hred}\lsLabel_{27}^{8}}}=\fig{-28.5pt}{1}{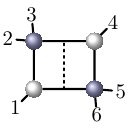}=\phantom{-}\fig{-28.5pt}{1}{ls1_2_234561}\hspace{-4pt}\times s_{123}}\vspace{-4pt}\label{sugra_ls_27_formula}\tag{${\color{hred}\lsLabel_{27}^{8}}$}}
\vspace{-4pt}\eq{\fwboxL{260pt}{\hspace{-80pt}\fwboxL{12.5pt}{{\color{hred}\lsLabel_{28}^{8}}}=\fig{-28.5pt}{1}{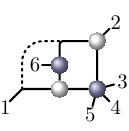}=\phantom{-}\fig{-28.5pt}{1}{ls1_1_345162}\hspace{-4pt}\times \frac{\langle1|6|1|2\rangle}{\ab{12}}}\vspace{-4pt}\label{sugra_ls_28_formula}\tag{${\color{hred}\lsLabel_{28}^{8}}$}}
\vspace{-4pt}\eq{\fwboxL{260pt}{\hspace{-80pt}\fwboxL{12.5pt}{{\color{hred}\lsLabel_{29}^{8}}}=\fig{-28.5pt}{1}{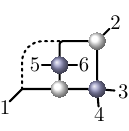}=\phantom{-}\fig{-28.5pt}{1}{ls1_2_341562}\hspace{-4pt}\times \frac{\langle1|(5\text{+}6)|1|2\rangle}{\ab{12}}}\vspace{-4pt}\label{sugra_ls_29_formula}\tag{${\color{hred}\lsLabel_{29}^{8}}$}}
\vspace{-4pt}\eq{\fwboxL{260pt}{\hspace{-80pt}\fwboxL{12.5pt}{{\color{hred}\lsLabel_{30}^{8}}}=\fig{-28.5pt}{1}{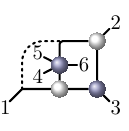}=\phantom{-}\fig{-28.5pt}{1}{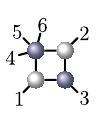}\hspace{-4pt}\times \frac{\langle2|1|(2\text{+}3)|1\rangle}{\ab{12}}}\vspace{-4pt}\label{sugra_ls_30_formula}\tag{${\color{hred}\lsLabel_{30}^{8}}$}}
\vspace{-4pt}\eq{\fwboxL{260pt}{\hspace{-80pt}\fwboxL{12.5pt}{{\color{hred}\lsLabel_{31}^{8}}}=\fig{-28.5pt}{1}{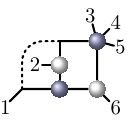}=-\fig{-28.5pt}{1}{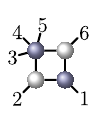}\hspace{-4pt}\times s_{12}}\vspace{-4pt}\label{sugra_ls_31_formula}\tag{${\color{hred}\lsLabel_{31}^{8}}$}}
\vspace{-4pt}\eq{\fwboxL{260pt}{\hspace{-80pt}\fwboxL{12.5pt}{{\color{hred}\lsLabel_{32}^{8}}}=\fig{-28.5pt}{1}{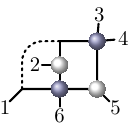}=-\fig{-28.5pt}{1}{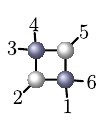}\hspace{-4pt}\times \frac{\langle2|1|(3\text{+}4)|5\rangle}{\ab{25}}}\vspace{-4pt}\label{sugra_ls_32_formula}\tag{${\color{hred}\lsLabel_{32}^{8}}$}}
\vspace{-4pt}\eq{\fwboxL{260pt}{\hspace{-80pt}\fwboxL{12.5pt}{{\color{hred}\lsLabel_{33}^{8}}}=\fig{-28.5pt}{1}{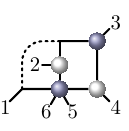}=-\fig{-28.5pt}{1}{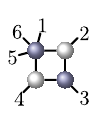}\hspace{-4pt}\times \frac{\langle2|1|3|4\rangle}{\ab{24}}}\vspace{-4pt}\label{sugra_ls_33_formula}\tag{${\color{hred}\lsLabel_{33}^{8}}$}}
\vspace{-4pt}\eq{\fwboxL{130pt}{\hspace{-80pt}\fwboxL{12.5pt}{{\color{hred}\lsLabel_{34}^{8}}}=\fig{-28.5pt}{1}{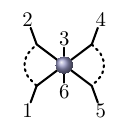}=\fig{-28.5pt}{1}{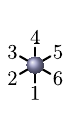}\times s_{12}s_{45}\;,}\fwboxL{130pt}{\hspace{-5pt}\fwboxL{12.5pt}{{\color{hred}\lsLabel_{35}^{8}}}=\fig{-28.5pt}{1}{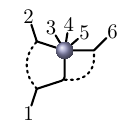}=\fig{-28.5pt}{1}{ls0}\times s_{12}s_{16}}\vspace{-4pt}\label{sugra_ls_34_35_formula}\tag{${\color{hred}\lsLabel_{34,35}^{8}}$}}
\vspace{-4pt}\eq{\fwboxL{130pt}{\hspace{-80pt}\fwboxL{12.5pt}{{\color{hred}\lsLabel_{36}^{8}}}=\fig{-28.5pt}{1}{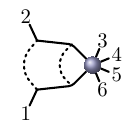}=\fig{-28.5pt}{1}{ls0}\times s_{12}^{2}\;,}\fwboxL{130pt}{\hspace{-5pt}\fwboxL{12.5pt}{{\color{hred}\lsLabel_{37}^{8}}}=\fig{-28.5pt}{1}{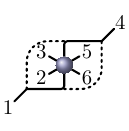}=\fig{-28.5pt}{1}{ls0}\times s_{14}^{2}}\vspace{-4pt}\label{sugra_ls_36_37_formula}\tag{${\color{hred}\lsLabel_{36,37}^{8}}$}}
\vspace{-4pt}\eq{\fwboxL{260pt}{\hspace{-80pt}\fwboxL{12.5pt}{{\color{hred}\lsLabel_{33}^{8}}}=\fig{-14.5pt}{1}{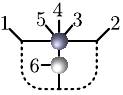}=\fig{-28.5pt}{1}{ls0}\hspace{-4pt}\times0}\vspace{-4pt}\label{sugra_ls_38_formula}\tag{${\color{hred}\lsLabel_{38}^{8}}$}}

%================================================================================================================
\vspace{-0pt}\subsection*{Poles at Infinity in Supergravity}\label{subsec:sugra_at_infinity}\vspace{-0pt}
%================================================================================================================

In $\mathcal{N}=8$ supergravity, there are three poles at infinity that are related to the failure of the following three residue theorems (that are satisfied in $\mathcal{N}=4$). Surprisingly, the value of the residue at infinity is the same in all three cases. The violated residue theorems can be obtained from the following hepta-cuts by making use of Cauchy's theorem in the remaining unfixed variable denoted by `$b$':
\begin{align}\Res_{{\color{hred}b}\to\infty}\fig{-39.65pt}{1}{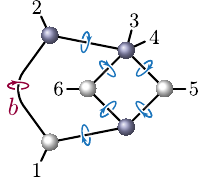}&=\Res_{{\color{hred}b}\to\infty}\fig{-39.65pt}{1}{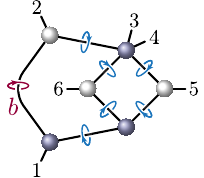}=\Res_{{\color{hred}b}\to\infty}\fig{-39.65pt}{1}{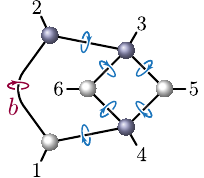}\nonumber\\
&=\frac{\sb{12}\sb{23}\sb{24}\sb{25}\sb{26}\ab{12}}{\ab{13}\ab{14}\ab{15}\ab{16}\ab{35}\ab{36}\ab{45}\ab{46}}\equivL{\color{hred}\lsLabel^8_{\infty}}\,.
\label{nonzero_infinity_grt_heptacuts}
\end{align}
A number of residues will correspond to regular factorization poles that have an on-shell diagrammatic interpretation. We will not write these terms out in a graphical notation, but summarize the contributing terms in the ${\color{hred}\lsLabel^{\mathcal{N}}_{i}}$ conventions of the previous indices and also indicate the relevant permutation of external legs:
\begin{align}
{\color{hred}\lsLabel^8_{\infty}}&={\color{hred}\lsLabel^8_{20}}(\hspace{-1pt}1,\!2,\!3,\!4,\!5,\!6\hspace{-1pt})\!-\!\hspace{-0pt}{\color{hred}\lsLabel^8_{6}}(\hspace{-1pt}1,\!2,\!3,\!4,\!5,\!6\hspace{-1pt})\!-\!{\color{hred}\lsLabel^8_{6}}(\hspace{-1pt}1,\!2,\!4,\!3,\!5,\!6\hspace{-1pt})\!-\!{\color{hred}\lsLabel^8_{11}}(\hspace{-1pt}5,\!3,\!4,\!2,\!1,\!6\hspace{-1pt})\!-\!{\color{hred}\lsLabel^8_{11}}(\hspace{-1pt}6,\!3,\!4,\!2,\!1,\!5\hspace{-1pt})\nonumber\\
&={\color{hred}\lsLabel^8_{6}}(\hspace{-1pt}3,\!2,\!1,\!4,\!5,\!6\hspace{-1pt})\!-\!{\color{hred}\lsLabel^8_{12}}(\hspace{-1pt}1,\!2,\!3,\!5,\!4,\!6\hspace{-1pt})\!-\!{\color{hred}\lsLabel^8_{12}}(\hspace{-1pt}1,\!2,\!3,\!6,\!4,\!5\hspace{-1pt})\!-\!{\color{hred}\lsLabel^8_{21}}(\hspace{-1pt}1,\!2,\!3,\!5,\!4,\!6\hspace{-1pt})\label{grts_for_poles_at_infinity_in_sugra}\\
&={\color{hred}\lsLabel^8_{20}}(\hspace{-1pt}2,\!1,\!3,\!4,\!5,\!6\hspace{-1pt})\!-\!{\color{hred}\lsLabel^8_{33}}(\hspace{-1pt}5,\!6,\!2,\!1,\!3,\!4\hspace{-1pt})\!-\!{\color{hred}\lsLabel^8_{33}}(\hspace{-1pt}6,\!5,\!2,\!1,\!3,\!4\hspace{-1pt})\,.\nonumber
\end{align}
Plugging in the explicit formulae for supergravity given above, one reproduces the equalities (\ref{nonzero_infinity_grt_heptacuts}). We have explicitly checked that these are the only residue theorems that are violated in supergravity, so that the only non-vanishing poles at infinity originate in supergravity are linked to the topologies in (\ref{grts_for_poles_at_infinity_in_sugra}).

\newpage
%================================================================================================================
%    Appendix C:  
%================================================================================================================
\vspace{-0pt}\section[\mbox{Organization of Ancillary Files}]{\mbox{\hspace{0pt}Organization of Ancillary Files}}\label{appendix:ancillary_files}\vspace{-0pt}

The main results described in this work are available as ancillary files. These files can be downloaded from the abstract page for this work on the {\tt arXiv}---linked to on the right-hand panel (below `Download'). Specifically, we have prepared three files for the interested reader:
\begin{itemize}
\item {\bf six\uscore point\uscore integrand\uscore data\uscore v2.dat}: a \emph{plain text} data file, consisting of (just) the details of our result described in these appendices. 
\item {\bf six\uscore point\uscore integrand\uscore tools\uscore v2.m}: a {\sc Mathematica} package file, consisting of code useful to analyze, understand, and work the the raw data.
\item {\bf six\uscore point\uscore integrand\uscore walkthrough.nb\uscore v2}: a {\sc Mathematica} notebook file illustrating our results and the main functionality of the codebase.
\end{itemize} 
These files are all well-documented within, but the key data structures in the data file are described below:\\

\subsection*{Main Results: Six-Point MHV Amplitude Integrands}

The ancillary file {\bf six\uscore point\uscore integrand\uscore data\uscore v2.dat} enumerates all of the ingredients required to use and verify the results described in this work. These ingredients consist of the objects:\\[-14pt]

%%%%%%%%%%%%%%%%%%%%%%%%%%%%%%%%%%%%%%%%%%%%%%%%%%%%%%%%%%%%%%%%%%%%%%%%%%%%%%%%%%%
\defnNA{chiralIntegrandSeedData}{}{a list of 38 permutation-seeds which generate the six-point two-loop MHV amplitude integrands for \emph{both} sYM and SUGRA as represented in the summand (\ref{schematic_loop_integrand_formula}). These contributions are encoded somewhat indirectly, and it is worthwhile to describe how each term is represented. \\[-12pt]

Each element of the list \fun{chiralIntegrandSeedData} consists of 5 elements:
\begin{enumerate}
\item \fun{f}{\tt[\var{i},\var{legList}]}: a symbol representing ${\color{hred}\lsLabel_{\var{i}}}$ in either sYM or SUGRA.
\item \fun{graphDialList}: for each vertex in the corresponding graph, a list of edges incident with it, with a sign $\mp1$ depending on whether edge has been oriented as incoming or outgoing, respectively. Edges for external momenta $p_i$ are labelled by $i\in\{1,...,6\}$; and internal edges are labelled $\{a,\ldots,h\}$---exactly matching the conventions of \mbox{appendix \ref{explicit_integrand_numerator_tables}}.
\item \fun{graphEdgeList}: a list of \emph{oriented} edges of for graph, encoded as pairs of $\{v_1,v_2\}$ for an edge connecting $v_1\mapsto v_2$. The {\tt Heads} of the vertices are chosen from the set $\{\fun{b},\fun{w},\fun{p},\fun{s}\}$ for MHV, $\overline{\text{MHV}}$, external, or `soft', respectively. This list of edges has duplicated entries in the case of soft (composite) leading singularities; with this duplication, the edges are always listed according to the assignments in \mbox{appendix \ref{explicit_integrand_numerator_tables}} of labels $\{a,\ldots,h,p_1,\ldots,p_6\}$.
\item \fun{loopRoutingRules}: one possible assignment of loop momenta $\ell_1,\ell_2$ to the edges of the graph consistent with momentum conservation and the graph's (arbitrary, but chosen) orientation. This is encoded as a list of \texttt{Rules} of the form $\{\edgeA\mapsto\ell_1,\cdots\}$. Again, we should emphasize that nothing about our result depends on this assignment of internal loop momenta. 
\item \fun{integrandNumerator}: the numerator for this integrand in the prescriptive basis, as enumerated in \mbox{appendix \ref{explicit_integrand_numerator_tables}}. 
\end{enumerate}
}

%%%%%%%%%%%%%%%%%%%%%%%%%%%%%%%%%%%%%%%%%%%%%%%%%%%%%%%%%%%%%%%%%%%%%%%%%%%%%%%%%%%
\defn{ymLS}{\vardef{i}}{an explicit formula for the leading singularity ${\color{hred}\lsLabel_{\var{i}}^{4}}$ in $\mathcal{N}=4$ sYM. These are given separately and explicitly so that they may be used by researchers without relying on any functionality of {\sc Mathematica}. They are expressed as sums of products of terms built from the (abstract) objects \fun{colorF}{\tt[}\var{a},\var{b},\var{c}{\tt]} and \fun{gamma}{\tt[}\var{a},\var{b},\var{c}{\tt]}, defined in equations (\ref{our_color_factors}) and (\ref{shuffle_sum_formulae}), respectively. 
}

%%%%%%%%%%%%%%%%%%%%%%%%%%%%%%%%%%%%%%%%%%%%%%%%%%%%%%%%%%%%%%%%%%%%%%%%%%%%%%%%%%%
\defn{sugraLS}{\vardef{i}}{an explicit formula for the leading singularity ${\color{hred}\lsLabel_{\var{i}}^{8}}$ in $\mathcal{N}=8$ SUGRA. These are given separately and explicitly so that they may be used by researchers without relying on any functionality of {\sc Mathematica}. They are expressed directly in terms of familiar spinor products $\ab{ij}\equivL$\fun{ab}{\tt[}\var{i},\var{j}{\tt]},$\sb{ij}\equivL$\fun{sb}{\tt[}\var{i},\var{j}{\tt]}, and etc.
}

%%%%%%%%%%%%%%%%%%%%%%%%%%%%%%%%%%%%%%%%%%%%%%%%%%%%%%%%%%%%%%%%%%%%%%%%%%%%%%%%%%%
\defnNA{grtGenerators}{}{a list of permutation-class representatives of residue theorems (`grt's) satisfied by the on-shell functions ${\color{hred}\lsLabel_{\var{i}}^{\mathcal{N}}}$. \emph{Nota bene}: the first three on this list correspond to those of (\ref{grts_for_poles_at_infinity_in_sugra}) which are not identities in $\mathcal{N}=8$ supergravity. 
}

%================================================================================================================
%    References
%================================================================================================================
\newpage

%\bibliographystyle{physics}
%\phantomsection
%\bibliography{amplitude_refs}

\begin{thebibliography}{100}

\bibitem{Bourjaily:2019gqu}
J.~L. Bourjaily, E.~Herrmann, C.~Langer, A.~J. McLeod, and J.~Trnka,
  ``{All-Multiplicity Non-Planar MHV Amplitudes in sYM at Two Loops},''
\href{http://arxiv.org/abs/1911.09106}{{ arXiv:1911.09106 [hep-th]}}.
%%CITATION = ARXIV:1911.09106;%%.

\bibitem{Bern:1994zx}
Z.~Bern, L.~J. Dixon, D.~C. Dunbar, and D.~A. Kosower, ``{One-Loop $n$-Point
  Gauge Theory Amplitudes, Unitarity and Collinear Limits},''
  \href{http://dx.doi.org/10.1016/0550-3213(94)90179-1}{{\em Nucl. Phys.} {\bf
  B425} (1994)  217--260},
\href{http://arxiv.org/abs/hep-ph/9403226}{{ arXiv:hep-ph/9403226}}.
%%CITATION = HEP-PH/9403226;%%.

\bibitem{Bern:1994cg}
Z.~Bern, L.~J. Dixon, D.~C. Dunbar, and D.~A. Kosower, ``{Fusing Gauge Theory
  Tree Amplitudes into Loop Amplitudes},''
  \href{http://dx.doi.org/10.1016/0550-3213(94)00488-Z}{{\em Nucl. Phys.} {\bf
  B435} (1995)  59--101},
\href{http://arxiv.org/abs/hep-ph/9409265}{{ arXiv:hep-ph/9409265}}.
%%CITATION = HEP-PH/9409265;%%.

\bibitem{Bern:1997sc}
Z.~Bern, L.~J. Dixon, and D.~A. Kosower, ``{One-Loop Amplitudes for $e^+\,e^-$
  to Four Partons},''
  \href{http://dx.doi.org/10.1016/S0550-3213(97)00703-7}{{\em Nucl. Phys.} {\bf
  B513} (1998)  3--86},
\href{http://arxiv.org/abs/hep-ph/9708239}{{ arXiv:hep-ph/9708239 [hep-ph]}}.
%%CITATION = HEP-PH/9708239;%%.

\bibitem{Britto:2004nc}
R.~Britto, F.~Cachazo, and B.~Feng, ``{Generalized Unitarity and One-Loop
  Amplitudes in $\mathcal{N}\!=\!4$ Super-Yang-Mills},''
  \href{http://dx.doi.org/10.1016/j.nuclphysb.2005.07.014}{{\em Nucl. Phys.}
  {\bf B725} (2005)  275--305},
\href{http://arxiv.org/abs/hep-th/0412103}{{ arXiv:hep-th/0412103}}.
%%CITATION = HEP-TH/0412103;%%.

\bibitem{Bern:2007ct}
Z.~Bern, J.~Carrasco, H.~Johansson, and D.~Kosower, ``{Maximally Supersymmetric
  Planar Yang-Mills Amplitudes at Five Loops},''
  \href{http://dx.doi.org/10.1103/PhysRevD.76.125020}{{\em Phys. Rev.} {\bf
  D76} (2007)  125020},
\href{http://arxiv.org/abs/0705.1864}{{ arXiv:0705.1864 [hep-th]}}.
%%CITATION = ARXIV:0705.1864;%%.

\bibitem{BCF}
R.~Britto, F.~Cachazo, and B.~Feng, ``{New Recursion Relations for Tree
  Amplitudes of Gluons},''
  \href{http://dx.doi.org/10.1016/j.nuclphysb.2005.02.030}{{\em Nucl.Phys.}
  {\bf B715} (2005)  499--522},
\href{http://arxiv.org/abs/hep-th/0412308}{{ arXiv:hep-th/0412308 [hep-th]}}.
%%CITATION = HEP-TH/0412308;%%.

\bibitem{BCFW}
R.~Britto, F.~Cachazo, B.~Feng, and E.~Witten, ``{Direct Proof of Tree-Level
  Recursion Relation in Yang- Mills Theory},''
  \href{http://dx.doi.org/10.1103/PhysRevLett.94.181602}{{\em Phys. Rev. Lett.}
  {\bf 94} (2005)  181602},
\href{http://arxiv.org/abs/hep-th/0501052}{{ arXiv:hep-th/0501052}}.
%%CITATION = HEP-TH/0501052;%%.

\bibitem{ArkaniHamed:2010kv}
N.~Arkani-Hamed, J.~L. Bourjaily, F.~Cachazo, S.~Caron-Huot, and J.~Trnka,
  ``{The All-Loop Integrand For Scattering Amplitudes in Planar
  $\mathcal{N}\!=\!4$ SYM},''
  \href{http://dx.doi.org/10.1007/JHEP01(2011)041}{{\em JHEP} {\bf 1101} (2011)
   041},
\href{http://arxiv.org/abs/1008.2958}{{ arXiv:1008.2958 [hep-th]}}.
%%CITATION = ARXIV:1008.2958;%%.

\bibitem{ArkaniHamed:2009dg}
N.~Arkani-Hamed, J.~Bourjaily, F.~Cachazo, and J.~Trnka, ``{Unification of
  Residues and Grassmannian Dualities},''
  \href{http://dx.doi.org/10.1007/JHEP01(2011)049}{{\em JHEP} {\bf 1101} (2011)
   049},
\href{http://arxiv.org/abs/0912.4912}{{ arXiv:0912.4912 [hep-th]}}.
%%CITATION = ARXIV:0912.4912;%%.

\bibitem{ArkaniHamed:2009dn}
N.~Arkani-Hamed, F.~Cachazo, C.~Cheung, and J.~Kaplan, ``{A Duality For The
  $S$-Matrix},'' \href{http://dx.doi.org/10.1007/JHEP03(2010)020}{{\em JHEP}
  {\bf 1003} (2010)  020},
\href{http://arxiv.org/abs/0907.5418}{{ arXiv:0907.5418 [hep-th]}}.
%%CITATION = ARXIV:0907.5418;%%.

\bibitem{ArkaniHamed:2009sx}
N.~Arkani-Hamed, J.~Bourjaily, F.~Cachazo, and J.~Trnka, ``{Local Spacetime
  Physics from the Grassmannian},''
  \href{http://dx.doi.org/10.1007/JHEP01(2011)108}{{\em JHEP} {\bf 1101} (2011)
   108},
\href{http://arxiv.org/abs/0912.3249}{{ arXiv:0912.3249 [hep-th]}}.
%%CITATION = ARXIV:0912.3249;%%.

\bibitem{ArkaniHamed:2009vw}
N.~Arkani-Hamed, F.~Cachazo, and C.~Cheung, ``{The Grassmannian Origin Of Dual
  Superconformal Invariance},''
  \href{http://dx.doi.org/10.1007/JHEP03(2010)036}{{\em JHEP} {\bf 1003} (2010)
   036},
\href{http://arxiv.org/abs/0909.0483}{{ arXiv:0909.0483 [hep-th]}}.
%%CITATION = ARXIV:0909.0483;%%.

\bibitem{ArkaniHamed:2010gg}
N.~Arkani-Hamed, J.~L. Bourjaily, F.~Cachazo, A.~Hodges, and J.~Trnka, ``{A
  {N}ote on {P}olytopes for {S}cattering {A}mplitudes},''
  \href{http://dx.doi.org/10.1007/JHEP04(2012)081}{{\em JHEP} {\bf 1204} (2012)
   081},
\href{http://arxiv.org/abs/1012.6030}{{ arXiv:1012.6030 [hep-th]}}.
%%CITATION = ARXIV:1012.6030;%%.

\bibitem{ArkaniHamed:2012nw}
N.~Arkani-Hamed, J.~L. Bourjaily, F.~Cachazo, A.~B. Goncharov, A.~Postnikov,
  and J.~Trnka, ``{Scattering Amplitudes and the Positive Grassmannian},''
\href{http://arxiv.org/abs/1212.5605}{{ arXiv:1212.5605 [hep-th]}}.
%%CITATION = ARXIV:1212.5605;%%.

\bibitem{Arkani-Hamed:2013jha}
N.~Arkani-Hamed and J.~Trnka, ``{The Amplituhedron},''
  \href{http://dx.doi.org/10.1007/JHEP10(2014)030}{{\em JHEP} {\bf 1410} (2014)
   30},
\href{http://arxiv.org/abs/1312.2007}{{ arXiv:1312.2007 [hep-th]}}.
%%CITATION = ARXIV:1312.2007;%%.

\bibitem{BCJ}
Z.~Bern, J.~Carrasco, and H.~Johansson, ``{New Relations for Gauge-Theory
  Amplitudes},'' \href{http://dx.doi.org/10.1103/PhysRevD.78.085011}{{\em Phys.
  Rev.} {\bf D78} (2008)  085011},
\href{http://arxiv.org/abs/0805.3993}{{ arXiv:0805.3993 [hep-ph]}}.
%%CITATION = ARXIV:0805.3993;%%.

\bibitem{Bern:2019prr}
Z.~Bern, J.~J. Carrasco, M.~Chiodaroli, H.~Johansson, and R.~Roiban, ``{The
  Duality Between Color and Kinematics and its Applications},''
\href{http://arxiv.org/abs/1909.01358}{{ arXiv:1909.01358 [hep-th]}}.
%%CITATION = ARXIV:1909.01358;%%.

\bibitem{Cachazo:2013gna}
F.~Cachazo, S.~He, and E.~Y. Yuan, ``{Scattering Equations and
  Kawai-Lewellen-Tye Orthogonality},''
  \href{http://dx.doi.org/10.1103/PhysRevD.90.065001}{{\em Phys. Rev.} {\bf
  D90} (2014) no. 6, 065001},
\href{http://arxiv.org/abs/1306.6575}{{ arXiv:1306.6575 [hep-th]}}.
%%CITATION = ARXIV:1306.6575;%%.

\bibitem{Cachazo:2013hca}
F.~Cachazo, S.~He, and E.~Y. Yuan, ``{Scattering of Massless Particles in
  Arbitrary Dimensions},''
  \href{http://dx.doi.org/10.1103/PhysRevLett.113.171601}{{\em Phys. Rev.
  Lett.} {\bf 113} (2014) no. 17, 171601},
\href{http://arxiv.org/abs/1307.2199}{{ arXiv:1307.2199 [hep-th]}}.
%%CITATION = ARXIV:1307.2199;%%.

\bibitem{Cachazo:2013iaa}
F.~Cachazo, S.~He, and E.~Y. Yuan, ``{Scattering in Three Dimensions from
  Rational Maps},'' \href{http://dx.doi.org/10.1007/JHEP10(2013)141}{{\em JHEP}
  {\bf 1310} (2013)  141},
\href{http://arxiv.org/abs/1306.2962}{{ arXiv:1306.2962 [hep-th]}}.
%%CITATION = ARXIV:1306.2962;%%.

\bibitem{Cachazo:2013iea}
F.~Cachazo, S.~He, and E.~Y. Yuan, ``{Scattering of Massless Particles:
  Scalars, Gluons and Gravitons},''
  \href{http://dx.doi.org/10.1007/JHEP07(2014)033}{{\em JHEP} {\bf 1407} (2014)
   033},
\href{http://arxiv.org/abs/1309.0885}{{ arXiv:1309.0885 [hep-th]}}.
%%CITATION = ARXIV:1309.0885;%%.

\bibitem{Bourjaily:2011hi}
J.~L. Bourjaily, A.~DiRe, A.~Shaikh, M.~Spradlin, and A.~Volovich, ``{The
  Soft-Collinear Bootstrap: $\mathcal{N}\!=\!4$ Yang-Mills Amplitudes at Six
  and Seven Loops},'' \href{http://dx.doi.org/10.1007/JHEP03(2012)032}{{\em
  JHEP} {\bf 1203} (2012)  032},
\href{http://arxiv.org/abs/1112.6432}{{ arXiv:1112.6432 [hep-th]}}.
%%CITATION = ARXIV:1112.6432;%%.

\bibitem{Dixon:2013eka}
L.~J. Dixon, J.~M. Drummond, M.~von Hippel, and J.~Pennington, ``{Hexagon
  Functions and the Three-Loop Remainder Function},''
  \href{http://dx.doi.org/10.1007/JHEP12(2013)049}{{\em JHEP} {\bf 1312} (2013)
   049},
\href{http://arxiv.org/abs/1308.2276}{{ arXiv:1308.2276 [hep-th]}}.
%%CITATION = ARXIV:1308.2276;%%.

\bibitem{Golden:2014pua}
J.~Golden and M.~Spradlin, ``{A Cluster Bootstrap for Two-Loop MHV
  Amplitudes},'' \href{http://dx.doi.org/10.1007/JHEP02(2015)002}{{\em JHEP}
  {\bf 1502} (2015)  002},
\href{http://arxiv.org/abs/1411.3289}{{ arXiv:1411.3289 [hep-th]}}.
%%CITATION = ARXIV:1411.3289;%%.

\bibitem{Drummond:2014ffa}
J.~M. Drummond, G.~Papathanasiou, and M.~Spradlin, ``{A Symbol of Uniqueness:
  The Cluster Bootstrap for the 3-Loop MHV Heptagon},''
  \href{http://dx.doi.org/10.1007/JHEP03(2015)072}{{\em JHEP} {\bf 03} (2015)
  072},
\href{http://arxiv.org/abs/1412.3763}{{ arXiv:1412.3763 [hep-th]}}.
%%CITATION = ARXIV:1412.3763;%%.

\bibitem{Bourjaily:2015bpz}
J.~L. Bourjaily, P.~Heslop, and V.-V. Tran, ``{Perturbation Theory at Eight
  Loops: Novel Structures and the Breakdown of Manifest Conformality in
  $\mathcal{N}\!=\!4$ Supersymmetric Yang-Mills Theory},''
  \href{http://dx.doi.org/10.1103/PhysRevLett.116.191602}{{\em Phys. Rev.
  Lett.} {\bf 116} (2016) no. 19, 191602},
\href{http://arxiv.org/abs/1512.07912}{{ arXiv:1512.07912 [hep-th]}}.
%%CITATION = ARXIV:1512.07912;%%.

\bibitem{Bourjaily:2016evz}
J.~L. Bourjaily, P.~Heslop, and V.-V. Tran, ``{Amplitudes and Correlators to
  Ten Loops Using Simple, Graphical Bootstraps},''
  \href{http://dx.doi.org/10.1007/JHEP11(2016)125}{{\em JHEP} {\bf 11} (2016)
  125},
\href{http://arxiv.org/abs/1609.00007}{{ arXiv:1609.00007 [hep-th]}}.
%%CITATION = ARXIV:1609.00007;%%.

\bibitem{Caron-Huot:2016owq}
S.~Caron-Huot, L.~J. Dixon, A.~McLeod, and M.~von Hippel, ``{Bootstrapping a
  Five-Loop Amplitude Using Steinmann Relations},''
  \href{http://dx.doi.org/10.1103/PhysRevLett.117.241601}{{\em Phys. Rev.
  Lett.} {\bf 117} (2016) no. 24, 241601},
\href{http://arxiv.org/abs/1609.00669}{{ arXiv:1609.00669 [hep-th]}}.
%%CITATION = ARXIV:1609.00669;%%.

\bibitem{Almelid:2017qju}
{\O}.~Almelid, C.~Duhr, E.~Gardi, A.~McLeod, and C.~D. White, ``{Bootstrapping
  the QCD Soft Anomalous Dimension},''
  \href{http://dx.doi.org/10.1007/JHEP09(2017)073}{{\em JHEP} {\bf 09} (2017)
  073},
\href{http://arxiv.org/abs/1706.10162}{{ arXiv:1706.10162}}.
%%CITATION = ARXIV:1706.10162;%%.

\bibitem{Caron-Huot:2018dsv}
S.~Caron-Huot, L.~J. Dixon, M.~von Hippel, A.~J. McLeod, and G.~Papathanasiou,
  ``{The Double Pentaladder Integral to All Orders},''
  \href{http://dx.doi.org/10.1007/JHEP07(2018)170}{{\em JHEP} {\bf 07} (2018)
  170},
\href{http://arxiv.org/abs/1806.01361}{{ arXiv:1806.01361 [hep-th]}}.
%%CITATION = ARXIV:1806.01361;%%.

\bibitem{Henn:2018cdp}
J.~Henn, E.~Herrmann, and J.~Parra-Martinez, ``{Bootstrapping Two-Loop Feynman
  Integrals for Planar $\mathcal{N}\!=\!4$ sYM},''
  \href{http://dx.doi.org/10.1007/JHEP10(2018)059}{{\em JHEP} {\bf 10} (2018)
  059},
\href{http://arxiv.org/abs/1806.06072}{{ arXiv:1806.06072 [hep-th]}}.
%%CITATION = ARXIV:1806.06072;%%.

\bibitem{Parke:1986gb}
S.~J. Parke and T.~R. Taylor, ``{An Amplitude for $n$-Gluon Scattering},''
\href{http://dx.doi.org/10.1103/PhysRevLett.56.2459}{{\em Phys. Rev. Lett.}
  {\bf 56} (1986)  2459}.
%%CITATION = PRLTA,56,2459;%%.

\bibitem{Bern:2005iz}
Z.~Bern, L.~J. Dixon, and V.~A. Smirnov, ``{Iteration of Planar Amplitudes in
  Maximally Supersymmetric Yang-Mills Theory at Three Loops and Beyond},''
  \href{http://dx.doi.org/10.1103/PhysRevD.72.085001}{{\em Phys. Rev.} {\bf
  D72} (2005)  085001},
\href{http://arxiv.org/abs/hep-th/0505205}{{ arXiv:hep-th/0505205}}.
%%CITATION = HEP-TH/0505205;%%.

\bibitem{Drummond:2006rz}
J.~Drummond, J.~Henn, V.~Smirnov, and E.~Sokatchev, ``{Magic Identities for
  Conformal Four-Point Integrals},''
  \href{http://dx.doi.org/10.1088/1126-6708/2007/01/064}{{\em JHEP} {\bf 0701}
  (2007)  064},
\href{http://arxiv.org/abs/hep-th/0607160}{{ arXiv:hep-th/0607160}}.
%%CITATION = HEP-TH/0607160;%%.

\bibitem{Drummond:2008vq}
J.~Drummond, J.~Henn, G.~Korchemsky, and E.~Sokatchev, ``{Dual Superconformal
  Symmetry of Scattering Amplitudes in $\mathcal{N}\!=\!4$ super Yang-Mills
  Theory},'' \href{http://dx.doi.org/10.1016/j.nuclphysb.2009.11.022}{{\em
  Nucl. Phys.} {\bf B828} (2010)  317--374},
\href{http://arxiv.org/abs/0807.1095}{{ arXiv:0807.1095 [hep-th]}}.
%%CITATION = ARXIV:0807.1095;%%.

\bibitem{Goncharov:2010jf}
A.~B. Goncharov, M.~Spradlin, C.~Vergu, and A.~Volovich, ``{Classical
  Polylogarithms for Amplitudes and Wilson Loops},''
  \href{http://dx.doi.org/10.1103/PhysRevLett.105.151605}{{\em Phys. Rev.
  Lett.} {\bf 105} (2010)  151605},
\href{http://arxiv.org/abs/1006.5703}{{ arXiv:1006.5703 [hep-th]}}.
%%CITATION = ARXIV:1006.5703;%%.

\bibitem{CaronHuot:2011ky}
S.~Caron-Huot, ``{Superconformal Symmetry and Two-Loop Amplitudes in Planar
  $\mathcal{N}\!=\!4$ Super Yang-Mills},''
  \href{http://dx.doi.org/10.1007/JHEP12(2011)066}{{\em JHEP} {\bf 1112} (2011)
   066},
\href{http://arxiv.org/abs/1105.5606}{{ arXiv:1105.5606 [hep-th]}}.
%%CITATION = ARXIV:1105.5606;%%.

\bibitem{Alday:2007hr}
L.~F. Alday and J.~M. Maldacena, ``{Gluon Scattering Amplitudes at Strong
  Coupling},'' \href{http://dx.doi.org/10.1088/1126-6708/2007/06/064}{{\em
  JHEP} {\bf 06} (2007)  064},
\href{http://arxiv.org/abs/0705.0303}{{ arXiv:0705.0303 [hep-th]}}.
%%CITATION = ARXIV:0705.0303;%%.

\bibitem{Drummond:2009fd}
J.~M. Drummond, J.~M. Henn, and J.~Plefka, ``{Yangian Symmetry of Scattering
  Amplitudes in $\mathcal{N}\!=\!4$ Super Yang-Mills Theory},''
  \href{http://dx.doi.org/10.1088/1126-6708/2009/05/046}{{\em JHEP} {\bf 05}
  (2009)  046},
\href{http://arxiv.org/abs/0902.2987}{{ arXiv:0902.2987 [hep-th]}}.
%%CITATION = 0902.2987;%%.

\bibitem{ArkaniHamed:2008gz}
N.~Arkani-Hamed, F.~Cachazo, and J.~Kaplan, ``{What is the Simplest Quantum
  Field Theory?},'' \href{http://dx.doi.org/10.1007/JHEP09(2010)016}{{\em JHEP}
  {\bf 1009} (2010)  016},
\href{http://arxiv.org/abs/0808.1446}{{ arXiv:0808.1446 [hep-th]}}.
%%CITATION = ARXIV:0808.1446;%%.

\bibitem{Bern:2012uc}
Z.~Bern, J.~Carrasco, H.~Johansson, and R.~Roiban, ``{The Five-Loop Four-Point
  Amplitude of $\mathcal{N}\!=\!4$ Super-Yang-Mills Theory},''
  \href{http://dx.doi.org/10.1103/PhysRevLett.109.241602}{{\em Phys. Rev.
  Lett.} {\bf 109} (2012)  241602},
\href{http://arxiv.org/abs/1207.6666}{{ arXiv:1207.6666 [hep-th]}}.
%%CITATION = ARXIV:1207.6666;%%.

\bibitem{Bern:2017ucb}
Z.~Bern, J.~J.~M. Carrasco, W.-M. Chen, H.~Johansson, R.~Roiban, and M.~Zeng,
  ``{The Five-Loop Four-Point Integrand of $\mathcal{N}\!=\!8$ Supergravity as
  a Generalized Double Copy},''
\href{http://arxiv.org/abs/1708.06807}{{ arXiv:1708.06807 [hep-th]}}.
%%CITATION = ARXIV:1708.06807;%%.

\bibitem{Carrasco:2011mn}
J.~J. Carrasco and H.~Johansson, ``{Five-Point Amplitudes in
  $\mathcal{N}\!=\!4$ Super-Yang-Mills Theory and $\mathcal{N}\!=\!8$
  Supergravity},'' \href{http://dx.doi.org/10.1103/PhysRevD.85.025006}{{\em
  Phys. Rev.} {\bf D85} (2012)  025006},
\href{http://arxiv.org/abs/1106.4711}{{ arXiv:1106.4711 [hep-th]}}.
%%CITATION = ARXIV:1106.4711;%%.

\bibitem{Bern:2015ple}
Z.~Bern, E.~Herrmann, S.~Litsey, J.~Stankowicz, and J.~Trnka, ``{Evidence for a
  Nonplanar Amplituhedron},''
  \href{http://dx.doi.org/10.1007/JHEP06(2016)098}{{\em JHEP} {\bf 06} (2016)
  098},
\href{http://arxiv.org/abs/1512.08591}{{ arXiv:1512.08591 [hep-th]}}.
%%CITATION = ARXIV:1512.08591;%%.

\bibitem{Bourjaily:2013mma}
J.~L. Bourjaily, S.~Caron-Huot, and J.~Trnka, ``{Dual-Conformal Regularization
  of Infrared Loop Divergences and the {\it Chiral} Box Expansion},''
  \href{http://dx.doi.org/10.1007/JHEP01(2015)001}{{\em JHEP} {\bf 1501} (2015)
   001},
\href{http://arxiv.org/abs/1303.4734}{{ arXiv:1303.4734 [hep-th]}}.
%%CITATION = ARXIV:1303.4734;%%.

\bibitem{Bourjaily:2015jna}
J.~L. Bourjaily and J.~Trnka, ``{Local Integrand Representations of All
  Two-Loop Amplitudes in Planar SYM},''
  \href{http://dx.doi.org/10.1007/JHEP08(2015)119}{{\em JHEP} {\bf 08} (2015)
  119},
\href{http://arxiv.org/abs/1505.05886}{{ arXiv:1505.05886 [hep-th]}}.
%%CITATION = ARXIV:1505.05886;%%.

\bibitem{Bourjaily:2017wjl}
J.~L. Bourjaily, E.~Herrmann, and J.~Trnka, ``{Prescriptive Unitarity},''
  \href{http://dx.doi.org/10.1007/JHEP06(2017)059}{{\em JHEP} {\bf 06} (2017)
  059},
\href{http://arxiv.org/abs/1704.05460}{{ arXiv:1704.05460 [hep-th]}}.
%%CITATION = ARXIV:1704.05460;%%.

\bibitem{Caron-Huot:2019vjl}
S.~Caron-Huot, L.~J. Dixon, F.~Dulat, M.~von Hippel, A.~J. McLeod, and
  G.~Papathanasiou, ``{Six-Gluon Amplitudes in Planar $\mathcal{N}\!=\!4 $
  super-Yang-Mills Theory at Six and Seven Loops},''
  \href{http://dx.doi.org/10.1007/JHEP08(2019)016}{{\em JHEP} {\bf 08} (2019)
  016},
\href{http://arxiv.org/abs/1903.10890}{{ arXiv:1903.10890 [hep-th]}}.
%%CITATION = ARXIV:1903.10890;%%.

\bibitem{Caron-Huot:2019bsq}
S.~Caron-Huot, L.~J. Dixon, F.~Dulat, M.~Von~Hippel, A.~J. McLeod, and
  G.~Papathanasiou, ``{The Cosmic Galois Group and Extended Steinmann Relations
  for Planar $\mathcal{N} = 4$ SYM Amplitudes},''
\href{http://arxiv.org/abs/1906.07116}{{ arXiv:1906.07116 [hep-th]}}.
%%CITATION = ARXIV:1906.07116;%%.

\bibitem{Dixon:2016nkn}
L.~J. Dixon, J.~Drummond, T.~Harrington, A.~J. McLeod, G.~Papathanasiou, and
  M.~Spradlin, ``{Heptagons from the Steinmann Cluster Bootstrap},''
  \href{http://dx.doi.org/10.1007/JHEP02(2017)137}{{\em JHEP} {\bf 02} (2017)
  137},
\href{http://arxiv.org/abs/1612.08976}{{ arXiv:1612.08976 [hep-th]}}.
%%CITATION = ARXIV:1612.08976;%%.

\bibitem{Drummond:2018caf}
J.~Drummond, J.~Foster, {\"O}.~G{\"u}rdo\v{g}an, and G.~Papathanasiou,
  ``{Cluster Adjacency and the Four-Loop NMHV Heptagon},''
  \href{http://dx.doi.org/10.1007/JHEP03(2019)087}{{\em JHEP} {\bf 03} (2019)
  087},
\href{http://arxiv.org/abs/1812.04640}{{ arXiv:1812.04640 [hep-th]}}.
%%CITATION = ARXIV:1812.04640;%%.

\bibitem{Ben-Israel:2018ckc}
R.~Ben-Israel, A.~G. Tumanov, and A.~Sever, ``{Scattering Amplitudes -- Wilson
  Loops Duality for the First non-Planar Correction},''
  \href{http://dx.doi.org/10.1007/JHEP08(2018)122}{{\em JHEP} {\bf 08} (2018)
  122},
\href{http://arxiv.org/abs/1802.09395}{{ arXiv:1802.09395 [hep-th]}}.
%%CITATION = ARXIV:1802.09395;%%.

\bibitem{Tourkine:2019ukp}
P.~Tourkine, ``{On Integrands and Loop Momentum in String and Field Theory},''
\href{http://arxiv.org/abs/1901.02432}{{ arXiv:1901.02432 [hep-th]}}.
%%CITATION = ARXIV:1901.02432;%%.

\bibitem{integrandBases}
J.~L. Bourjaily, E.~Herrmann, and J.~Trnka, ``{Building Bases of Loop
  Integrands}.'' To appear.

\bibitem{Arkani-Hamed:2014via}
N.~Arkani-Hamed, J.~L. Bourjaily, F.~Cachazo, and J.~Trnka, ``{Singularity
  Structure of Maximally Supersymmetric Scattering Amplitudes},''
  \href{http://dx.doi.org/10.1103/PhysRevLett.113.261603}{{\em Phys. Rev.
  Lett.} {\bf 113} (2014) no. 26, 261603},
\href{http://arxiv.org/abs/1410.0354}{{ arXiv:1410.0354 [hep-th]}}.
%%CITATION = ARXIV:1410.0354;%%.

\bibitem{Bern:2014kca}
Z.~Bern, E.~Herrmann, S.~Litsey, J.~Stankowicz, and J.~Trnka, ``{Logarithmic
  Singularities and Maximally Supersymmetric Amplitudes},''
  \href{http://dx.doi.org/10.1007/JHEP06(2015)202}{{\em JHEP} {\bf 06} (2015)
  202},
\href{http://arxiv.org/abs/1412.8584}{{ arXiv:1412.8584 [hep-th]}}.
%%CITATION = ARXIV:1412.8584;%%.

\bibitem{ArkaniHamed:2010gh}
N.~Arkani-Hamed, J.~L. Bourjaily, F.~Cachazo, and J.~Trnka, ``{Local Integrals
  for Planar Scattering Amplitudes},''
  \href{http://dx.doi.org/10.1007/JHEP06(2012)125}{{\em JHEP} {\bf 1206} (2012)
   125},
\href{http://arxiv.org/abs/1012.6032}{{ arXiv:1012.6032 [hep-th]}}.
%%CITATION = ARXIV:1012.6032;%%.

\bibitem{Cachazo:2008vp}
F.~Cachazo, ``{Sharpening The Leading Singularity},''
\href{http://arxiv.org/abs/0803.1988}{{ arXiv:0803.1988 [hep-th]}}.
%%CITATION = 0803.1988;%%.

\bibitem{Bourjaily:2018omh}
J.~L. Bourjaily, E.~Herrmann, and J.~Trnka, ``{Amplitudes at Infinity},''
  \href{http://dx.doi.org/10.1103/PhysRevD.99.066006}{{\em Phys. Rev.} {\bf
  D99} (2019) no. 6, 066006},
\href{http://arxiv.org/abs/1812.11185}{{ arXiv:1812.11185 [hep-th]}}.
%%CITATION = ARXIV:1812.11185;%%.

\bibitem{Abreu:2018aqd}
S.~Abreu, L.~J. Dixon, E.~Herrmann, B.~Page, and M.~Zeng, ``{The Two-Loop
  Five-Point Amplitude in $\mathcal{N}\!=\!4$ super-Yang-Mills Theory},''
  \href{http://dx.doi.org/10.1103/PhysRevLett.122.121603}{{\em Phys. Rev.
  Lett.} {\bf 122} (2019) no. 12, 121603},
\href{http://arxiv.org/abs/1812.08941}{{ arXiv:1812.08941 [hep-th]}}.
%%CITATION = ARXIV:1812.08941;%%.

\bibitem{Chicherin:2018old}
D.~Chicherin, T.~Gehrmann, J.~M. Henn, P.~Wasser, Y.~Zhang, and S.~Zoia, ``{All
  Master Integrals for Three-Jet Production at Next-to-Next-to-Leading
  Order},'' \href{http://dx.doi.org/10.1103/PhysRevLett.123.041603}{{\em Phys.
  Rev. Lett.} {\bf 123} (2019) no. 4, 041603},
\href{http://arxiv.org/abs/1812.11160}{{ arXiv:1812.11160 [hep-ph]}}.
%%CITATION = ARXIV:1812.11160;%%.

\bibitem{Chicherin:2018yne}
D.~Chicherin, T.~Gehrmann, J.~M. Henn, P.~Wasser, Y.~Zhang, and S.~Zoia,
  ``{Analytic Result for a Two-Loop Five-Particle Amplitude},''
  \href{http://dx.doi.org/10.1103/PhysRevLett.122.121602}{{\em Phys. Rev.
  Lett.} {\bf 122} (2019) no. 12, 121602},
\href{http://arxiv.org/abs/1812.11057}{{ arXiv:1812.11057 [hep-th]}}.
%%CITATION = ARXIV:1812.11057;%%.

\bibitem{Chicherin:2019xeg}
D.~Chicherin, T.~Gehrmann, J.~M. Henn, P.~Wasser, Y.~Zhang, and S.~Zoia, ``{The
  Two-Loop Five-Particle Amplitude in $\mathcal{N}\!=\!8$ Supergravity},''
  \href{http://dx.doi.org/10.1007/JHEP03(2019)115}{{\em JHEP} {\bf 03} (2019)
  115},
\href{http://arxiv.org/abs/1901.05932}{{ arXiv:1901.05932 [hep-th]}}.
%%CITATION = ARXIV:1901.05932;%%.

\bibitem{Abreu:2019rpt}
S.~Abreu, L.~J. Dixon, E.~Herrmann, B.~Page, and M.~Zeng, ``{The Two-Loop
  Five-Point Amplitude in $ \mathcal{N}\!=\!8$ Supergravity},''
  \href{http://dx.doi.org/10.1007/JHEP03(2019)123}{{\em JHEP} {\bf 03} (2019)
  123},
\href{http://arxiv.org/abs/1901.08563}{{ arXiv:1901.08563 [hep-th]}}.
%%CITATION = ARXIV:1901.08563;%%.

\bibitem{Henn:2013pwa}
J.~M. Henn, ``{Multiloop Integrals in Dimensional Regularization Made
  Simple},'' \href{http://dx.doi.org/10.1103/PhysRevLett.110.251601}{{\em Phys.
  Rev. Lett.} {\bf 110} (2013) no. 25, 251601},
\href{http://arxiv.org/abs/1304.1806}{{ arXiv:1304.1806 [hep-th]}}.
%%CITATION = ARXIV:1304.1806;%%.

\bibitem{Henn:2014qga}
J.~M. Henn, ``{Lectures on Differential Equations for Feynman Integrals},''
  \href{http://dx.doi.org/10.1088/1751-8113/48/15/153001}{{\em J. Phys.} {\bf
  A48} (2015) no. 15, 153001},
\href{http://arxiv.org/abs/1412.2296}{{ arXiv:1412.2296 [hep-ph]}}.
%%CITATION = ARXIV:1412.2296;%%.

\bibitem{Badger:2017jhb}
S.~Badger, C.~Br{\o}nnum-Hansen, H.~B. Hartanto, and T.~Peraro, ``{First Look
  at Two-Loop Five-Gluon Scattering in QCD},''
  \href{http://dx.doi.org/10.1103/PhysRevLett.120.092001}{{\em Phys. Rev.
  Lett.} {\bf 120} (2018) no. 9, 092001},
\href{http://arxiv.org/abs/1712.02229}{{ arXiv:1712.02229 [hep-ph]}}.
%%CITATION = ARXIV:1712.02229;%%.

\bibitem{Abreu:2017hqn}
S.~Abreu, F.~Febres~Cordero, H.~Ita, B.~Page, and M.~Zeng, ``{Planar Two-Loop
  Five-Gluon Amplitudes from Numerical Unitarity},''
  \href{http://dx.doi.org/10.1103/PhysRevD.97.116014}{{\em Phys. Rev.} {\bf
  D97} (2018) no. 11, 116014},
\href{http://arxiv.org/abs/1712.03946}{{ arXiv:1712.03946 [hep-ph]}}.
%%CITATION = ARXIV:1712.03946;%%.

\bibitem{Abreu:2018jgq}
S.~Abreu, F.~Febres~Cordero, H.~Ita, B.~Page, and V.~Sotnikov, ``{Planar
  Two-Loop Five-Parton Amplitudes from Numerical Unitarity},''
  \href{http://dx.doi.org/10.1007/JHEP11(2018)116}{{\em JHEP} {\bf 11} (2018)
  116},
\href{http://arxiv.org/abs/1809.09067}{{ arXiv:1809.09067 [hep-ph]}}.
%%CITATION = ARXIV:1809.09067;%%.

\bibitem{Gehrmann:2015bfy}
T.~Gehrmann, J.~M. Henn, and N.~A. Lo~Presti, ``{Analytic Form of the Two-Loop
  Planar Five-Gluon All-Plus-Helicity Amplitude in QCD},''
  \href{http://dx.doi.org/10.1103/PhysRevLett.116.189903,
  10.1103/PhysRevLett.116.062001}{{\em Phys. Rev. Lett.} {\bf 116} (2016) no.
  6, 062001}, \href{http://arxiv.org/abs/1511.05409}{{ arXiv:1511.05409
  [hep-ph]}}.
%%CITATION = ARXIV:1511.05409;%%.

\bibitem{Badger:2018enw}
S.~Badger, C.~Br{\o}nnum-Hansen, H.~B. Hartanto, and T.~Peraro, ``{Analytic
  Helicity Amplitudes for Two-Loop Five-Gluon Scattering: the Single-Minus
  Case},'' \href{http://dx.doi.org/10.1007/JHEP01(2019)186}{{\em JHEP} {\bf 01}
  (2019)  186},
\href{http://arxiv.org/abs/1811.11699}{{ arXiv:1811.11699 [hep-ph]}}.
%%CITATION = ARXIV:1811.11699;%%.

\bibitem{Abreu:2018zmy}
S.~Abreu, J.~Dormans, F.~Febres~Cordero, H.~Ita, and B.~Page, ``{Analytic Form
  of Planar Two-Loop Five-Gluon Scattering Amplitudes in QCD},''
  \href{http://dx.doi.org/10.1103/PhysRevLett.122.082002}{{\em Phys. Rev.
  Lett.} {\bf 122} (2019) no. 8, 082002},
\href{http://arxiv.org/abs/1812.04586}{{ arXiv:1812.04586 [hep-ph]}}.
%%CITATION = ARXIV:1812.04586;%%.

\bibitem{Bern:2017gdk}
Z.~Bern, M.~Enciso, H.~Ita, and M.~Zeng, ``{Dual Conformal Symmetry,
  Integration-by-Parts Reduction, Differential Equations and the Nonplanar
  Sector},'' \href{http://dx.doi.org/10.1103/PhysRevD.96.096017}{{\em Phys.
  Rev.} {\bf D96} (2017) no. 9, 096017},
\href{http://arxiv.org/abs/1709.06055}{{ arXiv:1709.06055 [hep-th]}}.
%%CITATION = ARXIV:1709.06055;%%.

\bibitem{Bern:2018oao}
Z.~Bern, M.~Enciso, C.-H. Shen, and M.~Zeng, ``{Dual Conformal Structure Beyond
  the Planar Limit},''
  \href{http://dx.doi.org/10.1103/PhysRevLett.121.121603}{{\em Phys. Rev.
  Lett.} {\bf 121} (2018) no. 12, 121603},
\href{http://arxiv.org/abs/1806.06509}{{ arXiv:1806.06509 [hep-th]}}.
%%CITATION = ARXIV:1806.06509;%%.

\bibitem{Chicherin:2018wes}
D.~Chicherin, J.~M. Henn, and E.~Sokatchev, ``{Implications of Nonplanar Dual
  Conformal Symmetry},'' \href{http://dx.doi.org/10.1007/JHEP09(2018)012}{{\em
  JHEP} {\bf 09} (2018)  012},
\href{http://arxiv.org/abs/1807.06321}{{ arXiv:1807.06321 [hep-th]}}.
%%CITATION = ARXIV:1807.06321;%%.

\bibitem{Chicherin:2017dob}
D.~Chicherin, J.~Henn, and V.~Mitev, ``{Bootstrapping Pentagon Functions},''
\href{http://arxiv.org/abs/1712.09610}{{ arXiv:1712.09610 [hep-th]}}.
%%CITATION = ARXIV:1712.09610;%%.

\bibitem{Dixon:1996wi}
L.~J. Dixon, ``{Calculating Scattering Amplitudes Efficiently},''
\href{http://arxiv.org/abs/hep-ph/9601359}{{ arXiv:hep-ph/9601359}}.
%%CITATION = HEP-PH/9601359;%%.

\bibitem{Bern:2011qt}
Z.~Bern and Y.-t. Huang, ``{Basics of Generalized Unitarity},''
  \href{http://dx.doi.org/10.1088/1751-8113/44/45/454003}{{\em J. Phys.} {\bf
  A44} (2011)  454003},
\href{http://arxiv.org/abs/1103.1869}{{ arXiv:1103.1869 [hep-th]}}.
%%CITATION = ARXIV:1103.1869;%%.

\bibitem{Elvang:2013cua}
H.~Elvang and Y.-t. Huang, ``{Scattering Amplitudes},''
\href{http://arxiv.org/abs/1308.1697}{{ arXiv:1308.1697 [hep-th]}}.
%%CITATION = ARXIV:1308.1697;%%.

\bibitem{Henn:2014yza}
J.~M. Henn and J.~C. Plefka, ``{Scattering Amplitudes in Gauge Theories},''
\href{http://dx.doi.org/978-3-642-54021-9, 10.1007/978-3-642-54022-6}{{\em
  Lect. Notes Phys.} {\bf 883} (2014)  1--195}.
%%CITATION = LNPHA,883,1;%%.

\bibitem{Dixon:2015der}
L.~J. Dixon, \href{http://dx.doi.org/10.1142/9789814678766_0002}{``{A Brief
  Introduction to Modern Amplitude Methods},''} in {\em {Proceedings,
  Theoretical Advanced Study Institute in Elementary Particle Physics: Journeys
  Through the Precision Frontier: Amplitudes for Colliders (TASI 2014):
  Boulder, Colorado, June 2-27, 2014}}, pp.~39--97.
\newblock
2015.
\newblock
%%CITATION = INSPIRE-1407759;%%.

\bibitem{Cutkosky:1960sp}
R.~E. Cutkosky, ``{Singularities and Discontinuities of Feynman Amplitudes},''
\href{http://dx.doi.org/10.1063/1.1703676}{{\em J. Math. Phys.} {\bf 1} (1960)
  429--433}.
%%CITATION = JMAPA,1,429;%%.

\bibitem{Baadsgaard:2015twa}
C.~Baadsgaard, N.~E.~J. Bjerrum-Bohr, J.~L. Bourjaily, S.~Caron-Huot, P.~H.
  Damgaard, and B.~Feng, ``{New Representations of the Perturbative
  $S$-Matrix},'' \href{http://dx.doi.org/10.1103/PhysRevLett.116.061601}{{\em
  Phys. Rev. Lett.} {\bf 116} (2016) no. 6, 061601},
\href{http://arxiv.org/abs/1509.02169}{{ arXiv:1509.02169 [hep-th]}}.
%%CITATION = ARXIV:1509.02169;%%.

\bibitem{ArkaniHamed:book}
N.~Arkani-Hamed, J.~L. Bourjaily, F.~Cachazo, A.~B. Goncharov, A.~Postnikov,
  and J.~Trnka, {\em {Grassmannian Geometry of Scattering Amplitudes}}.
\newblock Cambridge University Press, 2016.

\bibitem{Dixon:2010ik}
L.~J. Dixon, J.~M. Henn, J.~Plefka, and T.~Schuster, ``{All Tree-Level
  Amplitudes in Massless QCD},''
  \href{http://dx.doi.org/10.1007/JHEP01(2011)035}{{\em JHEP} {\bf 1101} (2011)
   035},
\href{http://arxiv.org/abs/1010.3991}{{ arXiv:1010.3991 [hep-ph]}}.
%%CITATION = ARXIV:1010.3991;%%.

\bibitem{Bourjaily:2010wh}
J.~L. Bourjaily, ``{Efficient Tree-Amplitudes in $\mathcal{N}\!=\!4$: Automatic
  BCFW Recursion in {\sc Mathematica}},''
\href{http://arxiv.org/abs/1011.2447}{{ arXiv:1011.2447 [hep-ph]}}.
%%CITATION = ARXIV:1011.2447;%%.

\bibitem{Bourjaily:2012gy}
J.~L. Bourjaily, ``{Positroids, Plabic Graphs, and Scattering Amplitudes in
  {\sc Mathematica}},''
\href{http://arxiv.org/abs/1212.6974}{{ arXiv:1212.6974 [hep-th]}}.
%%CITATION = ARXIV:1212.6974;%%.

\bibitem{Ochirov:2016ewn}
A.~Ochirov and B.~Page, ``{Full Colour for Loop Amplitudes in Yang-Mills
  Theory},'' \href{http://dx.doi.org/10.1007/JHEP02(2017)100}{{\em JHEP} {\bf
  02} (2017)  100},
\href{http://arxiv.org/abs/1612.04366}{{ arXiv:1612.04366 [hep-ph]}}.
%%CITATION = ARXIV:1612.04366;%%.

\bibitem{Ochirov:2019mtf}
A.~Ochirov and B.~Page, ``{Multi-Quark Colour Decompositions from Unitarity},''
\href{http://arxiv.org/abs/1908.02695}{{ arXiv:1908.02695 [hep-ph]}}.
%%CITATION = ARXIV:1908.02695;%%.

\bibitem{Buchbinder:2005wp}
E.~I. Buchbinder and F.~Cachazo, ``{Two-Loop Amplitudes of Gluons and Octa-Cuts
  in $\mathcal{N}\!=\!4$ super Yang-Mills},''
  \href{http://dx.doi.org/10.1088/1126-6708/2005/11/036}{{\em JHEP} {\bf 0511}
  (2005)  036},
\href{http://arxiv.org/abs/hep-th/0506126}{{ arXiv:hep-th/0506126}}.
%%CITATION = HEP-TH/0506126;%%.

\bibitem{Arkani-Hamed:2014bca}
N.~Arkani-Hamed, J.~L. Bourjaily, F.~Cachazo, A.~Postnikov, and J.~Trnka,
  ``{On-Shell Structures of MHV Amplitudes Beyond the Planar Limit},''
  \href{http://dx.doi.org/10.1007/JHEP06(2015)179}{{\em JHEP} {\bf 06} (2015)
  179},
\href{http://arxiv.org/abs/1412.8475}{{ arXiv:1412.8475 [hep-th]}}.
%%CITATION = ARXIV:1412.8475;%%.

\bibitem{Elvang:2014fja}
H.~Elvang, Y.-t. Huang, C.~Keeler, T.~Lam, T.~M. Olson, S.~B. Roland, and D.~E.
  Speyer, ``{Grassmannians for Scattering Amplitudes in $4d$
  $\mathcal{N}\!=\!4$ SYM and $3d$ ABJM},''
  \href{http://dx.doi.org/10.1007/JHEP12(2014)181}{{\em JHEP} {\bf 1412} (2014)
   181},
\href{http://arxiv.org/abs/1410.0621}{{ arXiv:1410.0621 [hep-th]}}.
%%CITATION = ARXIV:1410.0621;%%.

\bibitem{Huang:2013owa}
Y.-T. Huang and C.~Wen, ``{ABJM Amplitudes and the Positive Orthogonal
  Grassmannian},'' \href{http://dx.doi.org/10.1007/JHEP02(2014)104}{{\em JHEP}
  {\bf 1402} (2014)  104},
\href{http://arxiv.org/abs/1309.3252}{{ arXiv:1309.3252 [hep-th]}}.
%%CITATION = ARXIV:1309.3252;%%.

\bibitem{Bourjaily:2016mnp}
J.~L. Bourjaily, S.~Franco, D.~Galloni, and C.~Wen, ``{Stratifying On-Shell
  Cluster Varieties: the Geometry of Non-Planar On-Shell Diagrams},''
  \href{http://dx.doi.org/10.1007/JHEP10(2016)003}{{\em JHEP} {\bf 10} (2016)
  003},
\href{http://arxiv.org/abs/1607.01781}{{ arXiv:1607.01781 [hep-th]}}.
%%CITATION = ARXIV:1607.01781;%%.

\bibitem{Herrmann:2016qea}
E.~Herrmann and J.~Trnka, ``{Gravity On-Shell Diagrams},''
  \href{http://dx.doi.org/10.1007/JHEP11(2016)136}{{\em JHEP} {\bf 11} (2016)
  136},
\href{http://arxiv.org/abs/1604.03479}{{ arXiv:1604.03479 [hep-th]}}.
%%CITATION = ARXIV:1604.03479;%%.

\bibitem{Ita:2015tya}
H.~Ita, ``{Two-Loop Integrand Decomposition into Master Integrals and Surface
  Terms},'' \href{http://dx.doi.org/10.1103/PhysRevD.94.116015}{{\em Phys.
  Rev.} {\bf D94} (2016) no. 11, 116015},
\href{http://arxiv.org/abs/1510.05626}{{ arXiv:1510.05626 [hep-th]}}.
%%CITATION = ARXIV:1510.05626;%%.

\bibitem{Georgoudis:2015hca}
A.~Georgoudis and Y.~Zhang, ``{Two-loop Integral Reduction from Elliptic and
  Hyperelliptic Curves},''
  \href{http://dx.doi.org/10.1007/JHEP12(2015)086}{{\em JHEP} {\bf 12} (2015)
  086},
\href{http://arxiv.org/abs/1507.06310}{{ arXiv:1507.06310 [hep-th]}}.
%%CITATION = ARXIV:1507.06310;%%.

\bibitem{Bern:2006ew}
Z.~Bern, M.~Czakon, L.~J. Dixon, D.~A. Kosower, and V.~A. Smirnov, ``{The
  Four-Loop Planar Amplitude and Cusp Anomalous Dimension in Maximally
  Supersymmetric Yang-Mills Theory},''
  \href{http://dx.doi.org/10.1103/PhysRevD.75.085010}{{\em Phys. Rev.} {\bf
  D75} (2007)  085010},
\href{http://arxiv.org/abs/hep-th/0610248}{{ arXiv:hep-th/0610248 [hep-th]}}.
%%CITATION = HEP-TH/0610248;%%.

\bibitem{Bern:1996je}
Z.~Bern, L.~J. Dixon, and D.~A. Kosower, ``{Progress in One-Loop QCD
  Computations},'' \href{http://dx.doi.org/10.1146/annurev.nucl.46.1.109}{{\em
  Ann. Rev. Nucl. Part. Sci.} {\bf 46} (1996)  109--148},
\href{http://arxiv.org/abs/hep-ph/9602280}{{ arXiv:hep-ph/9602280}}.
%%CITATION = HEP-PH/9602280;%%.

\bibitem{Anastasiou:2006jv}
C.~Anastasiou, R.~Britto, B.~Feng, Z.~Kunszt, and P.~Mastrolia,
  ``{$D$-Dimensional Unitarity Cut Method},''
  \href{http://dx.doi.org/10.1016/j.physletb.2006.12.022}{{\em Phys. Lett.}
  {\bf B645} (2007)  213--216},
\href{http://arxiv.org/abs/hep-ph/0609191}{{ arXiv:hep-ph/0609191}}.
%%CITATION = HEP-PH/0609191;%%.

\bibitem{Herrmann:2019upk}
E.~Herrmann and J.~Parra-Martinez, ``{Logarithmic Forms and Differential
  Equations for Feynman Integrals},''
\href{http://arxiv.org/abs/1909.04777}{{ arXiv:1909.04777 [hep-th]}}.
%%CITATION = ARXIV:1909.04777;%%.

\bibitem{CaronHuot:2012ab}
S.~Caron-Huot and K.~J. Larsen, ``{Uniqueness of Two-Loop Master Contours},''
  \href{http://dx.doi.org/10.1007/JHEP10(2012)026}{{\em JHEP} {\bf 1210} (2012)
   026},
\href{http://arxiv.org/abs/1205.0801}{{ arXiv:1205.0801 [hep-ph]}}.
%%CITATION = ARXIV:1205.0801;%%.

\bibitem{Bourjaily:2017bsb}
J.~L. Bourjaily, A.~J. McLeod, M.~Spradlin, M.~von Hippel, and M.~Wilhelm,
  ``{Elliptic Double-Box Integrals: Massless Scattering Amplitudes beyond
  Polylogarithms},''
  \href{http://dx.doi.org/10.1103/PhysRevLett.120.121603}{{\em Phys. Rev.
  Lett.} {\bf 120} (2018) no. 12, 121603},
\href{http://arxiv.org/abs/1712.02785}{{ arXiv:1712.02785 [hep-th]}}.
%%CITATION = ARXIV:1712.02785;%%.

\bibitem{Adams:2013kgc}
L.~Adams, C.~Bogner, and S.~Weinzierl, ``{The Two-Loop Sunrise Graph with
  Arbitrary Masses},'' \href{http://dx.doi.org/10.1063/1.4804996}{{\em J. Math.
  Phys.} {\bf 54} (2013)  052303},
\href{http://arxiv.org/abs/1302.7004}{{ arXiv:1302.7004 [hep-ph]}}.
%%CITATION = ARXIV:1302.7004;%%.

\bibitem{Bloch:2014qca}
S.~Bloch, M.~Kerr, and P.~Vanhove, ``{A Feynman Integral via Higher Normal
  Functions},'' \href{http://dx.doi.org/10.1112/S0010437X15007472}{{\em Compos.
  Math.} {\bf 151} (2015) no. 12, 2329--2375},
\href{http://arxiv.org/abs/1406.2664}{{ arXiv:1406.2664 [hep-th]}}.
%%CITATION = ARXIV:1406.2664;%%.

\bibitem{Bloch:2016izu}
S.~Bloch, M.~Kerr, and P.~Vanhove, ``{Local Mirror Symmetry and the Sunset
  Feynman Integral},''
  \href{http://dx.doi.org/10.4310/ATMP.2017.v21.n6.a1}{{\em Adv. Theor. Math.
  Phys.} {\bf 21} (2017)  1373--1453},
\href{http://arxiv.org/abs/1601.08181}{{ arXiv:1601.08181 [hep-th]}}.
%%CITATION = ARXIV:1601.08181;%%.

\bibitem{Adams:2016xah}
L.~Adams, C.~Bogner, A.~Schweitzer, and S.~Weinzierl, ``{The Kite Integral to
  All Orders in Terms of Elliptic Polylogarithms},''
  \href{http://dx.doi.org/10.1063/1.4969060}{{\em J. Math. Phys.} {\bf 57}
  (2016) no. 12, 122302},
\href{http://arxiv.org/abs/1607.01571}{{ arXiv:1607.01571 [hep-ph]}}.
%%CITATION = ARXIV:1607.01571;%%.

\bibitem{Hidding:2017jkk}
M.~Hidding and F.~Moriello, ``{All Orders Structure and Efficient Computation
  of Linearly-Reducible Elliptic Feynman Integrals},''
  \href{http://dx.doi.org/10.1007/JHEP01(2019)169}{{\em JHEP} {\bf 01} (2019)
  169},
\href{http://arxiv.org/abs/1712.04441}{{ arXiv:1712.04441 [hep-ph]}}.
%%CITATION = ARXIV:1712.04441;%%.

\bibitem{Adams:2018bsn}
L.~Adams, E.~Chaubey, and S.~Weinzierl, ``{Planar Double Box Integral for Top
  Pair Production with a Closed Top Loop to all orders in the Dimensional
  Regularization Parameter},''
  \href{http://dx.doi.org/10.1103/PhysRevLett.121.142001}{{\em Phys. Rev.
  Lett.} {\bf 121} (2018) no. 14, 142001},
\href{http://arxiv.org/abs/1804.11144}{{ arXiv:1804.11144 [hep-ph]}}.
%%CITATION = ARXIV:1804.11144;%%.

\bibitem{Broedel:2019hyg}
J.~Br\"{o}del, C.~Duhr, F.~Dulat, B.~Penante, and L.~Tancredi, ``{Elliptic
  Polylogarithms and Feynman Parameter Integrals},''
  \href{http://dx.doi.org/10.1007/JHEP05(2019)120}{{\em JHEP} {\bf 05} (2019)
  120},
\href{http://arxiv.org/abs/1902.09971}{{ arXiv:1902.09971 [hep-ph]}}.
%%CITATION = ARXIV:1902.09971;%%.

\bibitem{Broedel:2019kmn}
J.~Br\"{o}del, C.~Duhr, F.~Dulat, R.~Marzucca, B.~Penante, and L.~Tancredi,
  ``{An Analytic Solution for the Equal-Mass Banana Graph},''
\href{http://arxiv.org/abs/1907.03787}{{ arXiv:1907.03787 [hep-th]}}.
%%CITATION = ARXIV:1907.03787;%%.

\bibitem{Sabry}
A.~Sabry, ``{Fourth Order Spectral Functions for the Electron Propagator},''
  {\em Nucl. Phys.} {\bf 33} (1962) no. 17, 401--430.

\bibitem{Broadhurst:1987ei}
D.~J. Broadhurst, ``{The Master Two Loop Diagram With Masses},''
\href{http://dx.doi.org/10.1007/BF01551921}{{\em Z. Phys.} {\bf C47} (1990)
  115--124}.
%%CITATION = ZEPYA,C47,115;%%.

\bibitem{Laporta:2004rb}
S.~Laporta and E.~Remiddi, ``{Analytic Treatment of the Two-Loop Equal Mass
  Sunrise Graph},''
  \href{http://dx.doi.org/10.1016/j.nuclphysb.2004.10.044}{{\em Nucl. Phys.}
  {\bf B704} (2005)  349--386},
\href{http://arxiv.org/abs/hep-ph/0406160}{{ arXiv:hep-ph/0406160}}.
%%CITATION = HEP-PH/0406160;%%.

\bibitem{Czakon:2008ii}
M.~Czakon and A.~Mitov, ``{Inclusive Heavy Flavor Hadroproduction in NLO QCD:
  The Exact Analytic Result},''
  \href{http://dx.doi.org/10.1016/j.nuclphysb.2009.08.020}{{\em Nucl. Phys.}
  {\bf B824} (2010)  111--135},
\href{http://arxiv.org/abs/0811.4119}{{ arXiv:0811.4119 [hep-ph]}}.
%%CITATION = ARXIV:0811.4119;%%.

\bibitem{Bourjaily:2018ycu}
J.~L. Bourjaily, Y.-H. He, A.~J. Mcleod, M.~Von~Hippel, and M.~Wilhelm,
  ``{Traintracks Through Calabi-Yaus: Amplitudes Beyond Elliptic
  Polylogarithms},''
  \href{http://dx.doi.org/10.1103/PhysRevLett.121.071603}{{\em Phys. Rev.
  Lett.} {\bf 121} (2018) no. 7, 071603},
\href{http://arxiv.org/abs/1805.09326}{{ arXiv:1805.09326 [hep-th]}}.
%%CITATION = ARXIV:1805.09326;%%.

\bibitem{Bourjaily:2018yfy}
J.~L. Bourjaily, A.~J. McLeod, M.~von Hippel, and M.~Wilhelm, ``{A (Bounded)
  Bestiary of Feynman Integral Calabi-Yau Geometries},''
  \href{http://dx.doi.org/10.1103/PhysRevLett.122.031601}{{\em Phys. Rev.
  Lett.} {\bf 122} (2019) no. 3, 031601},
\href{http://arxiv.org/abs/1810.07689}{{ arXiv:1810.07689 [hep-th]}}.
%%CITATION = ARXIV:1810.07689;%%.

\bibitem{Gehrmann:2001ck}
T.~Gehrmann and E.~Remiddi, ``{Two-Loop Master Integrals for
  $\gamma^*\!\!\to\!3$ Jets: the Nonplanar Topologies},''
  \href{http://dx.doi.org/10.1016/S0550-3213(01)00074-8}{{\em Nucl. Phys.} {\bf
  B601} (2001)  287--317},
\href{http://arxiv.org/abs/hep-ph/0101124}{{ arXiv:hep-ph/0101124}}.
%%CITATION = HEP-PH/0101124;%%.

\bibitem{Gehrmann:1999as}
T.~Gehrmann and E.~Remiddi, ``{Differential Equations for Two-Loop Four-Point
  Functions},'' \href{http://dx.doi.org/10.1016/S0550-3213(00)00223-6}{{\em
  Nucl. Phys.} {\bf B580} (2000)  485--518},
\href{http://arxiv.org/abs/hep-ph/9912329}{{ arXiv:hep-ph/9912329 [hep-ph]}}.
%%CITATION = HEP-PH/9912329;%%.

\bibitem{Badger:2013gxa}
S.~Badger, H.~Frellesvig, and Y.~Zhang, ``{A Two-Loop Five-Gluon Helicity
  Amplitude in QCD},'' \href{http://dx.doi.org/10.1007/JHEP12(2013)045}{{\em
  JHEP} {\bf 12} (2013)  045},
\href{http://arxiv.org/abs/1310.1051}{{ arXiv:1310.1051 [hep-ph]}}.
%%CITATION = ARXIV:1310.1051;%%.

\bibitem{Badger:2015lda}
S.~Badger, G.~Mogull, A.~Ochirov, and D.~O'Connell, ``{A Complete Two-Loop,
  Five-Gluon Helicity Amplitude in Yang-Mills Theory},''
  \href{http://dx.doi.org/10.1007/JHEP10(2015)064}{{\em JHEP} {\bf 10} (2015)
  064},
\href{http://arxiv.org/abs/1507.08797}{{ arXiv:1507.08797 [hep-ph]}}.
%%CITATION = ARXIV:1507.08797;%%.

\bibitem{Badger:2016ozq}
S.~Badger, G.~Mogull, and T.~Peraro, ``{Local Integrands for Two-Loop All-Plus
  Yang-Mills Amplitudes},''
  \href{http://dx.doi.org/10.1007/JHEP08(2016)063}{{\em JHEP} {\bf 08} (2016)
  063},
\href{http://arxiv.org/abs/1606.02244}{{ arXiv:1606.02244 [hep-ph]}}.
%%CITATION = ARXIV:1606.02244;%%.

\bibitem{Drummond:2010mb}
J.~M. Drummond and J.~M. Henn, ``{Simple Loop Integrals and Amplitudes in
  $\mathcal{N}\!=\!4$ SYM},''
  \href{http://dx.doi.org/10.1007/JHEP05(2011)105}{{\em JHEP} {\bf 1105} (2011)
   105},
\href{http://arxiv.org/abs/1008.2965}{{ arXiv:1008.2965 [hep-th]}}.
%%CITATION = ARXIV:1008.2965;%%.

\bibitem{GriffithsHarris}
P.~Griffiths and J.~Harris, {\em {Principles of Algebraic Geometry}}.
\newblock Wiley Classics Library. John Wiley \& Sons Inc., New York, 1978.

\bibitem{Anastasiou:2018rib}
C.~Anastasiou and G.~Sterman, ``{Removing infrared divergences from two-loop
  integrals},'' \href{http://dx.doi.org/10.1007/JHEP07(2019)056}{{\em JHEP}
  {\bf 07} (2019)  056},
\href{http://arxiv.org/abs/1812.03753}{{ arXiv:1812.03753 [hep-ph]}}.
%%CITATION = ARXIV:1812.03753;%%.

\bibitem{Edison:2019ovj}
A.~Edison, E.~Herrmann, J.~Parra-Martinez, and J.~Trnka, ``{Gravity Loop
  Integrands from the Ultraviolet},''
\href{http://arxiv.org/abs/1909.02003}{{ arXiv:1909.02003 [hep-th]}}.
%%CITATION = ARXIV:1909.02003;%%.

\bibitem{vanderWaerden:1929}
B.~L. van~der Waerden, ``{Spinoranalyse},'' {\em Nach. Ges. Wiss. G\"{o}ttingen
  Math.-Phys.} {\bf 1} (1929)  100--109.

\bibitem{DelDuca:1999rs}
V.~Del~Duca, L.~J. Dixon, and F.~Maltoni, ``{New Color Decompositions for Gauge
  Amplitudes at Tree and Loop Level},''
  \href{http://dx.doi.org/10.1016/S0550-3213(99)00809-3}{{\em Nucl. Phys.} {\bf
  B571} (2000)  51--70},
\href{http://arxiv.org/abs/hep-ph/9910563}{{ arXiv:hep-ph/9910563 [hep-ph]}}.
%%CITATION = HEP-PH/9910563;%%.

\bibitem{Hodges:2011wm}
A.~Hodges, ``{New Expressions for Gravitational Scattering Amplitudes},''
  \href{http://dx.doi.org/10.1007/JHEP07(2013)075}{{\em JHEP} {\bf 07} (2013)
  075},
\href{http://arxiv.org/abs/1108.2227}{{ arXiv:1108.2227 [hep-th]}}.
%%CITATION = ARXIV:1108.2227;%%.

\bibitem{Hodges:2012ym}
A.~Hodges, ``{A Simple Formula for Gravitational MHV Amplitudes},''
\href{http://arxiv.org/abs/1204.1930}{{ arXiv:1204.1930 [hep-th]}}.
%%CITATION = ARXIV:1204.1930;%%.

\end{thebibliography}
%\clearpage
%
%\end{document}

\providecommand{\href}[2]{#2}\begingroup\raggedright\endgroup
\end{document}